\documentclass[aps,prd,preprint,groupedaddress,showpacs,superscriptaddress,floatfix]{revtex4-1}

\usepackage{amsmath}
\usepackage{graphicx}
\usepackage{amssymb}
\usepackage{bm}
\usepackage{array}
\usepackage[colorlinks,linkcolor=blue,urlcolor=blue,anchorcolor=blue,citecolor=blue]{hyperref}
\usepackage{slashed}
\usepackage{braket}

\begin{document}

\title{Quark Wigner distributions in a light-cone spectator model}

\newcommand*{\PKU}{School of Physics and State Key Laboratory of Nuclear Physics and
Technology, Peking University, Beijing 100871,
China}\affiliation{\PKU}
\newcommand*{\CICQM}{Collaborative Innovation
Center of Quantum Matter, Beijing, China}\affiliation{\CICQM}
\newcommand*{\CHEP}{Center for High Energy
Physics, Peking University, Beijing 100871,
China}\affiliation{\CHEP}

\author{Tianbo Liu}\affiliation{\PKU}
\author{Bo-Qiang Ma}\email{mabq@pku.edu.cn}\affiliation{\PKU}\affiliation{\CICQM}\affiliation{\CHEP}


\begin{abstract}

We investigate the quark Wigner distributions in a light-cone spectator model. The Wigner distribution, as a quasi-distribution function, provides the most general one-parton information in a hadron. Combining the polarization configurations, unpolarized, longitudinal polarized or transversal polarized, of the quark and the proton, we can define 16 independent Wigner distributions at leading twist. We calculate all these Wigner distributions for the $u$ quark and the $d$ quark respectively. In our calculation, both the scalar and the axial-vector spectators are included, and the Melosh-Wigner rotation effects for both the quark and the axial-vector spectator are taken into account. The results provide us a very rich picture of the quark structure in the proton.

\end{abstract}

\pacs{12.39.Ki, 14.20.Dh, 12.38.-t}

\maketitle

\section{Introduction}

One of the main goals of the particle physics is to unravel the quark and gluon structure of hadrons, which are fundamentally described by the quantum chromodynamics (QCD) in the framework of the Yang-Mills gauge theory. Due to the nonperturbative nature of the QCD at the hadron scale, it is almost impossible to calculate all the properties of hadrons directly from the QCD at present. Though the Euclidean lattice method provides a first-principle numerical simulation~\cite{Wilson:1974sk}, it is challenged by the enormous computational complexity and the unavoidable multi-hadron thresholds~\cite{Fodor:2012gf}. The Dyson--Schwinger equation method also leads to important insights on the nonperturbative properties of QCD~\cite{Cornwall:1981zr}. Besides, the remarkable connection between the quantum gauge field theory in four dimensions and the classical gravity theory in five dimensions, developed in last fifteen years and known as the holographic dual, sheds light on the confinement dynamics in QCD~\cite{Maldacena:1997re}.

The parton model formulated by Feynman and formalized by Bjorken and Paschos~\cite{Feynman:1969ej,Bjorken:1969ja} is proved successful in explaining the high energy hadronic scattering experiments. Based on the partonic picture and factorization theorem~\cite{Collins:1987pm}, a general process independent function, referred to as the parton distribution function (PDF), is defined to describe the light-cone longitudinal momentum fractions carried by the partons in a nucleon, and it is a powerful tool on data analyses of high energy hadronic scattering experiments. With the development of the polarization techniques, one is able to obtain spin-related partonic information in the nucleon from the helicity and transversity. To describe the three-dimensional structure of the nucleon, the transverse momentum dependent parton distributions (TMDs)~\cite{Collins:1981uk} and the generalized parton distributions (GPDs)~\cite{Mueller:1998fv} are introduced and proven to be useful tools. The TMD contains three-dimensional momentum distributions of the partons, while the GPD contains longitudinal momentum and two-dimensional transverse coordinates of the partons through the impact parameter dependent densities (IPDs).

As a further generalization, the Wigner distributions are defined as a position-momentum joint distributions to understand the partonic structure of the nucleon. The Wigner distribution is a kind of phase-space distribution which contains the most general one-parton information in a nucleon. In fact, it is not a new concept and has been widely applied in many physical areas, such as the quantum information, quantum molecular dynamics, optics, nonlinear dynamics and so on~\cite{Balazs:1983hk}, and is even directly measurable in some experiments~\cite{Vogel:1989zz}. The Wigner distribution was first explored in QCD as a six-dimensional (three positions and three momentums) function~\cite{Ji:2003ak,Belitsky:2003nz}, where the relativistic effects were neglected. Then a five-dimensional (two transverse positions and three momentums) Wigner distribution~\cite{Lorce:2011kd} was proposed in the light-cone formalism or in the infinite momentum frame (IMF), where the parton language is well-defined. The simple QCD vacuum structure in the light-cone form allows an unambiguous definition of the constituents of a hadron~\cite{Brodsky:1997de}. The five-dimensional Wigner distributions will reduce to the TMDs and IPDs, when the transverse coordinates or the transverse momentums are integrated. Thus possible relations between the TMDs and GPDs may be established via the Wigner distributions. Unlike the TMDs and the IPDs, the Wigner distributions do not have probability interpretations because of the Heisenberg uncertainty principle in quantum theories~\cite{aHeisenberg:1927zz}, but at twist-two they can be expressed in terms of light-cone wave functions and the semiclassical interpretations are still possible~\cite{Lorce:2011kd}.

In this paper, we focus on quark Wigner distributions in the proton. With the combination of the polarization configurations (unpolarized or polarized along three spatial directions) of the quark and the proton, 16 Wigner distributions are defined at the leading twist. We investigate all these Wigner distributions in a light-cone spectator model. Both the scalar and the axial-vector spectators are included, and the axial-vector one is necessary for flavor separation. This model is proven successful in calculating the structure functions~\cite{Feynman:1973xc}. With the Melosh-Wigner rotation effect~\cite{Wigner:1939cj} taking into account, this kind of models has been applied to investigate many physical quantities, such as quark unpolarized, helicity and transversity distributions~\cite{Ma:1996np,Ma:1997gy}, form factors~\cite{Ma:2002ir}, TMDs~\cite{Lu:2004au,She:2009jq,Bacchetta:2008af,Lu:2012gu} and GPDs~\cite{Burkardt:2003je,Chakrabarti:2005zm,Hwang:2007tb}, and the results are comparable with experiments. Therefore, the investigation of the Wigner distributions in such a simple model may still provide us some valuable information to help us understand the structure of the nucleon. The paper is organized as follows. We briefly review the light-cone spectator model in Sect. II, and then calculate the Wigner distributions in Sect. III. The numerical results are shown in Sect. IV including discussions. A summary is drawn in the last section.

\section{The light-cone spectator model}

The proton is an eigenstate of the light-cone hamiltonian $H_{\mathrm{LC}}=P^+P^--\bm{P}_\perp^2$ and the eigenvalue is the invariant mass square. In the light-cone (or front) form relativistic dynamics~\cite{Dirac:1949cp}, the fields are quantized at fixed light-cone time $\tau=t+z$ instead of the ordinary time $t$ in the instant form. The light-cone longitudinal momentum $k^+=k^0+k^3$ of any massive particle is restricted to be positive definite, and thus the Fock space vacuum state $|0\rangle$, {\it i.e.} the free vacuum, is exactly the QCD vacuum, if the possibility of the color-singlet states built on massless gluon quanta with zero momentum is ignored~\cite{Brodsky:1997de}. Therefore, one may have unambiguous definition of the constituents of the hadron. Then the proton state can be expanded on the complete basis of free multiparticle Fock states as
\begin{equation}
\begin{split}
\left|\Psi:P^+,\bm{P}_\perp,S_z\right\rangle=&\sum_{n,\{\lambda_i\}}\prod_{i=1}^N\int\frac{dx_id^2\bm{k}_{\perp i}}{2\sqrt{x_i}(2\pi)^3}(16\pi^3)\delta(1-\sum_{j=1}^N x_j)\delta^{(2)}(\sum_{j=1}^N \bm{k}_{\perp j})\\
&\times\psi_{n}(\{x_j\},\{\bm{k}_{\perp j}\},\{\lambda_j\})\left|n:\{x_iP^+\},\{x_i\bm{P}_\perp+\bm{k}_{\perp i}\},\{\lambda_i\}\right\rangle,
\end{split}
\end{equation}
where $N$ is the number of the components in the Fock state $|n\rangle$, $x_i=k_i^+/P^+$ is the light-cone longitudinal momentum fraction carried by the $i$-th component, and $\bm{k}_{\perp i}$ and $\lambda_i$ are its intrinsic transverse momentum and light-cone helicity. The projection of the proton state on the Fock state $|n\rangle$ is the so-called light-cone wave function $\psi_{n}$, which is frame independent.

The proton state is normalized as
\begin{equation}
\left\langle\Psi:P'^+,\bm{P}'_\perp,S'_z|\Psi:P^+,\bm{P}_\perp,S_z\right\rangle=2P^+(2\pi)^3\delta(P^+-P'^+)\delta^{(2)}(\bm{P}_\perp-\bm{P}'_\perp)\delta_{S_zS'_z},
\end{equation}
and the light-cone wave functions are correspondingly normalized as
\begin{equation}
\begin{split}
&\sum_{n,\{\lambda_i\}}\prod_{i=1}^N\int\frac{dx_id^2\bm{k}_{\perp i}}{2(2\pi)^3}(16\pi^3)\delta(1-\sum_{j=1}^N x_j)\delta^{(2)}(\sum_{j=1}^N \bm{k}_{\perp j})\\
&\quad\quad\quad\quad\times|\psi_{n}(\{x_j\},\{\bm{k}_{\perp j}\},\{\lambda_j\})|^2=1.
\end{split}
\end{equation}
The one-particle Fock state is defined as $|p\rangle=\sqrt{2p^+}\,a^\dagger(p)|0\rangle$ with the commutation or anticommutation relations:
\begin{equation}
[a(p),a^\dagger(p')]=\{b(p),b^\dagger(p')\}=(2\pi)^3\delta(p^+-p'^+)\delta^{(2)}(\bm{p}_\perp-\bm{p}'_\perp),
\end{equation}
where the $a(p)$, $a^\dagger(p)$, $b(p)$ and $b^\dagger(p)$ are annihilation and creation operators for bosons and fermions respectively.

In the spectator model, the proton is viewed as a struck quark and a spectator that contains the remaining part of the proton. Considering the fact that the leading term in the Fock states expansion for the proton is the valence three quark state $|uud\rangle$, we approximately regard the proton state as a valence quark and a spectator that carries the diquark quantum numbers, and some nonperturbative effects between the quarks and gluons in the spectator part from higher Fock states are effectively absorbed into the mass of the spectator. This kind of picture has been applied to study many physical observables, and is supported by some light-front holographic model~\cite{deTeramond:2013it}. However this picture is actually simplistic, and some effects cannot be realistically absorbed into the spectator mass~\cite{Cates:2011pz}. More realistic analyses will provide better descriptions of the nucleon~\cite{Roberts:2007jh,Cloet:2008re,Cloet:2012cy}. Here we only adopt this crude model to have some qualitative results.

Constained by the quantum numbers of the proton and the quark, the spectator can only be either a scalar or an axial-vector, and the latter one is necessary for flavor separation. Therefore, we express the proton state as
\begin{equation}
\left|\Psi\right\rangle=\sin\theta\,\phi_S|qS\rangle+\cos\theta\,\phi_V|qV\rangle,
\end{equation}
where $S$ and $V$ stand for the scalar and the axial-vector spectators respectively, $\phi_S$ and $\phi_V$ are the momentum space light-cone wave functions and $\theta$ is an angle to describe the $SU(6)$ spin-flavor symmetry breaking. In this paper, we choose the isospin symmetric case $\theta=\pi/4$ in our calculations. This choice is also adopted in the calculations of helicity and transversity distributions~\cite{Ma:1996np,Ma:1997gy}, form factors~\cite{Ma:2002ir} and single spin asymmetries~\cite{Zhu:2011ir}, and the results therein are comparable with the data. One might alternatively regard it as a parameter to fit data.

The spin space wave function of the quark-spectator state is acquired from the $SU(6)$ quark model in the instant form~\cite{Ma:1996np}:
\begin{equation}
\begin{split}
|qS\rangle^{\uparrow/\downarrow}=&u_T^{\uparrow/\downarrow}S(ud),\\
|qV\rangle^{\uparrow/\downarrow}=&\pm\frac{1}{3}[u_T^{\uparrow/\downarrow}V_T^0(ud)-\sqrt{2}u_T^{\downarrow/\uparrow}V_T^{\pm1}(ud)\\
&-\sqrt{2}d_T^{\uparrow/\downarrow}V_T^0(uu)+2d_T^{\downarrow/\uparrow}V_T^{\pm1}(uu)].
\end{split}
\end{equation}
Then we transform the instant form spinors to the light-cone form spinors through the Melosh--Wigner rotation~\cite{Wigner:1939cj}:
\begin{equation}\label{mw1}
\begin{split}
\chi_T^\uparrow&=w[(k^++m)\chi_F^\uparrow-(k^1+ik^2)\chi_F^\downarrow],\\
\chi_T^\downarrow&=w[(k^++m)\chi_F^\downarrow+(k^1-ik^2)\chi_F^\uparrow],
\end{split}
\end{equation}
which plays an important role in understanding the ``proton spin puzzle''~\cite{Ma:1991xq}. The factor $w=1/\sqrt{2k^+(k^0+m)}$, and the subscripts $T$ and $F$ stand for the instant form and the light-cone form spinors respectively. This transformation procedure is consistant with the results directly derived from the light-cone field theory~\cite{Xiao:2003wf}. For the spectator, the scalar one does not transform since it has spin-zero, and the axial-vector one transforms as~\cite{Ahluwalia:1993xa}
\begin{equation}\label{mw2}
\begin{split}
V_T^{+1}=&w^2[(k^++m)^2V_F^{+1}-\sqrt{2}(k^++m)(k^1+ik^2)V_F^0\\
&+(k^1+ik^2)^2V_F^{-1}],\\
V_T^{0\ }=&w^2[\sqrt{2}(k^++m)(k^1-ik^2)V_F^{+1}\\
&+2(k^+(k^0+m)-(k^1-ik^2)(k^1+ik^2))V_F^0\\
&-\sqrt{2}(k^++m)(k^1+ik^2)V_F^{-1}],\\
V_T^{-1}=&w^2[(k^1-ik^2)^2V_F^{+1}+\sqrt{2}(k^++m)(k^1-ik^2)V_F^0\\
&+(k^++m)^2V_F^{-1}].
\end{split}
\end{equation}

For the momentum space light-cone wave function, we assume the Brodsky--Huang--Lepage (BHL) prescription~\cite{Brodsky:1980vj}:
\begin{equation}\label{bhl}
\phi_D(x,\bm{k}_\perp)=A_D\exp\left\{-\frac{1}{8\beta_D^2}\left[\frac{m_q^2+\bm{k}_\perp^2}{x}+\frac{m_D^2+\bm{k}_\perp^2}{1-x}\right]\right\},
\end{equation}
where the subscript $D$ represents the type of the spectator with $S$ for the scalar one and $V$ for the axial-vector one, $m_q$ and $m_D$ are the masses of the quark and the spectator, $\beta_D$ is the harmonic oscillator scale parameter and $A_D$ is the normalization factor. We simply adopt $m=330$\,MeV, $\beta_D=330$\,MeV as often adopted in the literature~\cite{Schlumpf:1993rm,Weber:1990fx,Huang:1994dy}. In principle, the spectator masses have a spectrum that might be extracted from experiments, and some modern understandings on diquark correlations in the nucleon are explored in~\cite{Maris:2002yu}. Here, for simplicity, we regard them as parameters and choose the values as $m_S=600$\,MeV and $m_V=800$\,MeV. Apart from the BHL prescription, some other forms were proposed in literatures, such as the Terentev--Karmanov (TK) prescription~\cite{Terentev:1976jk}, Chung--Coaster--Polyzou (CCP) prescription~\cite{Chung:1988mu} and Vega--Schmidt--Gutsche--Lyubovitskij (VSGL) prescription~\cite{Vega:2013bxa}. The choice of the momentum space wave function only affects the results quantitatively, the qualitative properties are essentially determined by the spin structures.

\section{Quark Wigner distributions in the proton}

The Wigner distribution is a quantum phase-space distribution first introduced by Wigner in quantum mechanics~\cite{Wigner:1932eb}. It is in general not positive definite and has no probability interpretations because of the Heisenberg uncertainty principle in quantum theories. Whereas, at certain situation, semiclassical interpretations are still possible, and thus it can be viewed as a quasi-distribution function. Similar as the quark-quark correlation operater, the Hermitian Wigner operator for quarks at a fixed light-cone time is defined as~\cite{Lorce:2011kd}
\begin{equation}\label{wigneroperator}
\hat{W}^{[\Gamma]}(\bm{b}_\perp,\bm{k}_\perp,x)=\frac{1}{2}\int\frac{dz_+d^2\bm{z}_\perp}{(2\pi)^3}e^{ik\cdot z}
\left.\bar{\psi}(y-\frac{z}{2})\Gamma\mathcal{W}\psi(y+\frac{z}{2})\right|_{z^+=0},
\end{equation}
where $y=(0,0,\bm{b}_\perp)$, and $\mathcal{W}$ is the gauge link Wilson line connecting the points $y-z/2$ and $y+z/2$ to ensure the $SU(3)$ color gauge invariance of the Wigner operator. The $\Gamma$ represents a Dirac $\gamma$-matrix, and at twist-two it is $\gamma^+$, $\gamma^+\gamma_5$ or $i\sigma^{j+}\gamma_5$ with $j=1$ or $2$, corresponding to unpolarized, longitudinal polarized or $j$-direction transverse polarized quark respectively. The $\bm{b}_\perp$ and $\bm{k}_\perp$ are intrinsic transverse position and transverse momentum of the quark. They are not Fourier conjugate variables as demonstrated in~\cite{Lorce:2011kd}, but the $b_j$ and $k_j$ along the same direction are still protected by the uncertainty principle, since they are not commutative with each other.

Interpolating the Wigner operator (\ref{wigneroperator}) between the initial and final proton states with a transverse momentum $\bm{\Delta}_\perp$ transferred:
\begin{equation}\label{wignerdistr}
\begin{split}
\rho^{[\Gamma]}(\bm{b}_\perp,\bm{k}_\perp,x,\bm{S})
=\int\frac{d^2\bm{\Delta}_\perp}{(2\pi)^2}\Big\langle P^+,\frac{\bm{\Delta}_\perp}{2},\bm{S}\Big|\hat{W}^{[\Gamma]}(\bm{b}_\perp,\bm{k}_\perp,x)\Big|P^+,-\frac{\bm{\Delta}_\perp}{2},\bm{S}\Big\rangle,
\end{split}
\end{equation}
where $\bm{S}$ is the spin of the proton state, the five-dimensional Wigner distribution is defined. Here the Drell--Yan--West frame ($\Delta^+=0$)~\cite{Drell:1969km}, which is widely used in the calculation of form factors~\cite{Xiao:2002iv}, is adopted, and with this choice one may have semiclassical probability interpretations~\cite{Lorce:2011kd}. Including a longitudinal momentum transfer, one can define a six-dimensional Wigner distribution $\rho^{[\Gamma]}(\bm{b}_\perp,\xi,\bm{k}_\perp,x,\bm{S})$, but will lose the quasi-probability interpretation. Combining the polarization configurations, unpolarized (U), longitudinal polarized (L) and transverse polarized (T), of the proton and the quark, one can define 16 independent twist-two quark Wigner distributions. They are the unpolarized Wigner distribution
\begin{equation}\label{wuu}
\begin{split}
\rho_{_\mathrm{UU}}(\bm{b}_\perp,\bm{k}_\perp,x)
=\frac{1}{2}\left[\rho^{[\gamma^+]}(\bm{b}_\perp,\bm{k}_\perp,x,\hat{\bm{e}}_z)+\rho^{[\gamma^+]}(\bm{b}_\perp,\bm{k}_\perp,x,-\hat{\bm{e}}_z)\right],
\end{split}
\end{equation}
the unpol-longitudinal Wigner distribution
\begin{equation}\label{wul}
\begin{split}
\rho_{_\mathrm{UL}}(\bm{b}_\perp,\bm{k}_\perp,x)
=\frac{1}{2}\left[\rho^{[\gamma^+\gamma_5]}(\bm{b}_\perp,\bm{k}_\perp,x,\hat{\bm{e}}_z)+\rho^{[\gamma^+\gamma_5]}(\bm{b}_\perp,\bm{k}_\perp,x,-\hat{\bm{e}}_z)\right],
\end{split}
\end{equation}
the unpol-transverse Wigner distribution
\begin{equation}\label{wut}
\begin{split}
\rho^j_{_\mathrm{UT}}(\bm{b}_\perp,\bm{k}_\perp,x)
=\frac{1}{2}\left[\rho^{[i\sigma^{j+}\gamma_5]}(\bm{b}_\perp,\bm{k}_\perp,x,\hat{\bm{e}}_z)+\rho^{[i\sigma^{j+}\gamma_5]}(\bm{b}_\perp,\bm{k}_\perp,x,-\hat{\bm{e}}_z)\right],
\end{split}
\end{equation}
the longi-unpolarized Wigner distribution
\begin{equation}\label{wlu}
\begin{split}
\rho_{_\mathrm{LU}}(\bm{b}_\perp,\bm{k}_\perp,x)
=\frac{1}{2}\left[\rho^{[\gamma^+]}(\bm{b}_\perp,\bm{k}_\perp,x,\hat{\bm{e}}_z)-\rho^{[\gamma^+]}(\bm{b}_\perp,\bm{k}_\perp,x,-\hat{\bm{e}}_z)\right],
\end{split}
\end{equation}
the longitudinal Wigner distribution
\begin{equation}\label{wll}
\begin{split}
\rho_{_\mathrm{LL}}(\bm{b}_\perp,\bm{k}_\perp,x)
=\frac{1}{2}\left[\rho^{[\gamma^+\gamma_5]}(\bm{b}_\perp,\bm{k}_\perp,x,\hat{\bm{e}}_z)-\rho^{[\gamma^+\gamma_5]}(\bm{b}_\perp,\bm{k}_\perp,x,-\hat{\bm{e}}_z)\right],
\end{split}
\end{equation}
the longi-transverse Wigner distribution
\begin{equation}\label{wlt}
\begin{split}
\rho^j_{_\mathrm{LT}}(\bm{b}_\perp,\bm{k}_\perp,x)
=\frac{1}{2}\left[\rho^{[i\sigma^{j+}\gamma_5]}(\bm{b}_\perp,\bm{k}_\perp,x,\hat{\bm{e}}_z)-\rho^{[i\sigma^{j+}\gamma_5]}(\bm{b}_\perp,\bm{k}_\perp,x,-\hat{\bm{e}}_z)\right],
\end{split}
\end{equation}
the trans-unpolarized Wigner distribution
\begin{equation}\label{wtu}
\begin{split}
\rho^i_{_\mathrm{TU}}(\bm{b}_\perp,\bm{k}_\perp,x)
=\frac{1}{2}\left[\rho^{[\gamma^+]}(\bm{b}_\perp,\bm{k}_\perp,x,\hat{\bm{e}}_i)-\rho^{[\gamma^+]}(\bm{b}_\perp,\bm{k}_\perp,x,-\hat{\bm{e}}_i)\right],
\end{split}
\end{equation}
the trans-longitudinal Wigner distribution
\begin{equation}\label{wtl}
\begin{split}
\rho^i_{_\mathrm{TL}}(\bm{b}_\perp,\bm{k}_\perp,x)
=\frac{1}{2}\left[\rho^{[\gamma^+\gamma_5]}(\bm{b}_\perp,\bm{k}_\perp,x,\hat{\bm{e}}_i)-\rho^{[\gamma^+\gamma_5]}(\bm{b}_\perp,\bm{k}_\perp,x,-\hat{\bm{e}}_i)\right],
\end{split}
\end{equation}
the transverse Wigner distribution
\begin{equation}\label{wtt}
\begin{split}
\rho_{_\mathrm{TT}}(\bm{b}_\perp,\bm{k}_\perp,x)
=\frac{1}{2}\delta_{ij}\left[\rho^{[i\sigma^{j+}\gamma_5]}(\bm{b}_\perp,\bm{k}_\perp,x,\hat{\bm{e}}_i)-\rho^{[i\sigma^{j+}\gamma_5]}(\bm{b}_\perp,\bm{k}_\perp,x,-\hat{\bm{e}}_i)\right],
\end{split}
\end{equation}
and the pretzelous Wigner distribution
\begin{equation}\label{wttp}
\begin{split}
\rho_{_\mathrm{TT}}^\perp(\bm{b}_\perp,\bm{k}_\perp,x)
=\frac{1}{2}\epsilon_{ij}\left[\rho^{[i\sigma^{j+}\gamma_5]}(\bm{b}_\perp,\bm{k}_\perp,x,\hat{\bm{e}}_i)-\rho^{[i\sigma^{j+}\gamma_5]}(\bm{b}_\perp,\bm{k}_\perp,x,-\hat{\bm{e}}_i)\right].
\end{split}
\end{equation}
The first subscript stands for the proton polarization and the second one stands for the quark polarization. The names are given by considering the polarization configuration of the quark, and then a prefix is added to describe the proton polarization, unless it is parallel polarized with the quark. The pretzelous Wigner distribution is named after the pretzelocity TMD to describe the case that both the quark and the proton are transversely polarized but along two orthogonal directions.

The Wigner distributions have direct connection with the generalized parton correlation functions (GPCFs)~\cite{Meissner:2009ww}, which are the fully unintegrated off-diagonal quark-quark correlator for a nucleon. The generalized transverse momentum dependent parton distributions (GTMDs) can be defined by integrating the GPCFs over the light-cone energy. Then the Wigner distributions can be regarded as the transverse Fourier transformation of the GTMDs. Since all the TMDs and GPDs can be obtained from the GTMDs, the GTMD is viewed as the so-called mother function. Therefore, the Wigner distribution is a connection between the TMDs and IPDs. Since the TMDs and IPDs are in general independent functions, the relations between them, if exist, should be established at the level of Wigner distributions. Hence, the Wigner distribution is a useful tool to study the nucleon structure, although there are at present no clear methods to extract it from experiments.

The Wilson line $\mathcal{W}$ that connects the quark fields at two points $y-z/2$ and $y+z/2$ plays an important role in understanding the naive time-reversal odd (T-odd) TMDs~\cite{Brodsky:2002cx}. The path is process dependent, and a sign change of quark Sivers and Boer--Mulders functions in the semi-inclusive deeply inelastic scattering (SIDIS) and Drell--Yan (DY) processes is predicted on the basis of this understanding. Here we choose a staple-like path for the Wilson line as~\cite{Meissner:2009ww}
\begin{equation}
y-\frac{z}{2}=(0,-\frac{z^-}{2},\bm{b}_\perp-\frac{\bm{z}_\perp}{2})\rightarrow(0,\infty,\bm{b}_\perp-\frac{\bm{z}_\perp}{2})\rightarrow(0,\infty,\bm{b}_\perp+\frac{\bm{z}_\perp}{2})\rightarrow(0,\frac{z^-}{2},\bm{b}_\perp+\frac{\bm{z}_\perp}{2})=y+\frac{z}{2}\nonumber
\end{equation}
in order to obtain the appropriate Wilson line when taking the TMD and IPD limits. Although the two segments along the light-cone direction vanish in the light-cone gauge $A^+=0$, the transverse segment at the light-cone infinity is still nontrivial~\cite{Ji:2002aa}. Here we take a crude truncation of the Wilson line at the first order for simplicity, and it reduces to the unit operator $\bm{1}$. At the leading twist, there are 16 independent GTMDs, which are in general complex-valued functions, while the Wigner distributions are always real-valued. If separating the real part and the imaginary part of the GTMDs, one can define 32 real-valued GTMDs, and 16 of them are T-even, while the other 16 are T-odd. Our truncation of the Wilson line essentially means to neglect the T-odd ones, {\it i.e.} the contribution from the imaginary part of the GTMDs.

Quantized at the fixed light-cone time, the quark field operator in Eq. (\ref{wigneroperator}) is expressed as
\begin{equation}\label{quarkfield}
\begin{split}
\psi(y+\frac{z}{2})=\sum_\lambda\int\frac{d\ell^+}{\sqrt{2\ell^+}}\frac{d^2\bm{\ell}_\perp}{(2\pi)^3}\big[b_\lambda(\ell)u(\ell,\lambda)e^{-i\ell\cdot(y+\frac{z}{2})}+d_\lambda^\dagger(\ell)v(\ell,\lambda)e^{i\ell\cdot(y+\frac{z}{2})}\big],
\end{split}
\end{equation}
and similar for $\bar{\psi}(y-z/2)$. The twist-two Wigner operators are expressed as
{\allowdisplaybreaks
\begin{align}
\hat{W}^{[\gamma^+]}&(\bm{b}_\perp,\bm{k}_\perp,x)=\frac{1}{2}\int\frac{d\ell'^+d^2\bm{\ell}'_\perp}{(2\pi)^3}\int\frac{d\ell^+d^2\bm{\ell}_\perp}{(2\pi)^3}\nonumber\\ \label{wigneru}
&\times\bigg\{\big[b_\uparrow^\dagger(\ell')b_\uparrow(\ell)+b_\downarrow^\dagger(\ell')b_\downarrow(\ell)\big]e^{-i(\bm{\ell}'_\perp-\bm{\ell}_\perp)\cdot\bm{b}_\perp}\delta(k^+-\frac{\ell'^++\ell^+}{2})\delta^{(2)}(\bm{k}_\perp-\frac{\bm{\ell}'_\perp+\bm{\ell}_\perp}{2})\nonumber\\
&-\big[d_\uparrow^\dagger(\ell)d_\uparrow(\ell')+d_\downarrow^\dagger(\ell)d_\downarrow(\ell')\big]e^{i(\bm{\ell}'_\perp-\bm{\ell}_\perp)\cdot\bm{b}_\perp}\delta(k^++\frac{\ell'^++\ell^+}{2})\delta^{(2)}(\bm{k}_\perp+\frac{\bm{\ell}'_\perp+\bm{\ell}_\perp}{2})\bigg\},\\
\hat{W}^{[\gamma^+\gamma_5]}&(\bm{b}_\perp,\bm{k}_\perp,x)=\frac{1}{2}\int\frac{d\ell'^+d^2\bm{\ell}'_\perp}{(2\pi)^3}\int\frac{d\ell^+d^2\bm{\ell}_\perp}{(2\pi)^3}\nonumber\\ \label{wignerl}
&\times\bigg\{\big[b_\uparrow^\dagger(\ell')b_\uparrow(\ell)-b_\downarrow^\dagger(\ell')b_\downarrow(\ell)\big]e^{-i(\bm{\ell}'_\perp-\bm{\ell}_\perp)\cdot\bm{b}_\perp}\delta(k^+-\frac{\ell'^++\ell^+}{2})\delta^{(2)}(\bm{k}_\perp-\frac{\bm{\ell}'_\perp+\bm{\ell}_\perp}{2})\nonumber\\
&+\big[d_\uparrow^\dagger(\ell)d_\uparrow(\ell')-d_\downarrow^\dagger(\ell)d_\downarrow(\ell')\big]e^{i(\bm{\ell}'_\perp-\bm{\ell}_\perp)\cdot\bm{b}_\perp}\delta(k^++\frac{\ell'^++\ell^+}{2})\delta^{(2)}(\bm{k}_\perp+\frac{\bm{\ell}'_\perp+\bm{\ell}_\perp}{2})\bigg\},\\
\hat{W}^{[i\sigma^{1+}\gamma_5]}&(\bm{b}_\perp,\bm{k}_\perp,x)=\frac{1}{2}\int\frac{d\ell'^+d^2\bm{\ell}'_\perp}{(2\pi)^3}\int\frac{d\ell^+d^2\bm{\ell}_\perp}{(2\pi)^3}\nonumber\\ \label{wignert}
&\times\bigg\{\big[b_\uparrow^\dagger(\ell')b_\downarrow(\ell)+b_\downarrow^\dagger(\ell')b_\uparrow(\ell)\big]e^{-i(\bm{\ell}'_\perp-\bm{\ell}_\perp)\cdot\bm{b}_\perp}\delta(k^+-\frac{\ell'^++\ell^+}{2})\delta^{(2)}(\bm{k}_\perp-\frac{\bm{\ell}'_\perp+\bm{\ell}_\perp}{2})\nonumber\\
&-\big[d_\uparrow^\dagger(\ell)d_\downarrow(\ell')+d_\downarrow^\dagger(\ell)d_\uparrow(\ell')\big]e^{i(\bm{\ell}'_\perp-\bm{\ell}_\perp)\cdot\bm{b}_\perp}\delta(k^++\frac{\ell'^++\ell^+}{2})\delta^{(2)}(\bm{k}_\perp+\frac{\bm{\ell}'_\perp+\bm{\ell}_\perp}{2})\bigg\},
\end{align}
}
where we use the Lepage--Brodsky (LB) conventions for the properties of the light-cone spinors~\cite{Lepage:1980fj}. In the spectator model, the proton state in Eq. (\ref{wignerdistr}) is expressed as
\begin{equation}
\begin{split}\label{qs}
\Big|P^+,-\frac{\bm{\Delta}_\perp}{2},\bm{S}\Big\rangle=&\sum_{\sigma,s}\int\frac{dx_q}{2\sqrt{x_q}}\frac{d^2\bm{k}_{q\perp}}{(2\pi)^3}\int\frac{dx_D}{2\sqrt{x_D}}\frac{d^2\bm{k}_{D\perp}}{(2\pi)^3}16\pi^3\delta(1-x_q-x_D)\delta^{(2)}(\bm{k}_{q\perp}+\bm{k}_{D\perp})\\
&\times\psi^{S}_{\sigma s}(x_q,\bm{k}_{q\perp})\sqrt{2k_q^+}\sqrt{2k_D^+}b_\sigma^\dagger(x_q,\bm{k}_{q\perp}-x_q\frac{\bm{\Delta}_\perp}{2})a_s^\dagger(x_D,\bm{k}_{D\perp}-x_D\frac{\bm{\Delta}_\perp}{2})\big|0\big\rangle,
\end{split}
\end{equation}
where $\bm{k}_{q\perp}$ and $\bm{k}_{D\perp}$ are the intrinsic transverse momentum for the quark and the spectator, and $x_q=k_q^+/P^+$ and $x_D=k_D^+/P^+$ are the light-cone momentum fractions carried by the quark and the spectator. The arguments for the creation (annihilation) operators are the frame dependent transverse momentum in the one-particle Fock state definition. Whereas, the light-cone wave function $\psi^S_{\sigma s}(x,\bm{k}_{q\perp})$, which describes the amplitude of the quark-spectator state in the proton, is frame independent, and only depends on boost invariant variables $x$ and $\bm{k}_{q\perp}$. The light-cone wave function can be factored into a spin space part and a momentum space part. The spin space part is given by the Melosh--Wigner rotations (\ref{mw1}) and (\ref{mw2}), and the momentum space part is given by the BHL prescription (\ref{bhl}). Then, sandwiching the Wigner operators in Eqs. (\ref{wigneru})-(\ref{wignert}) between the proton states with different polarization configures, we can obtain the expressions of all the 16 quark Wigner distributions in term of the overlap of light-cone wave functions, as shown in the Appendix.

\section{Numerical results and discussions}

The Wigner distribution is a five-dimensional function of $b_x$, $b_y$, $k_x$, $k_y$ and $x$. As a phase-space distribution, what we concern most is the behavior in the transverse coordinate and transverse momentum space. Therefore, we integrate over the light-cone momentum fraction $x$, and display its behavior in the remaining four dimentions, {\it i.e.} the so-called transverse Wigner distributions~\cite{Lorce:2011kd}. We plot the Wigner distribution in the transverse coordinate space with definite transverse momentum and in the transverse momentum space with definite coordinate. Apart from the TMDs and IPDs which can all be obtained from the Wigner distributions, one may define the mixing distributions by integrating over a transverse coordinate and a transverse momentum along two perpendicular transverse directions as
\begin{align}
\tilde{\rho}_{_\mathrm{\Lambda\lambda}}(b_x,k_y,x)&=\int db_ydk_x\rho_{_\mathrm{\Lambda\lambda}}(\bm{b}_\perp,\bm{k}_\perp,x),\label{mixing1}\\
\bar{\rho}_{_\mathrm{\Lambda\lambda}}(b_y,k_x,x)&=\int db_xdk_y\rho_{_\mathrm{\Lambda\lambda}}(\bm{b}_\perp,\bm{k}_\perp,x).\label{mixing2}
\end{align}
Unlike the Wigner distributions, the mixing distributions are real distribution functions since the remaining variables are not protected by the uncertainty principle, and they describe the correlation of quark transverse coordinate and transverse momentum in orthogonal directions. Both Wigner distributions and mixing distributions are dimensionless quantities in natural units. In this section, we display the numerical results of quark Wigner distributions and mixing distributions.

\subsection{Unpolarized proton}

\begin{figure}
\includegraphics[width=0.25\textwidth]{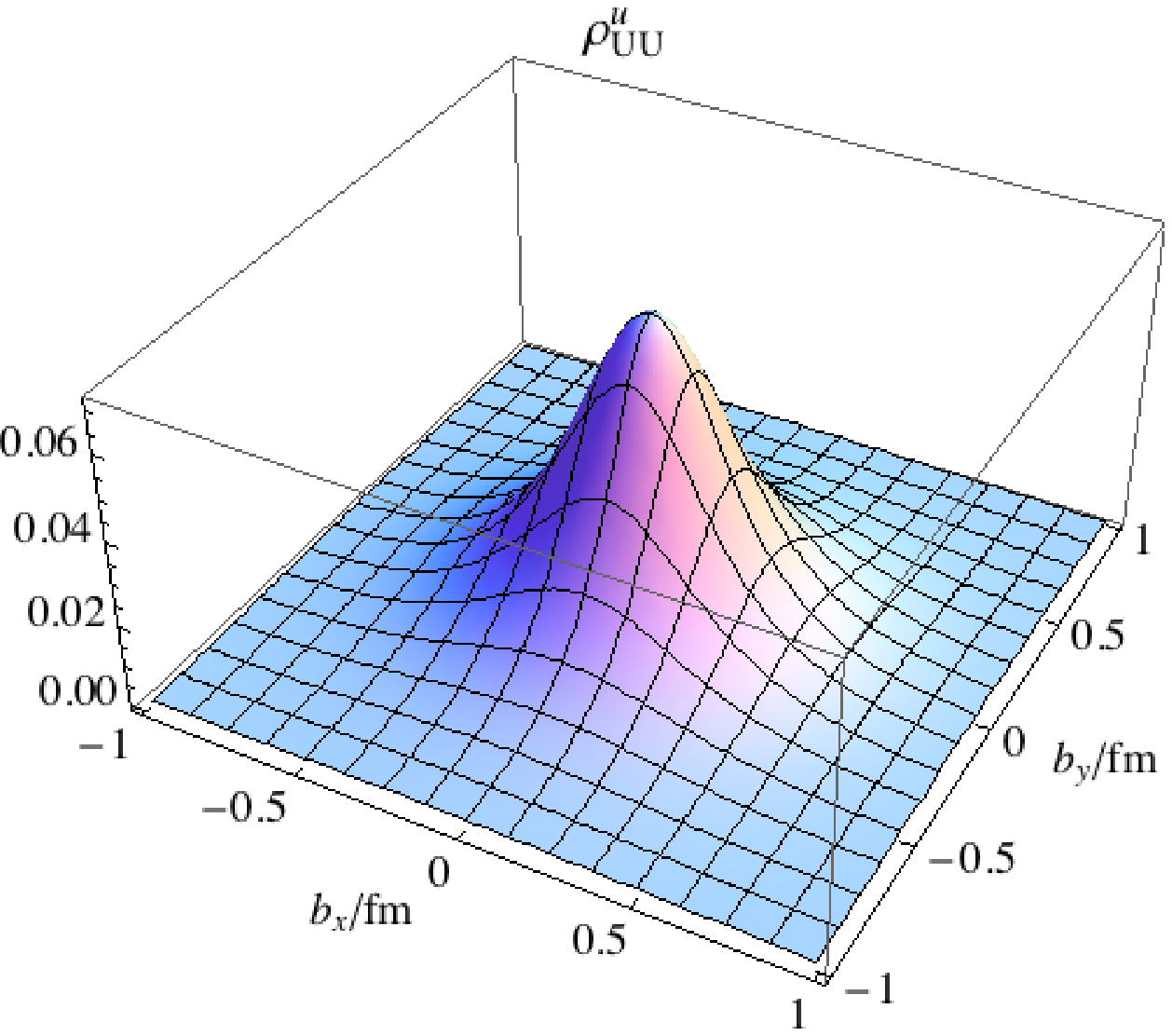}
\includegraphics[width=0.25\textwidth]{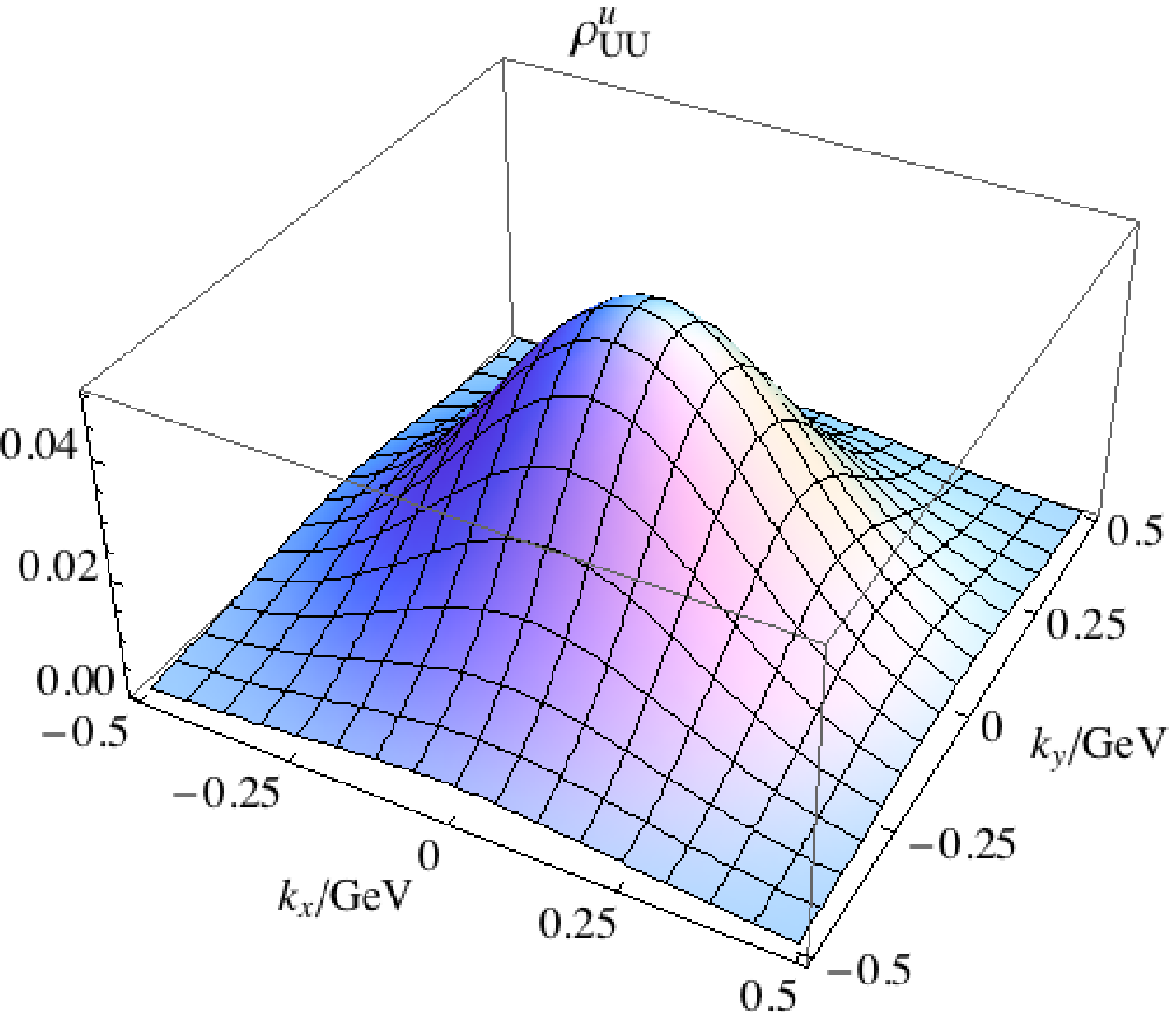}
\includegraphics[width=0.25\textwidth]{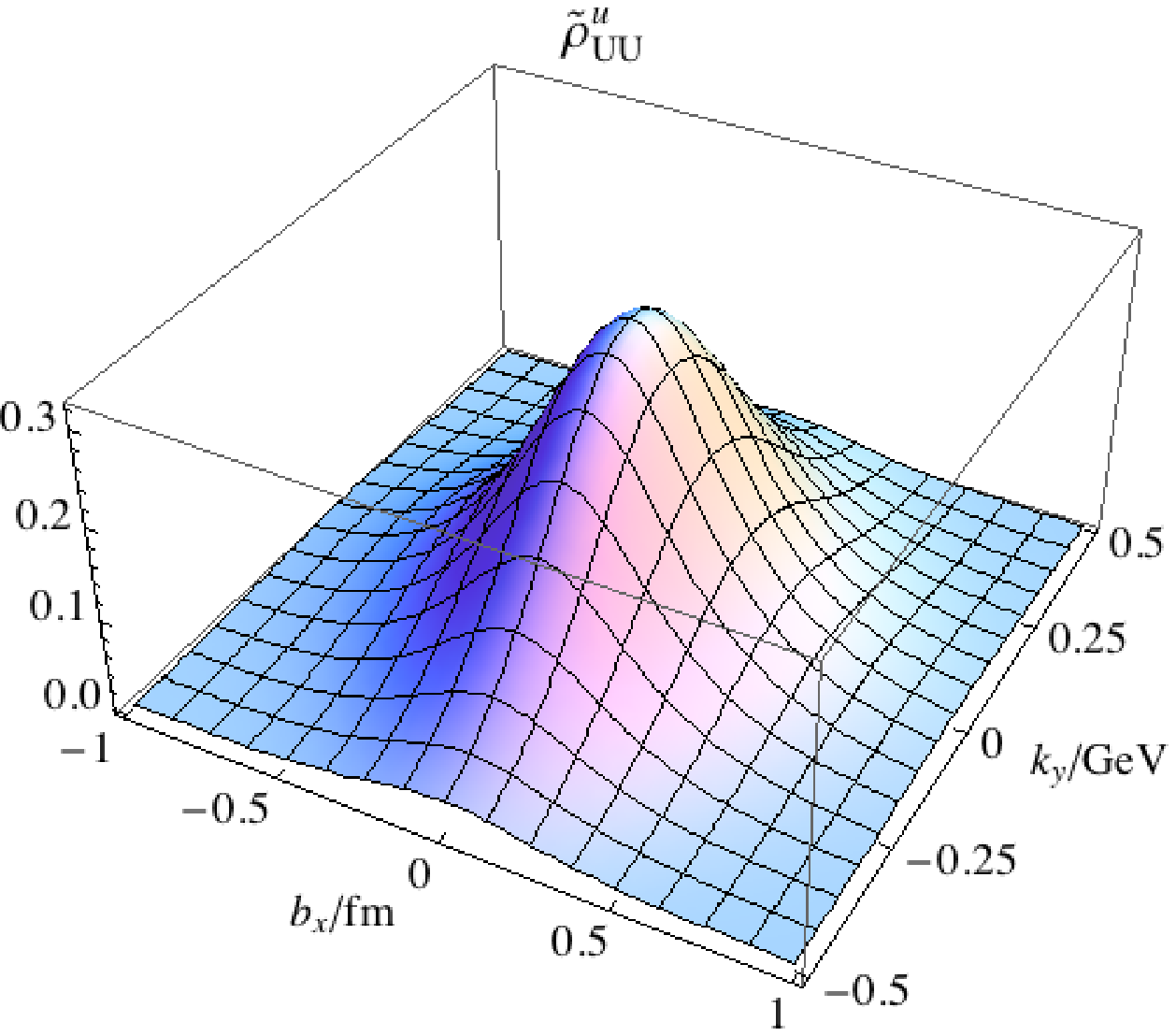}
\includegraphics[width=0.25\textwidth]{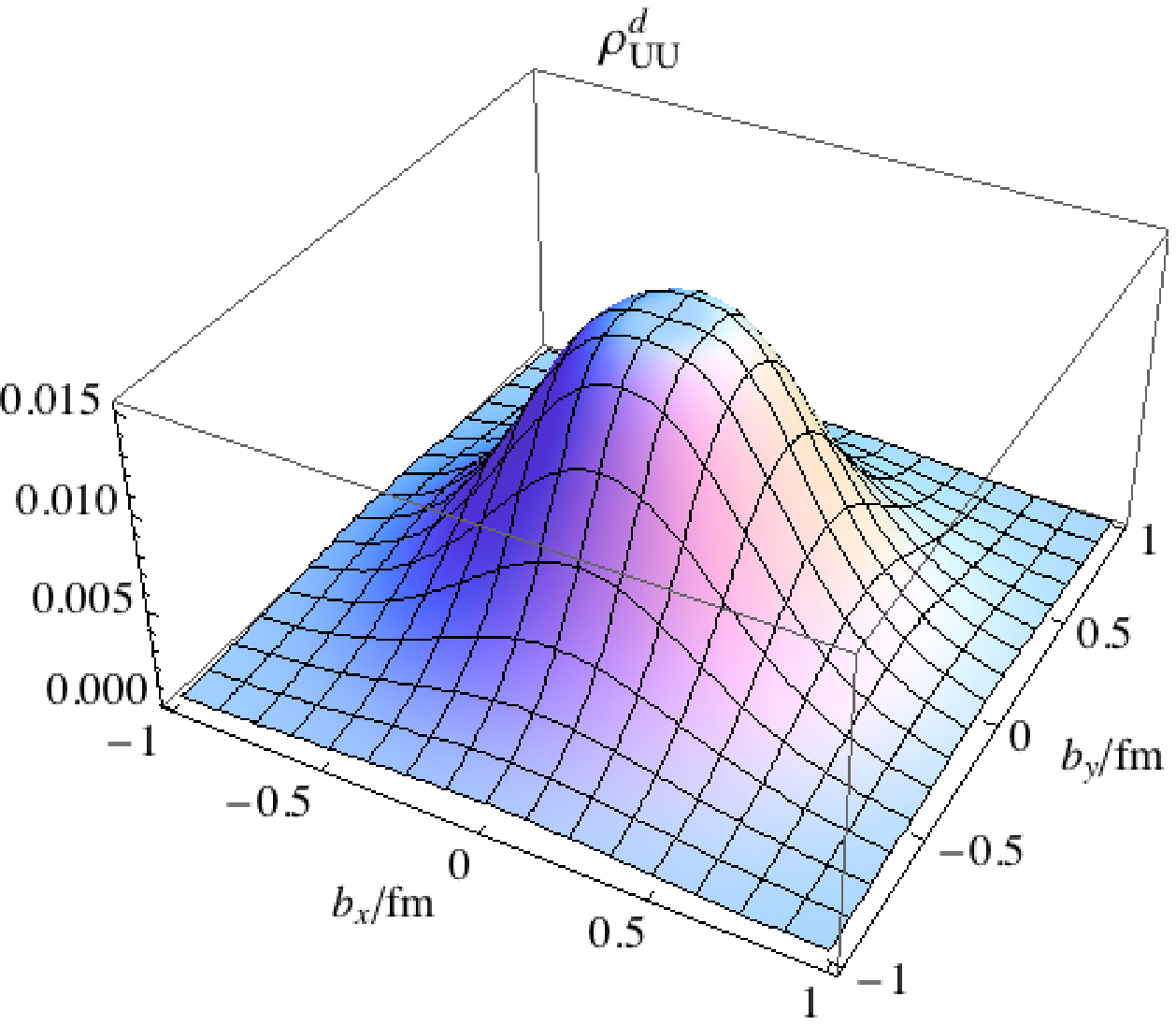}
\includegraphics[width=0.25\textwidth]{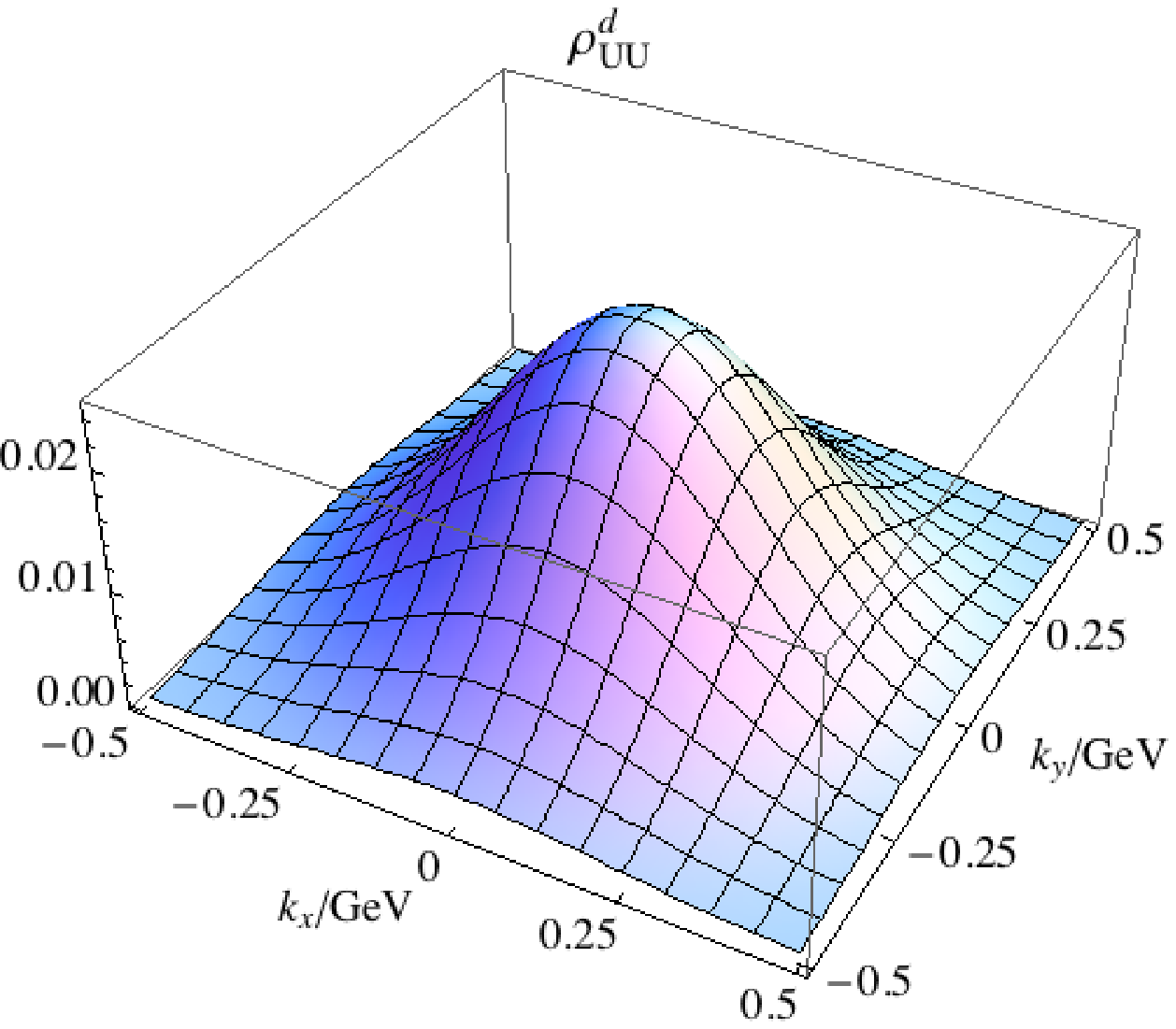}
\includegraphics[width=0.25\textwidth]{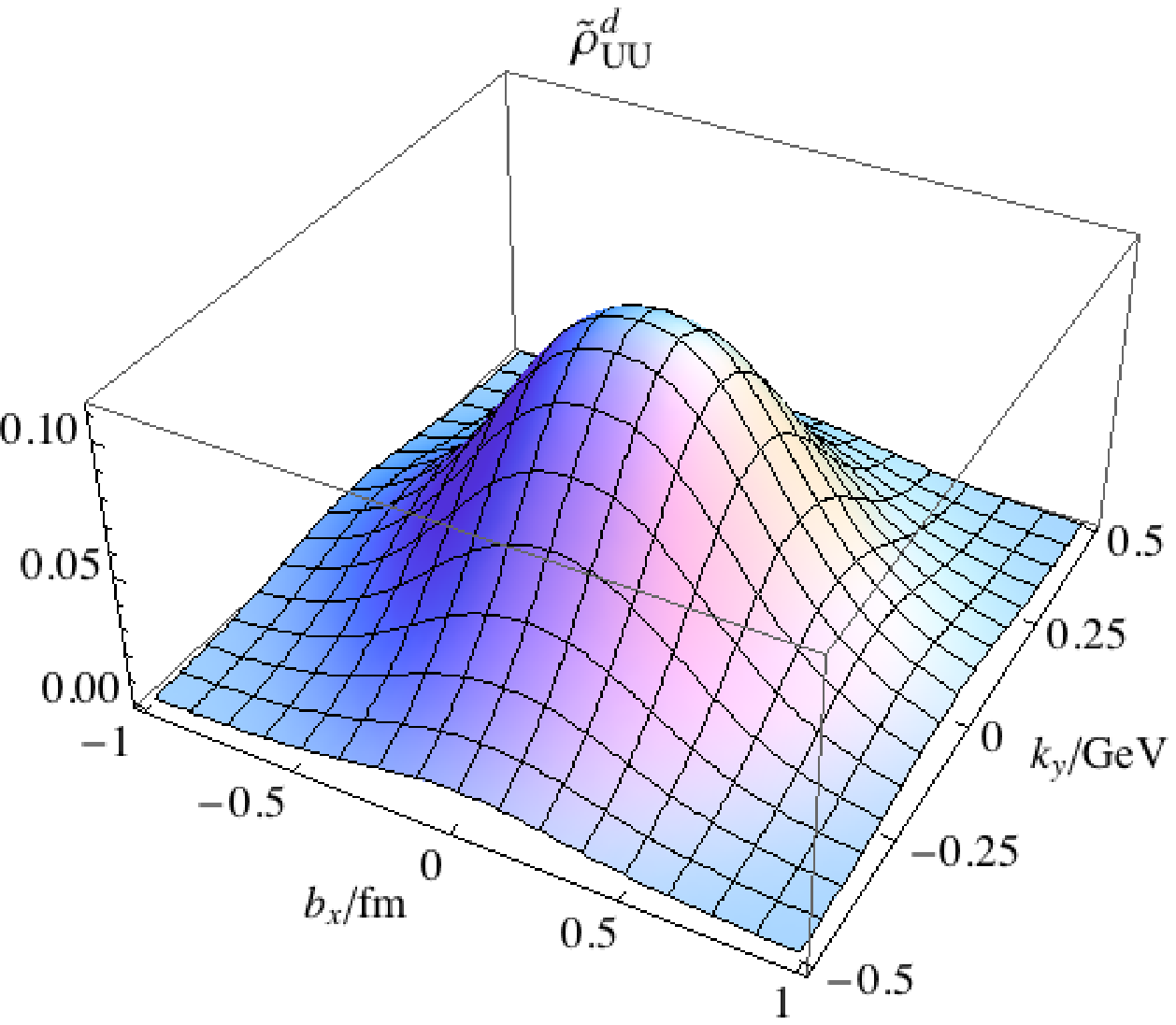}
\caption{(Color online). Unpolarized Wigner distributions $\rho_{_\mathrm{UU}}$ and mixing distributions $\tilde{\rho}_{_\mathrm{UU}}$ for the $u$ quark (upper) and the $d$ quark (lower). The first column the Wigner distributions in transverse coordinate space with definite transverse momentum $\bm{k}_\perp=0.3\,\textrm{GeV}\,\hat{\bm{e}}_y$. The second column are the Wigner distributions in transverse momentum space with definite transverse coordinate $\bm{b}_\perp=0.4\,\textrm{fm}\,\hat{\bm{e}}_y$. The third column are the mixing distributions $\tilde{\rho}_{_\mathrm{UU}}$. \label{rhouu}}
\end{figure}
\begin{figure}
\includegraphics[width=0.25\textwidth]{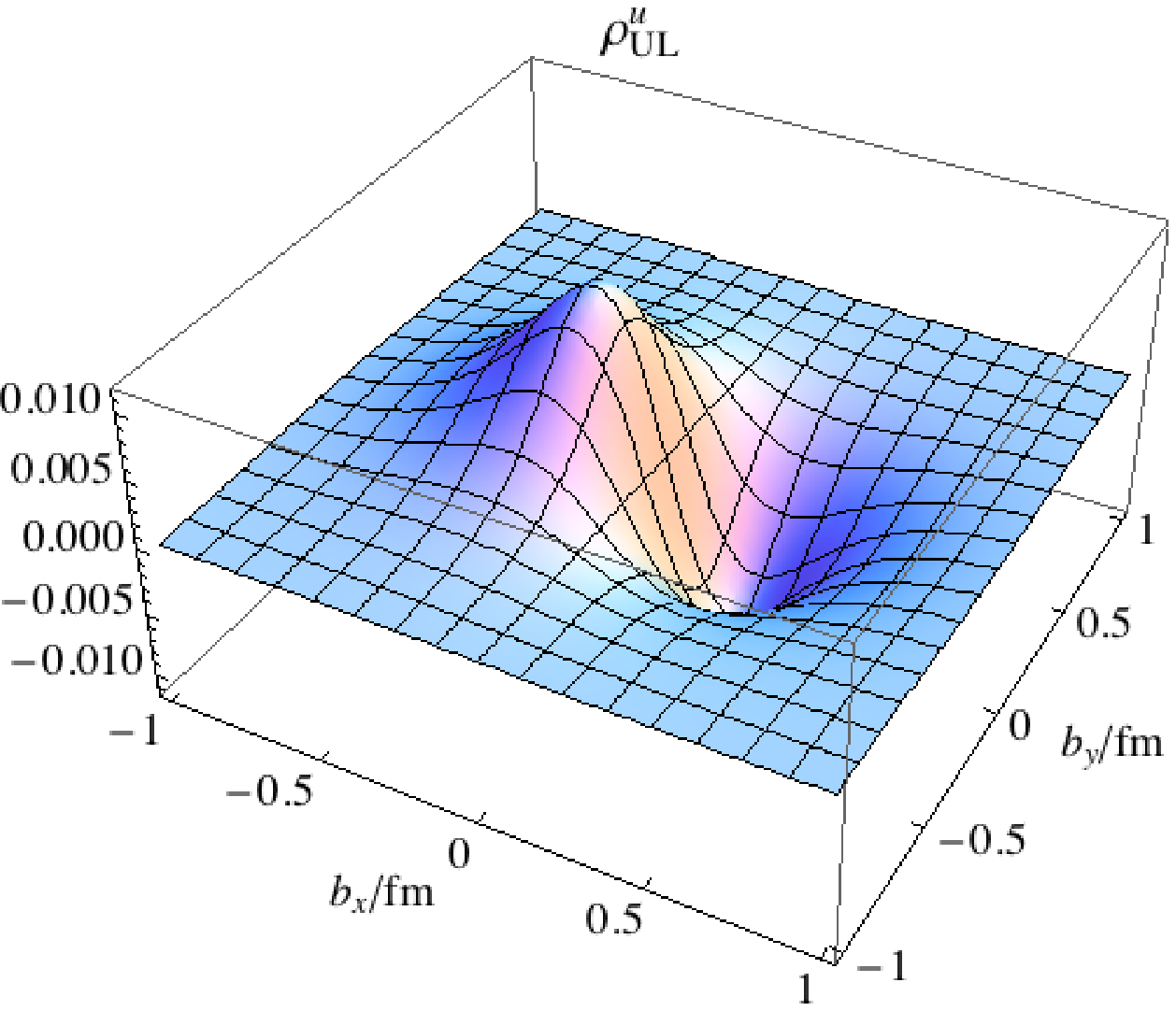}
\includegraphics[width=0.25\textwidth]{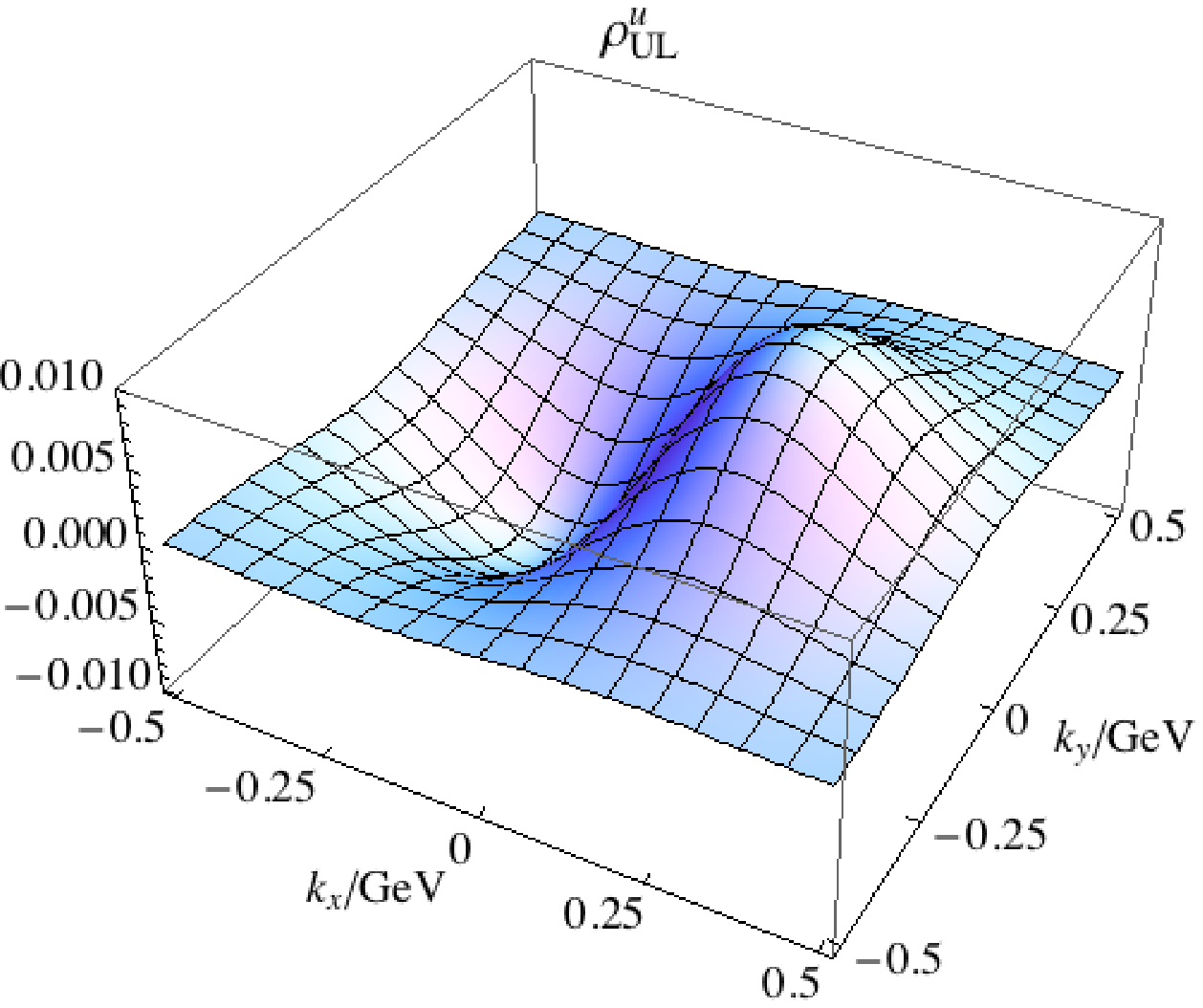}
\includegraphics[width=0.25\textwidth]{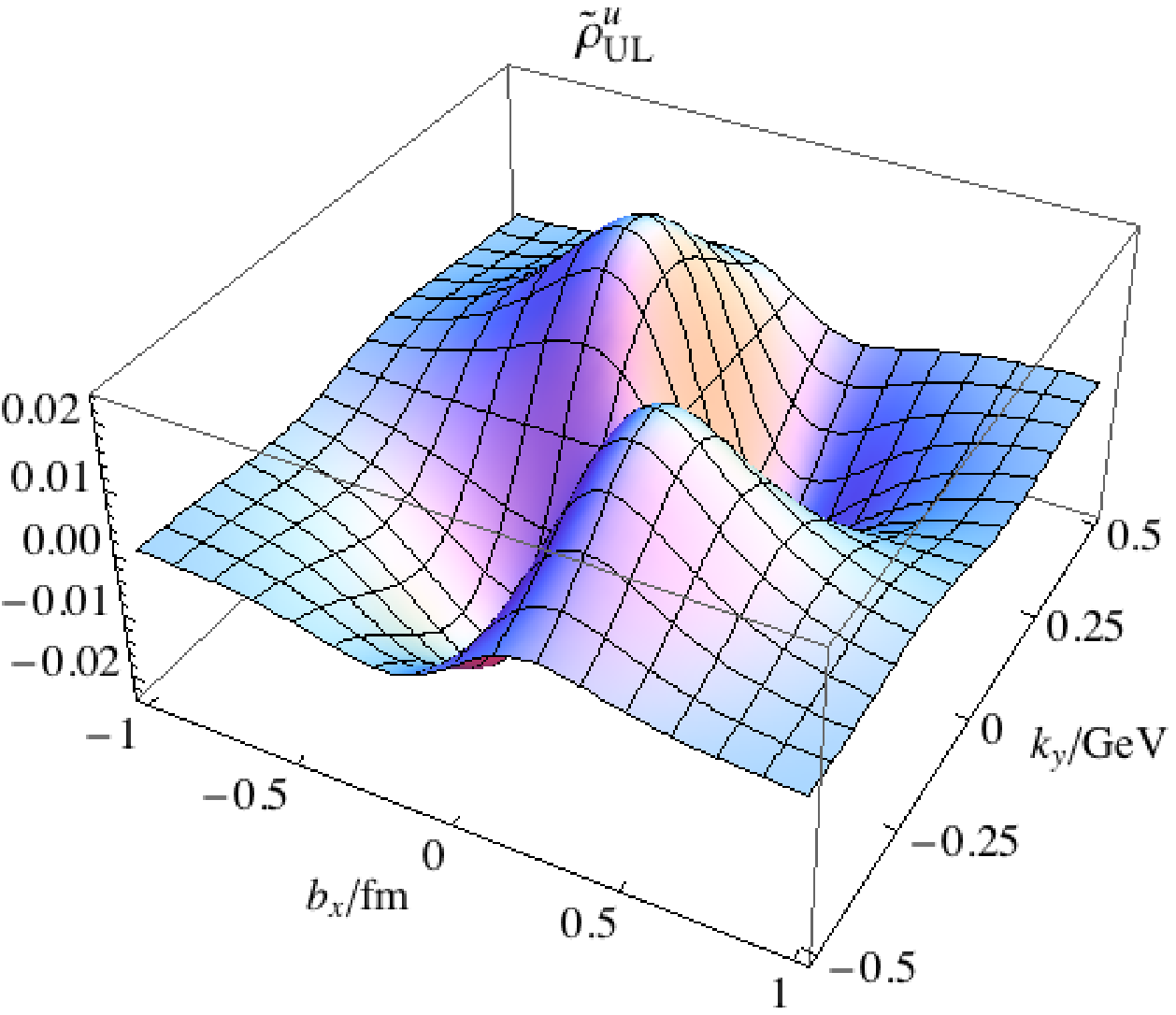}
\includegraphics[width=0.25\textwidth]{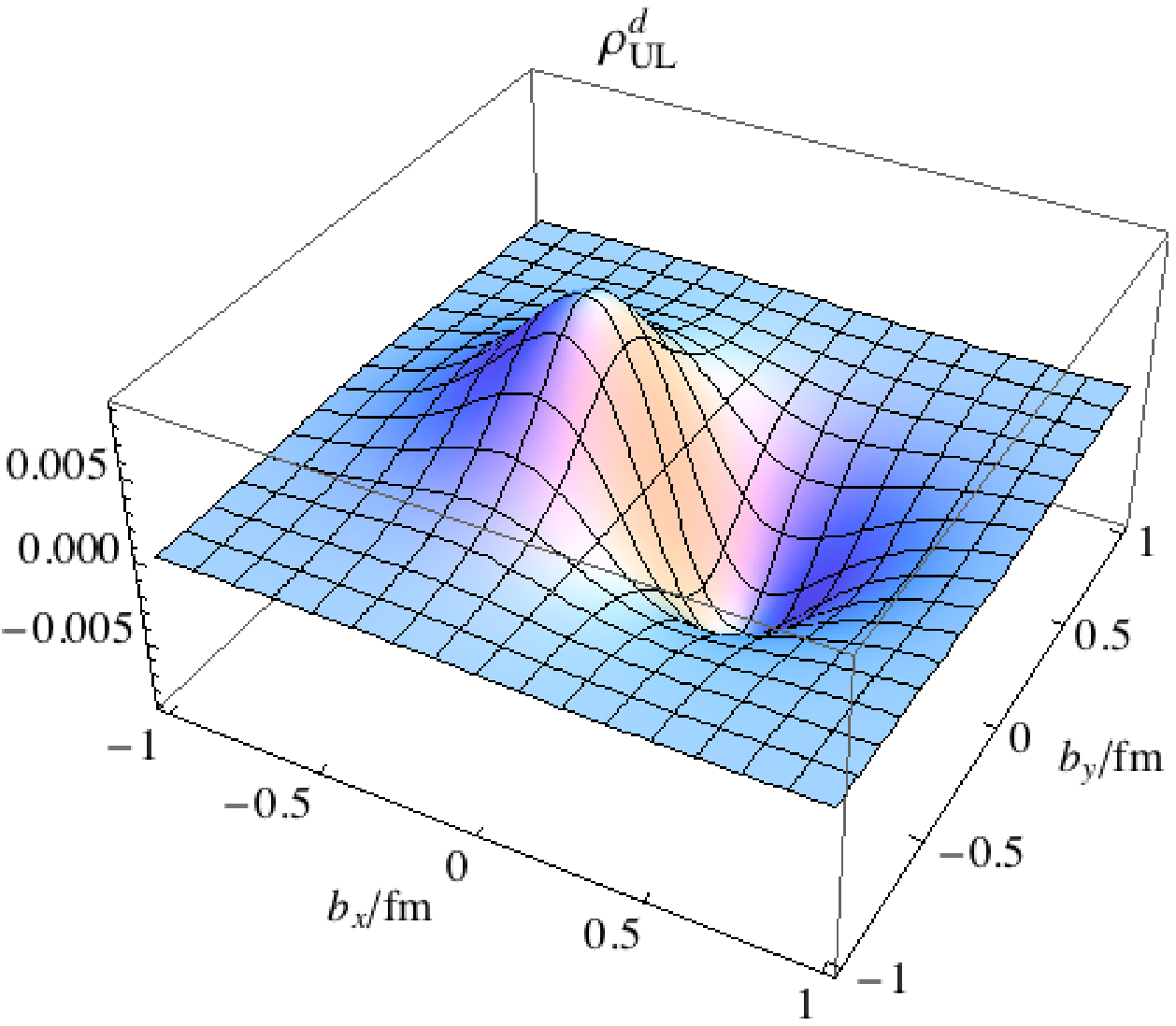}
\includegraphics[width=0.25\textwidth]{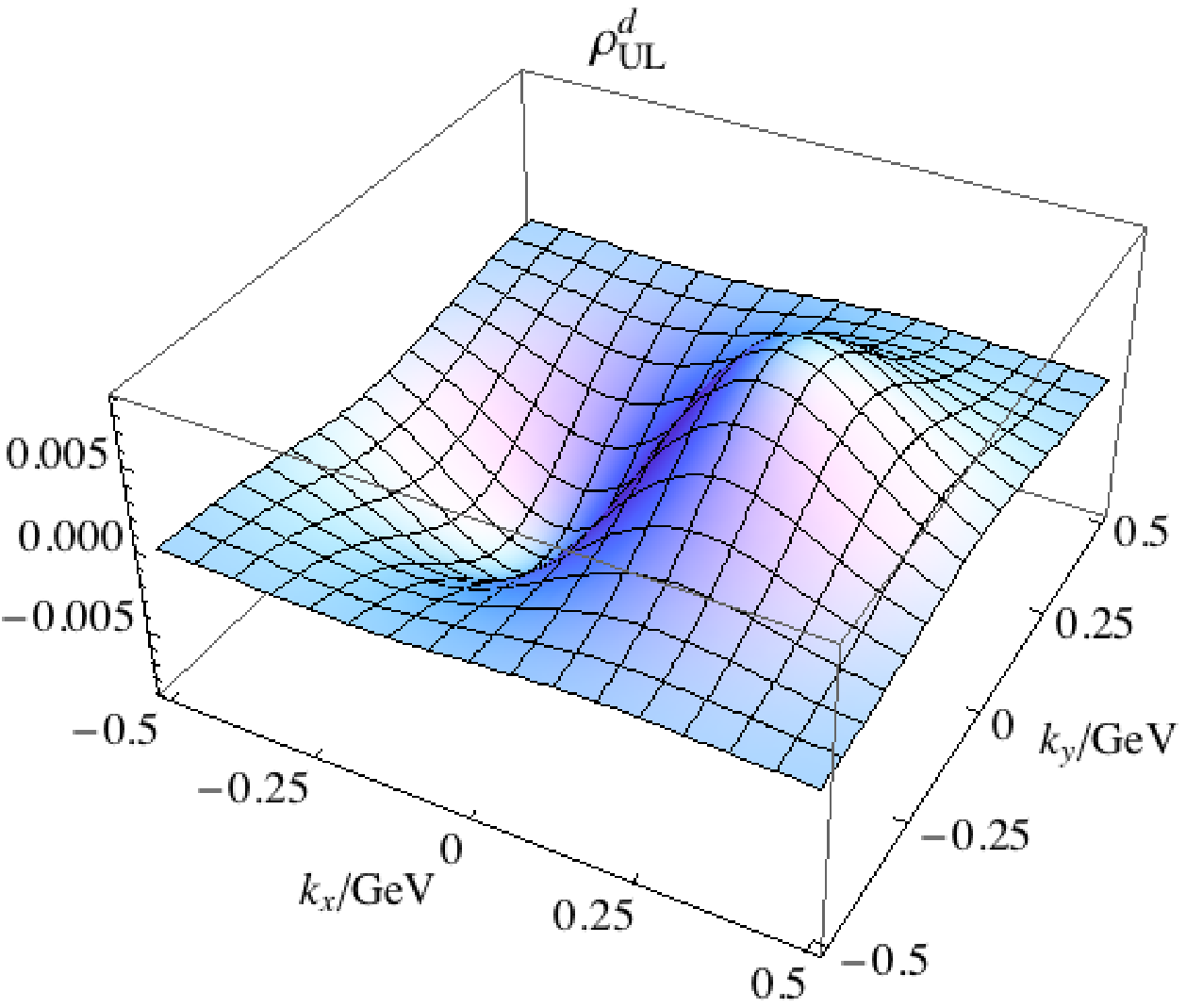}
\includegraphics[width=0.25\textwidth]{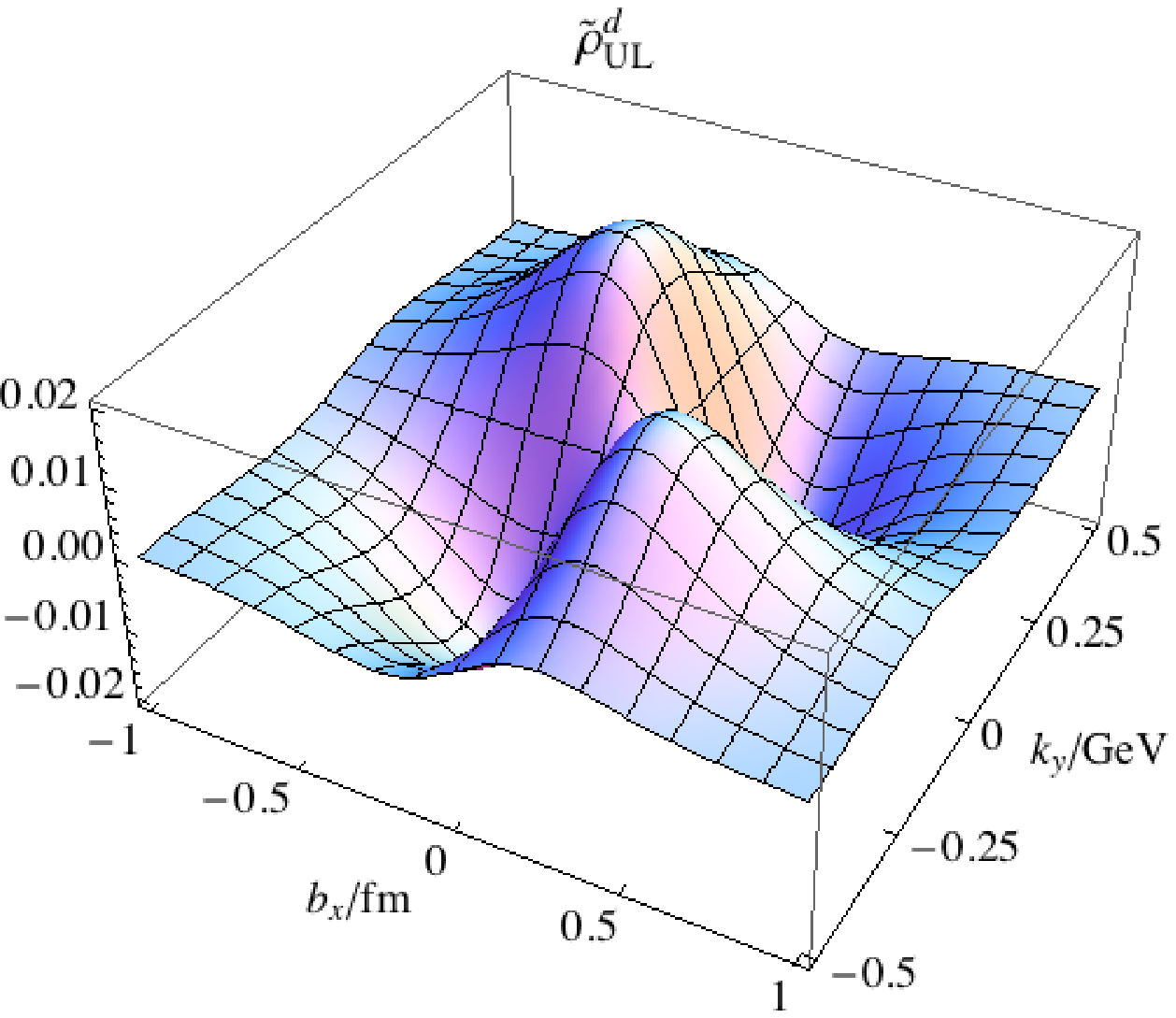}
\caption{(Color online). Unpol-longitudinal Wigner distributions $\rho_{_\mathrm{UL}}$ and mixing distributions $\tilde{\rho}_{_\mathrm{UL}}$ for the $u$ quark (upper) and the $d$ quark (lower). The first column the Wigner distributions in transverse coordinate space with definite transverse momentum $\bm{k}_\perp=0.3\,\textrm{GeV}\,\hat{\bm{e}}_y$. The second column are the Wigner distributions in transverse momentum space with definite transverse coordinate $\bm{b}_\perp=0.4\,\textrm{fm}\,\hat{\bm{e}}_y$. The third column are the mixing distributions $\tilde{\rho}_{_\mathrm{UL}}$. \label{rhoul}}
\end{figure}
\begin{figure}
\includegraphics[width=0.23\textwidth]{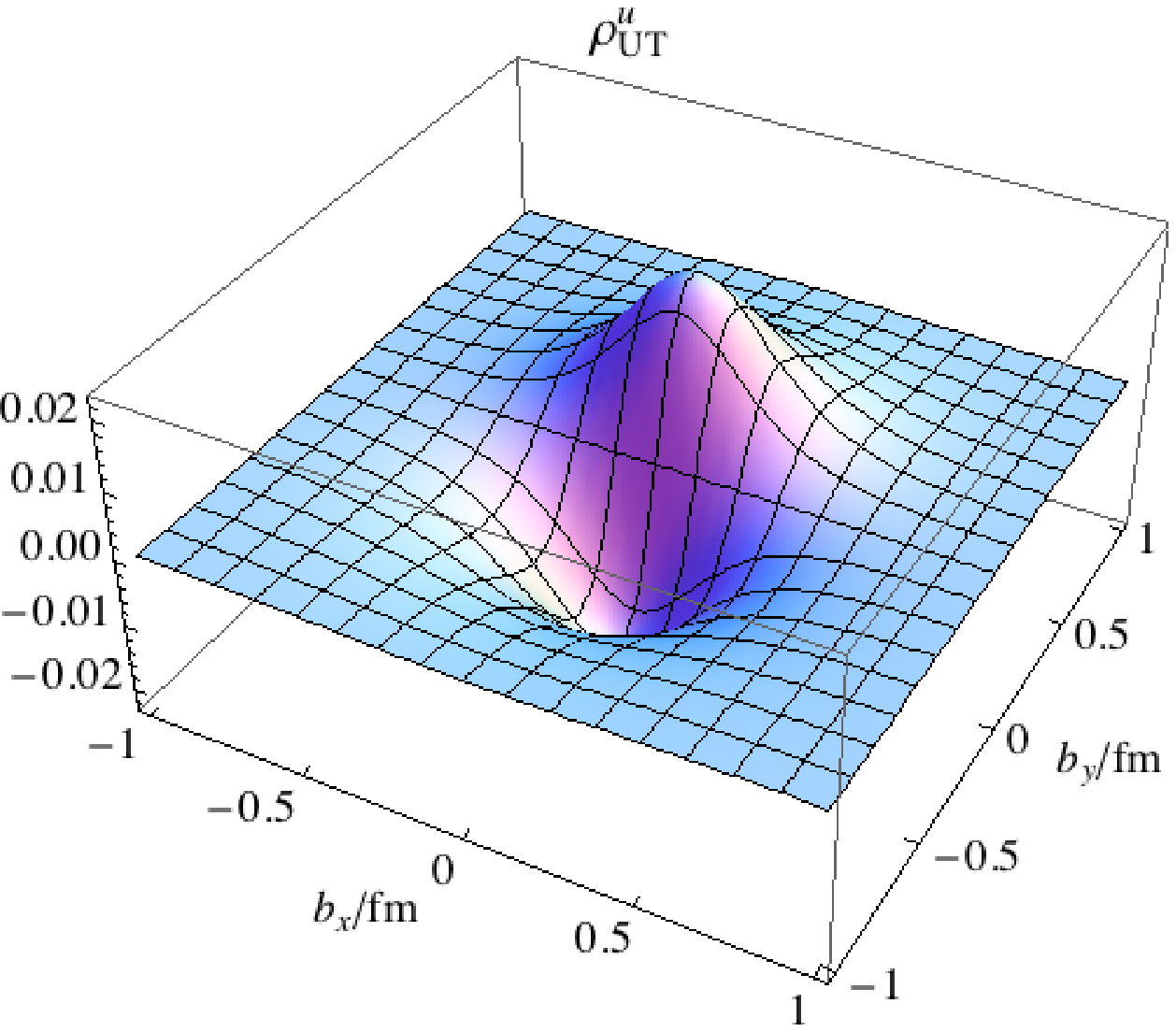}
\includegraphics[width=0.23\textwidth]{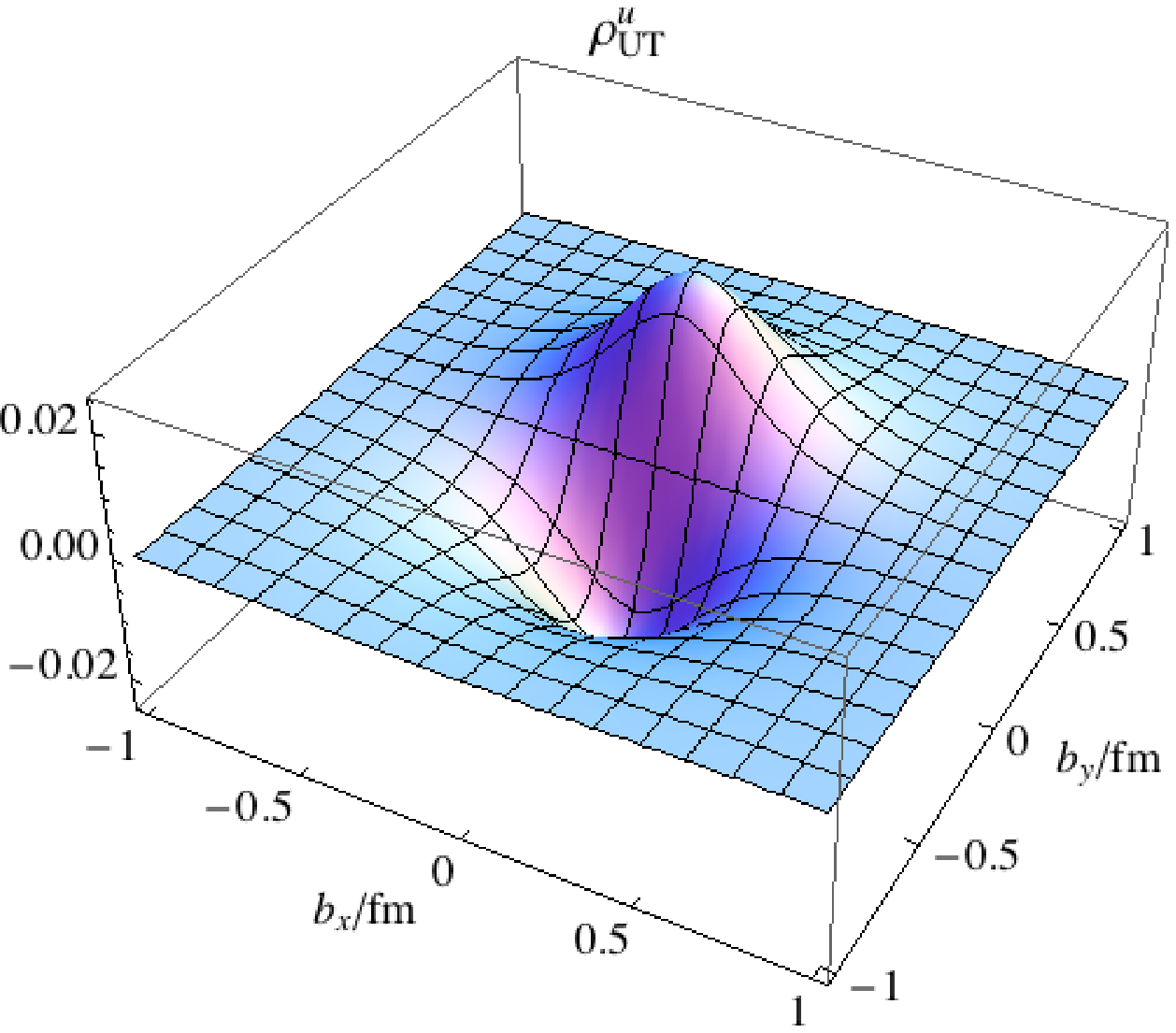}
\includegraphics[width=0.23\textwidth]{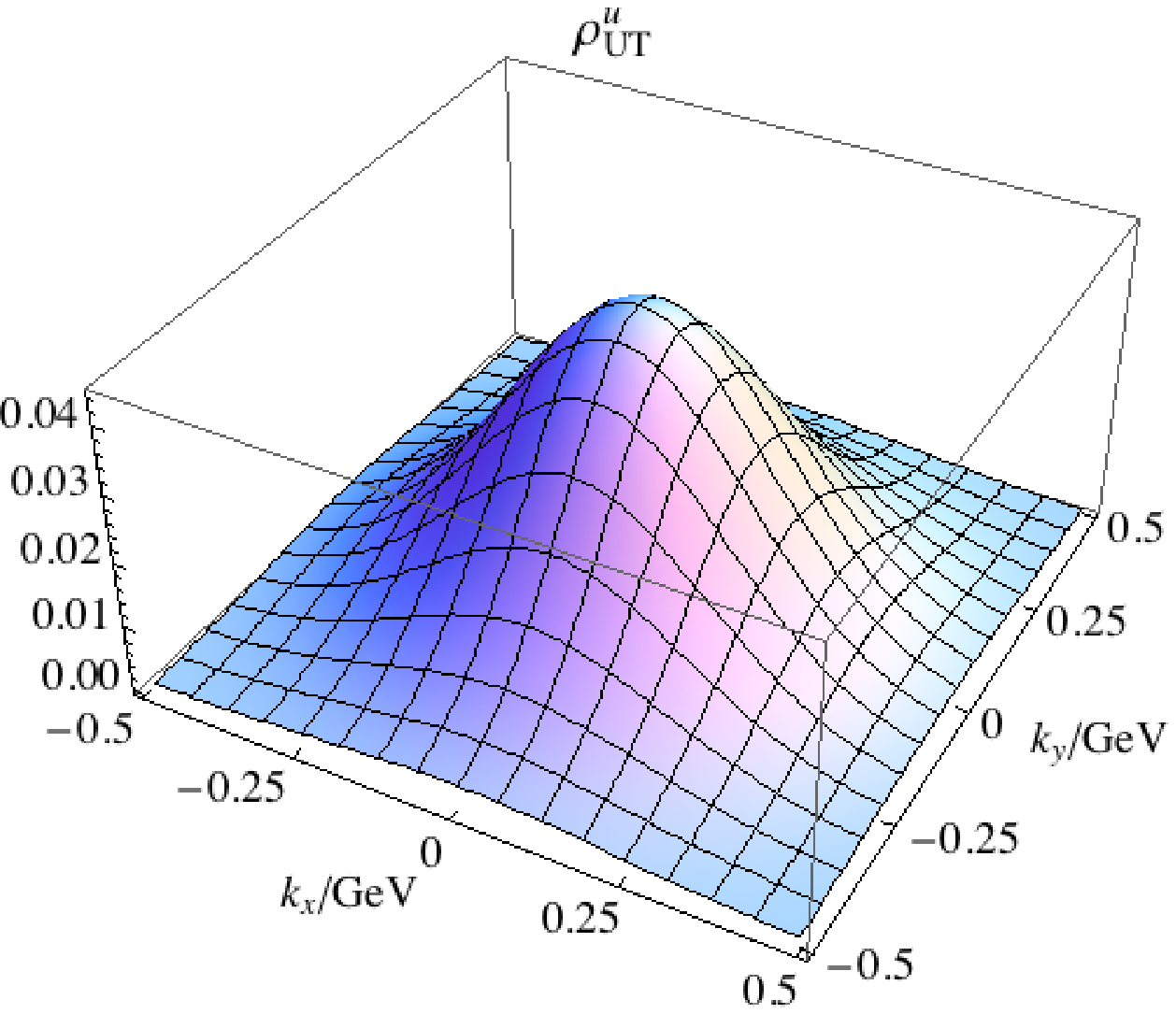}
\includegraphics[width=0.23\textwidth]{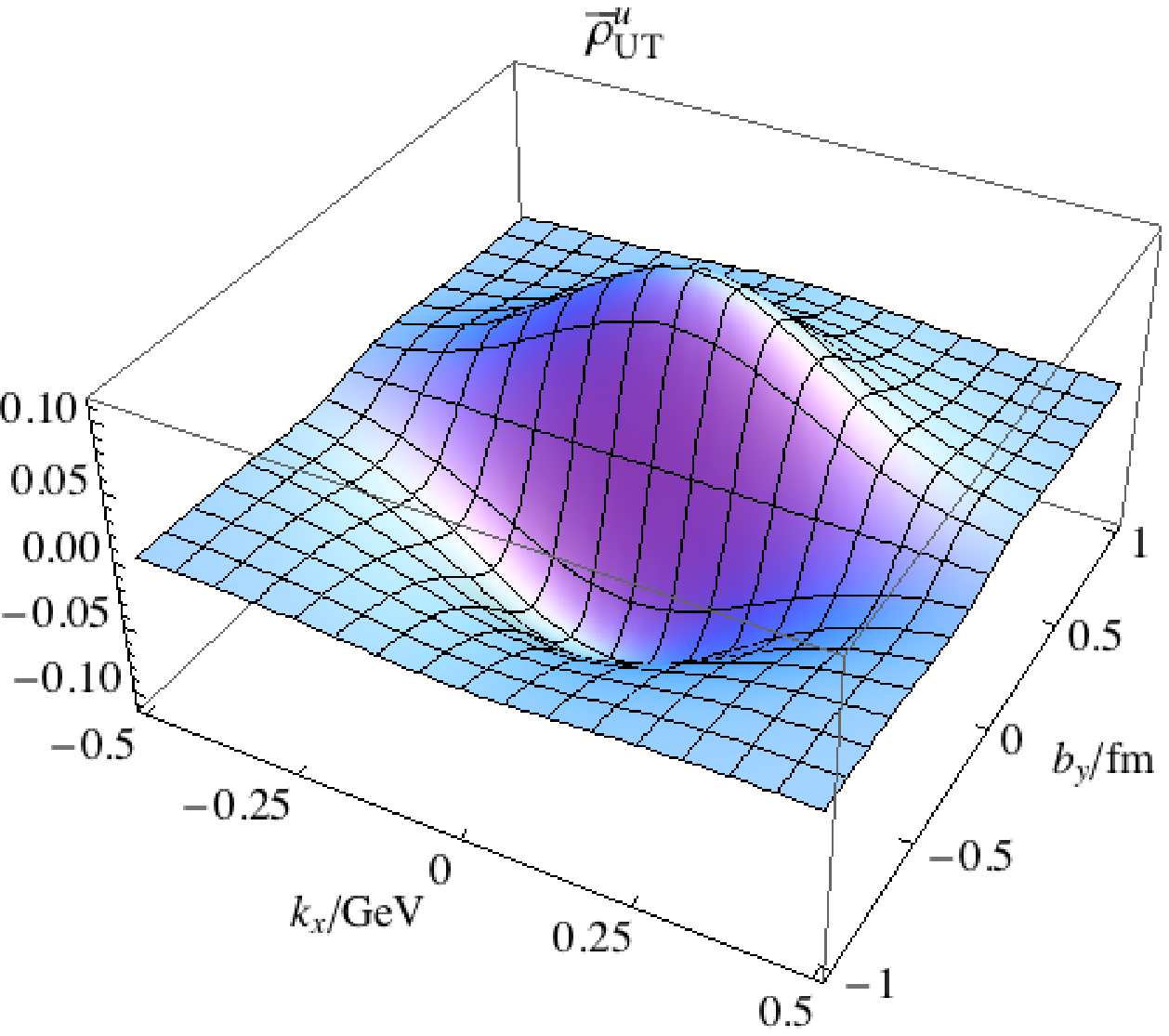}
\includegraphics[width=0.23\textwidth]{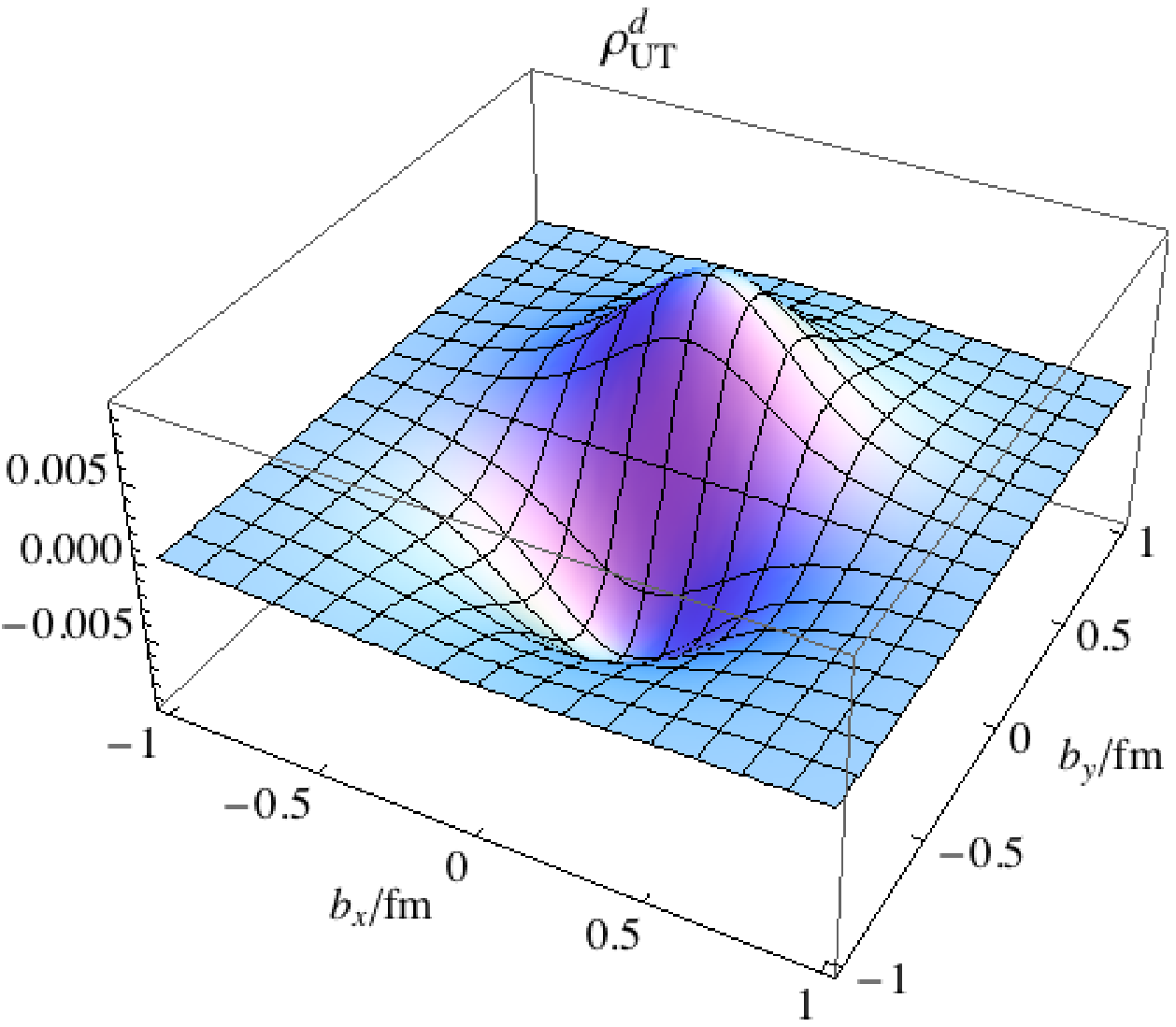}
\includegraphics[width=0.23\textwidth]{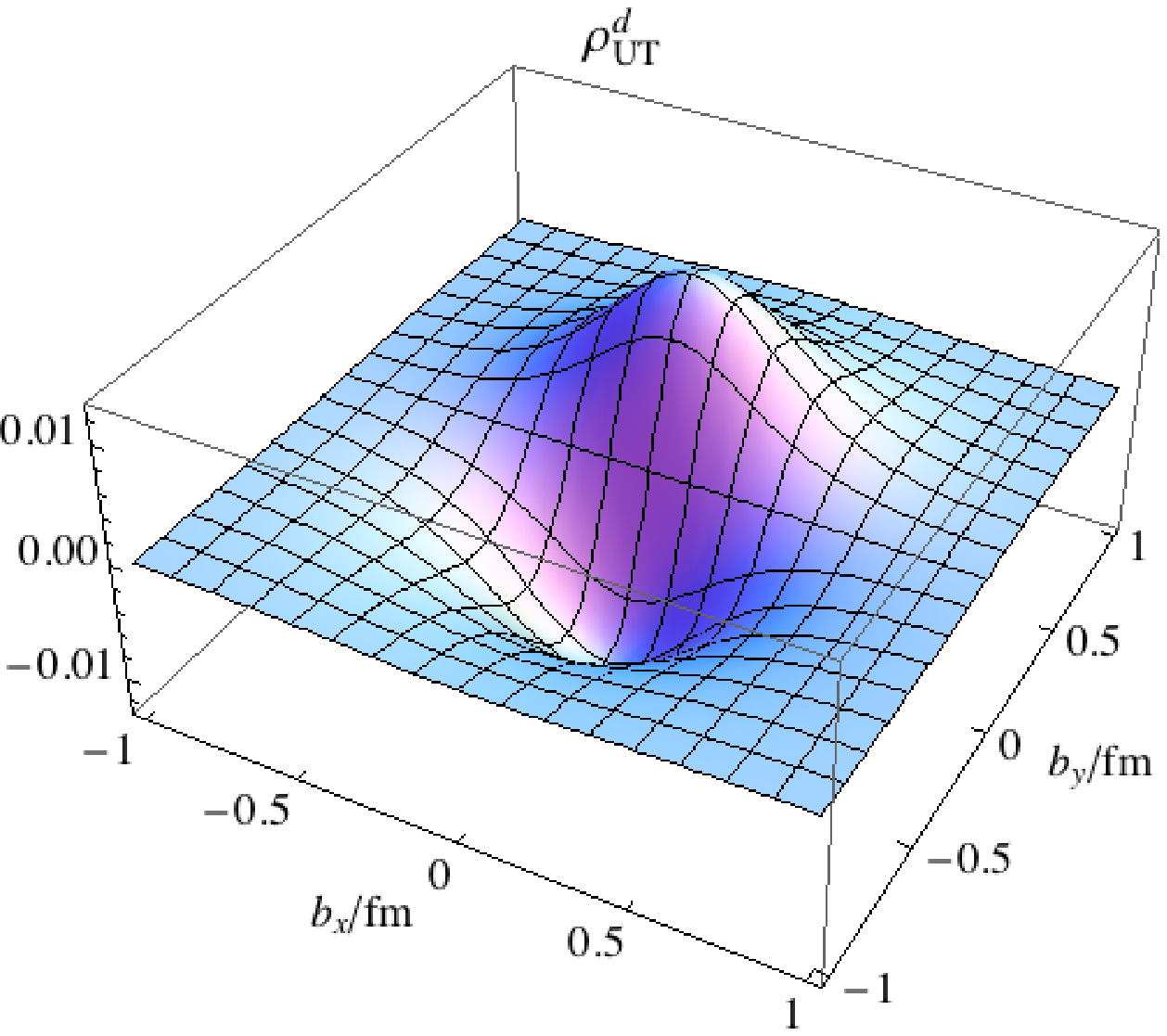}
\includegraphics[width=0.23\textwidth]{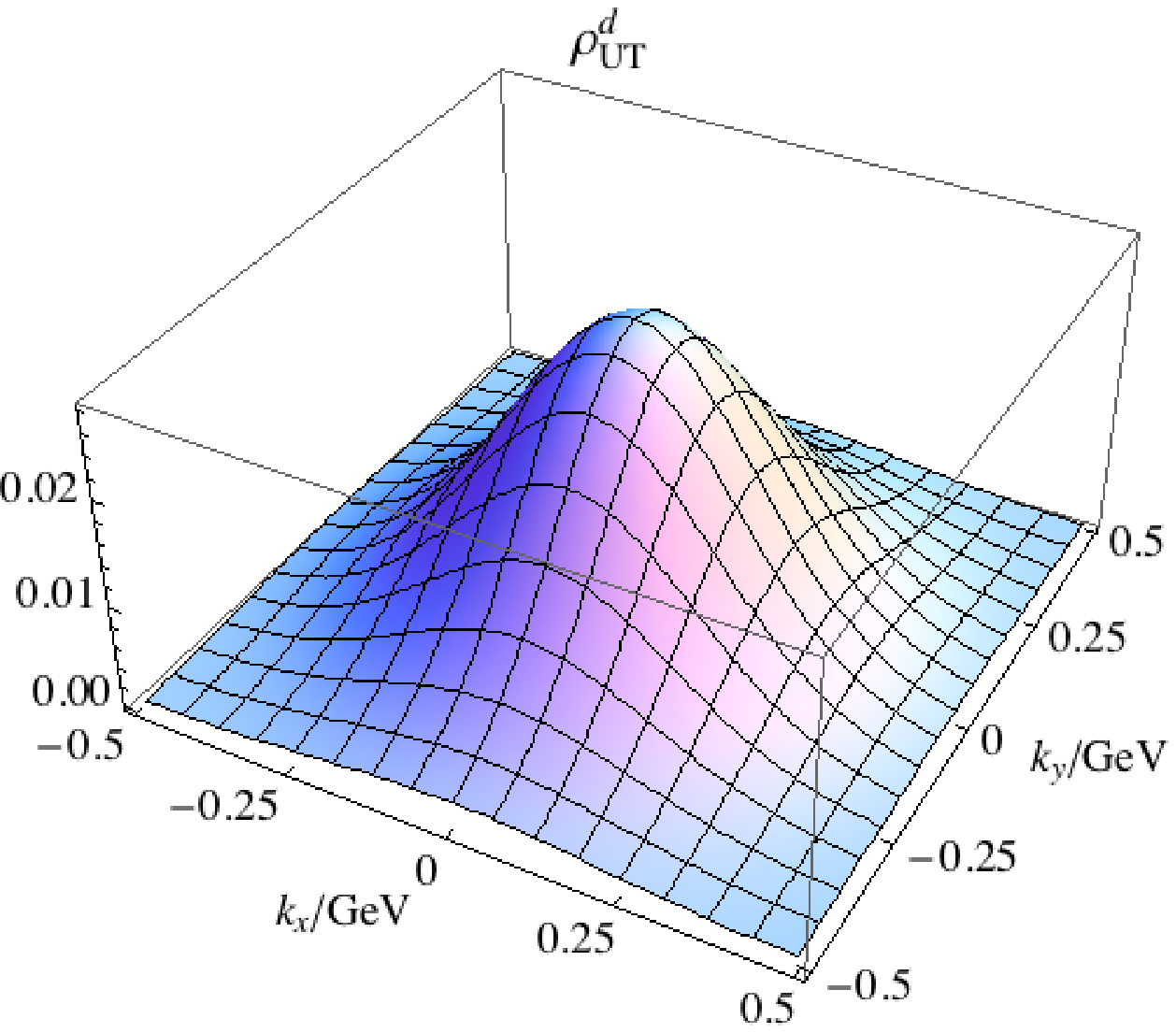}
\includegraphics[width=0.23\textwidth]{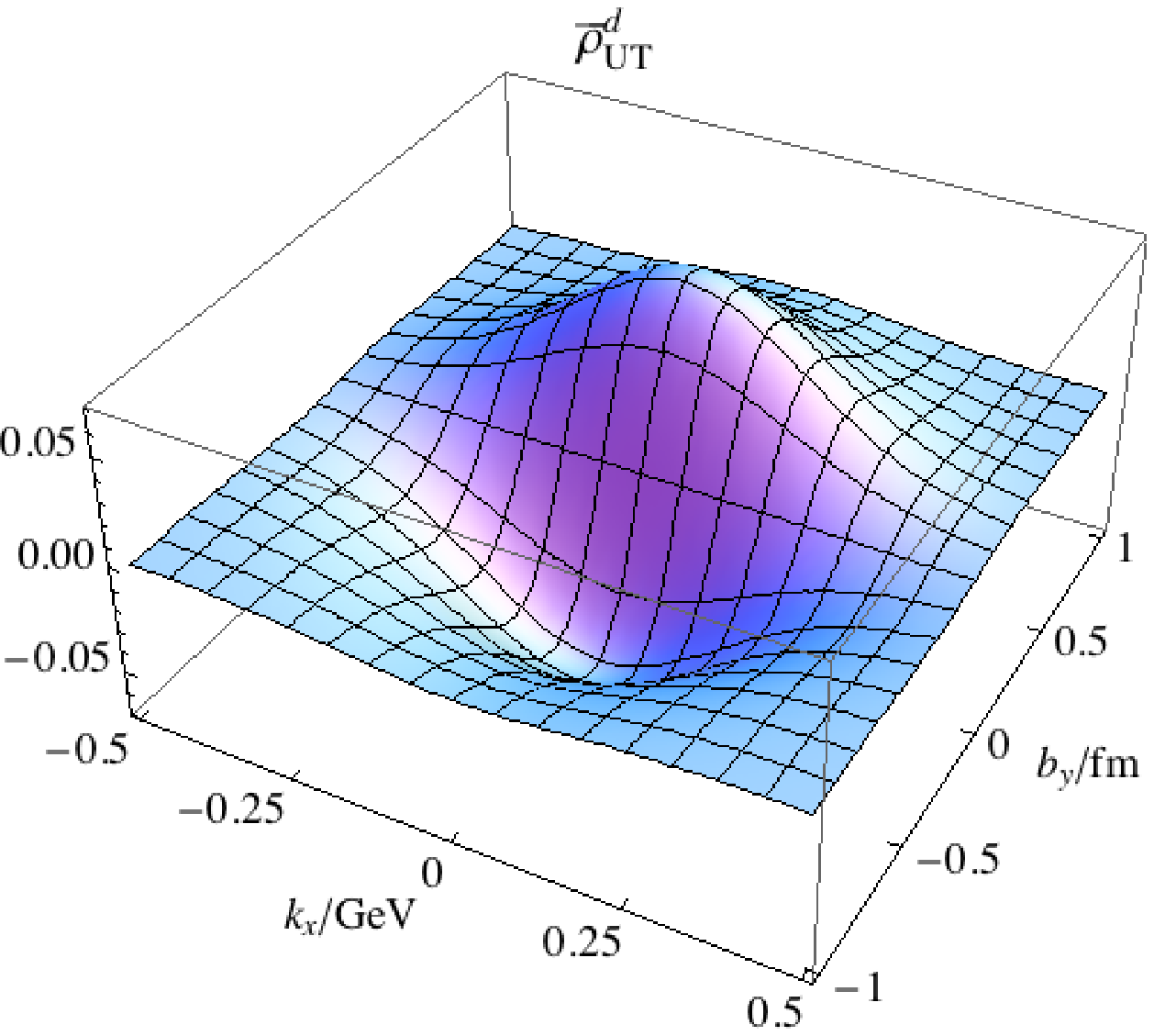}
\caption{(Color online). Unpol-transverse Wigner distributions $\rho_{_\mathrm{UT}}$ and mixing distributions $\bar{\rho}_{_\mathrm{UT}}$ for $u$ quark (upper) and $d$ quark (lower). The first column are the distributions in transverse coordinate space with fixed transverse momentum $\bm{k}_\perp=0.3\,\textrm{GeV}\,\hat{\bm{e}}_x$ parallel to the quark polarization, and the second column are those with fixed transverse momentum $\bm{k}_\perp=0.3\,\textrm{GeV}\,\hat{\bm{e}}_y$ perpendicular to the quark polarization. The third column are the distributions in transverse momentrum space with fixed transverse coordinate $\bm{b}_\perp=0.4\,\textrm{fm}\,\hat{\bm{e}}_y$ perpendicular to the quark polarization. The fourth column are the mixing distributions. \label{rhout}}
\end{figure}

In Fig. \ref{rhouu}, we plot the unpolarized Wigner distributions $\rho_{_\mathrm{UU}}$ and mixing distributions $\tilde{\rho}_{_\mathrm{UU}}$. The unpolarized Wigner distributions represent the transverse phase-space distribution of the unpolarized quark in an unpolarized proton. Its behaviors in transverse coordinate space and transverse momentum space are plotted with fixed transverse momentum and transverse coordinate separately. We denote the direction of the fixed momentum or the fixed coordinate as the $y$-direction, since there is no priviledged transverse direction because of the SO(2) symmetry in the transverse plane. The same case also occurs for $\rho_{_\mathrm{UL}}$, $\rho_{_\mathrm{LU}}$ and $\rho_{_\mathrm{LL}}$ where neither the quark nor the proton is transverse polarized.

As general properties, the left ($-\hat{\bm{e}}_x$)-right ($\hat{\bm{e}}_x$) symmetry of the unpolarized Wigner distributions is a direct result of the space symmetry, and reflects the fact that the proton viewed from any direction is the same. This property is also observed in the mixing distributions where one may clearly find the quark has no preference on moving either clockwise or anticlockwise. It is consistant with our topological knowledge that any preference must break the SO(3) symmetry. The up ($\hat{\bm{e}}_y$)-down ($-\hat{\bm{e}}_y$) symmetry of the Wigner distributions reflects the nature that the quarks in the proton is in equilibrium. The two classes of mixing distributions defined in (\ref{mixing1}) and (\ref{mixing2}) have a simple relation
\begin{equation}
\tilde{\rho}_{_\mathrm{UU}}(b_x,k_y,x)=\bar{\rho}_{_\mathrm{UU}}(b_y,-k_x,x),
\end{equation}
and this relation also survives for $\tilde{\rho}_{_\mathrm{UL}}$, $\tilde{\rho}_{_\mathrm{LU}}$ and $\tilde{\rho}_{_\mathrm{LL}}$.

Comparing the behaviors of the $u$ quark and the $d$ quark, we find in our model that the $u$ quark favors concentrating in the center relative to the $d$ quark in the coordinate space, while they have similar spread behaviors in the momentum space. Though it is a common result to find the $u$ concentrates in the center in quark models, this property does have dependence on the parameters in the model. The similar behavior in the momentum space is due to the same choice of the $\beta$ parameter in the scalar and axial-vector cases. Besides, as found in some constituent quark model and chiral quark soliton model~\cite{Lorce:2011kd}, the property that the distributions decrease faster in $y$-direction than $x$-direction is not obvious in our model.

In Fig. \ref{rhoul}, we plot the unpol-longitudinal Wigner distributions $\rho_{_\mathrm{UL}}$ and mixing distributions $\tilde{\rho}_{_\mathrm{UL}}$. They describe the longitudinal polarized quark in an unpolarized proton. This distribution will vanish in either TMD or IPD limit, and therefore no general relations at twist-two are found between the TMDs or IPDs and the unpol-longitudinal Wigner distributions. Hence the information contained in this distribution cannot be observed from the leading twist TMDs or IPDs, though some effects might be found at subleading twist.

Due to the priviledged direction defined by the quark polarization, dipole structures are found in the unpol-longitudinal Wigner distributions, and correspondingly quadrupole structures appear in the mixing distributions. Considering the physical interpretations, these distributions essentially reflect quark spin-orbit correlations. This correlation can be clearly observed from the quadrupole structure of the unpol-longitudinal mixing distributions, and no correlation case corresponds to vanishing mixing distributions.

We find in our model that both the $u$ quark and the $d$ quark have negative spin-orbit correlations. This result has little dependence on the choice of momentum space wave functions or the parameters in the model. It is mainly determined by the spin structures of the quark and the spectator. However, it is indeed model dependent, and our result is opposite to that from some constituent quark model~\cite{Lorce:2011kd}. Therefore realistic analyses in QCD on this distribution may help us clarify this issue.

In Fig. \ref{rhout}, we plot the unpol-transverse Wigner distributions $\rho_{_\mathrm{UT}}$ and mixing distributions $\bar{\rho}_{_\mathrm{UT}}$. They describe the transverse polarized quark in an unpolarized proton. The SO(2) symmetry in the transverse space is explicity broken by the quark polarization direction, referred to as $x$-direction.

At the TMD limit, the unpol-transverse Wigner distribution will reduce to the Boer--Mulders function $h_1^{\perp}$, while at the IPD limit, it is related to the $\tilde{H}_T$ together with some other distributions. Whereas, the $h_1^\perp$ corresponds to the T-odd part, while the $\tilde{H}_T$ corresponds to the T-even part. Since the T-odd part is neglected in this paper, the results of $\rho_{_\textrm{UT}}$ here will vanish at the TMD limit.

As a general property determined by the spin structures, the unpol-transverse Wigner distributions vanish when the quark intrinsic transverse coordinate is parallel to the polarization. In other words, quark transverse spin has no correlation with its parallel transverse coordinate. Besides, the results displayed here have no dependence on the direction of quark transverse momentum, but this behavior may be changed by a nontrivial Wilson line.

\subsection{Longitudinal polarized proton}

\begin{figure}
\includegraphics[width=0.25\textwidth]{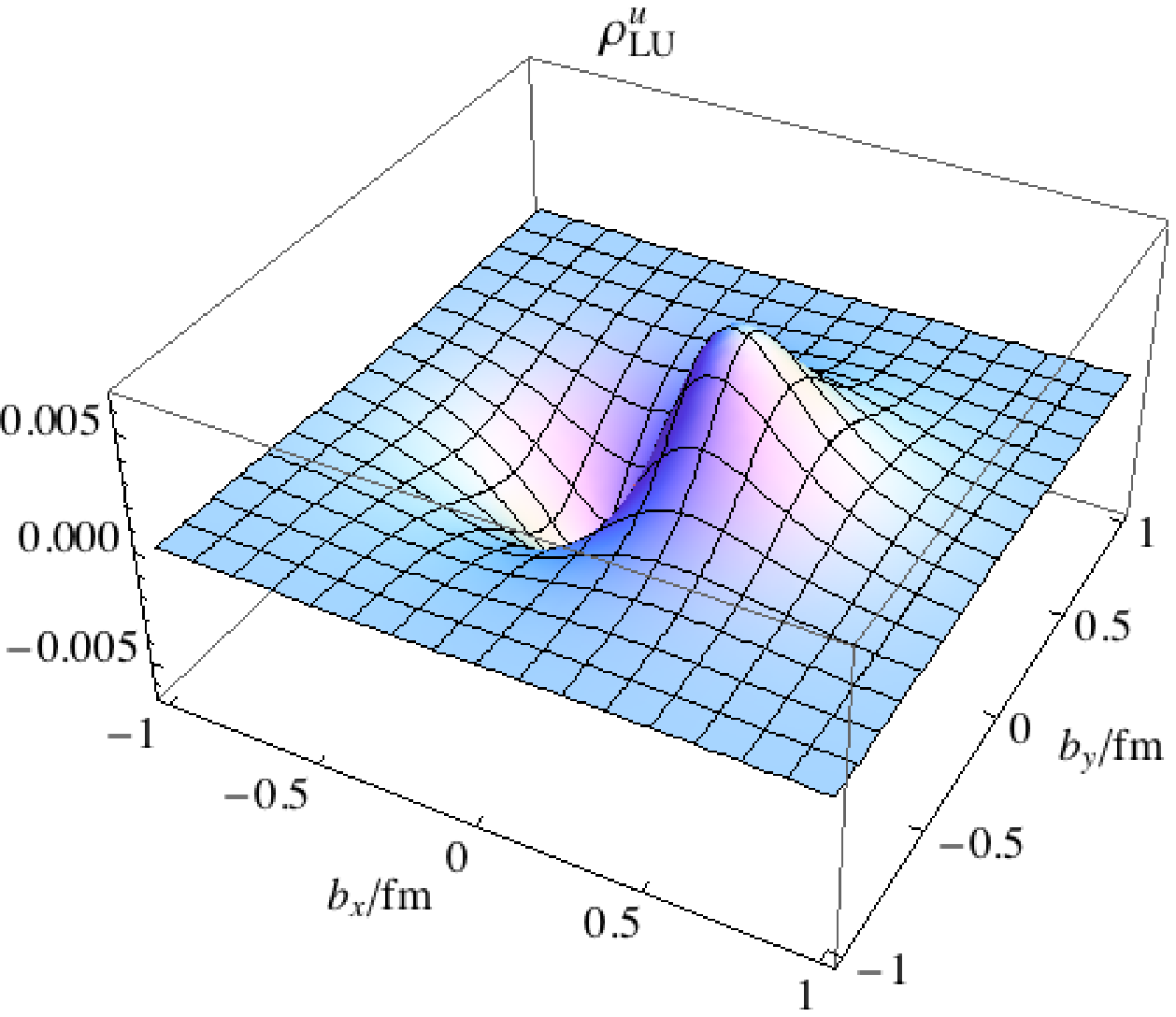}
\includegraphics[width=0.25\textwidth]{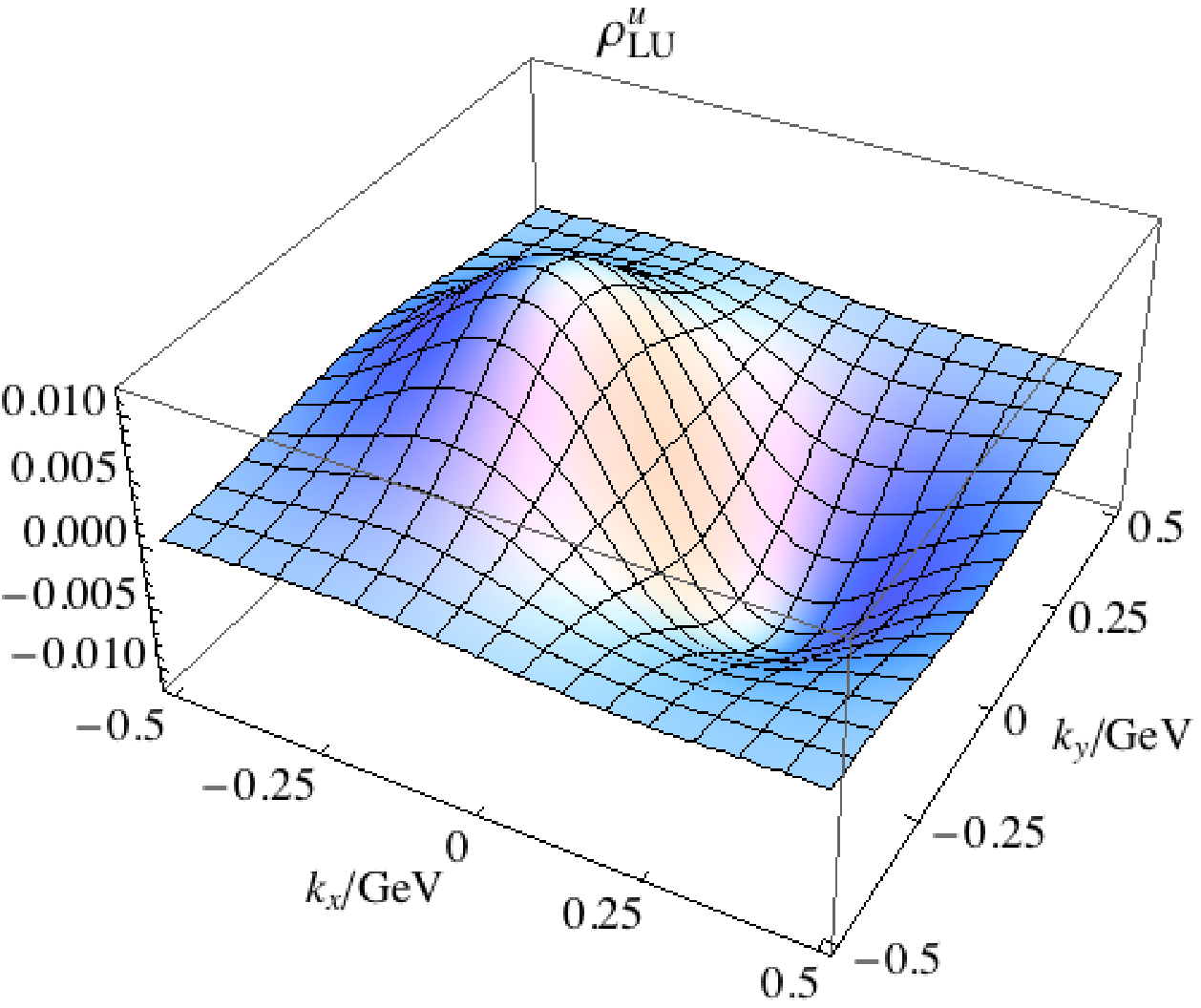}
\includegraphics[width=0.25\textwidth]{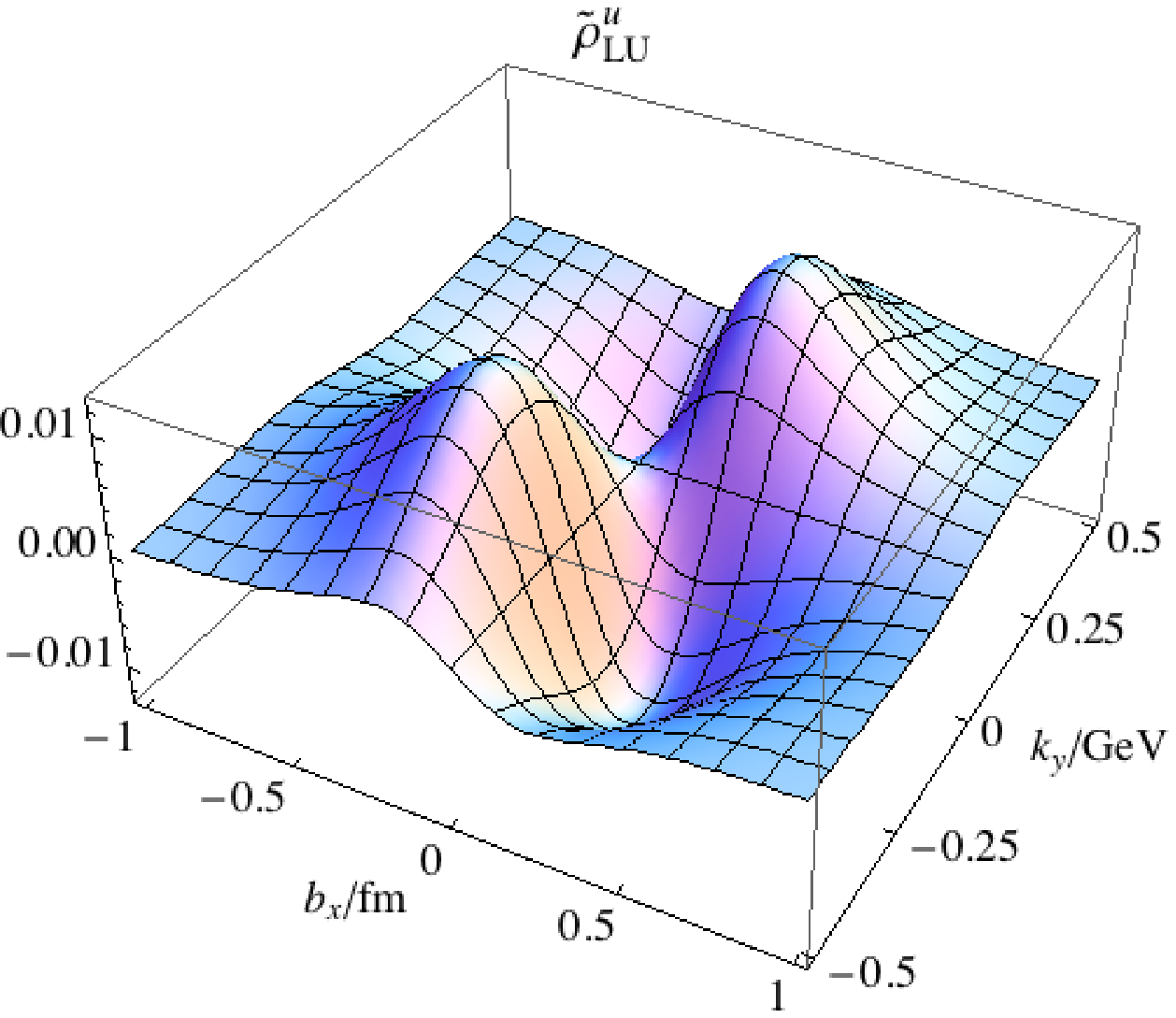}
\includegraphics[width=0.25\textwidth]{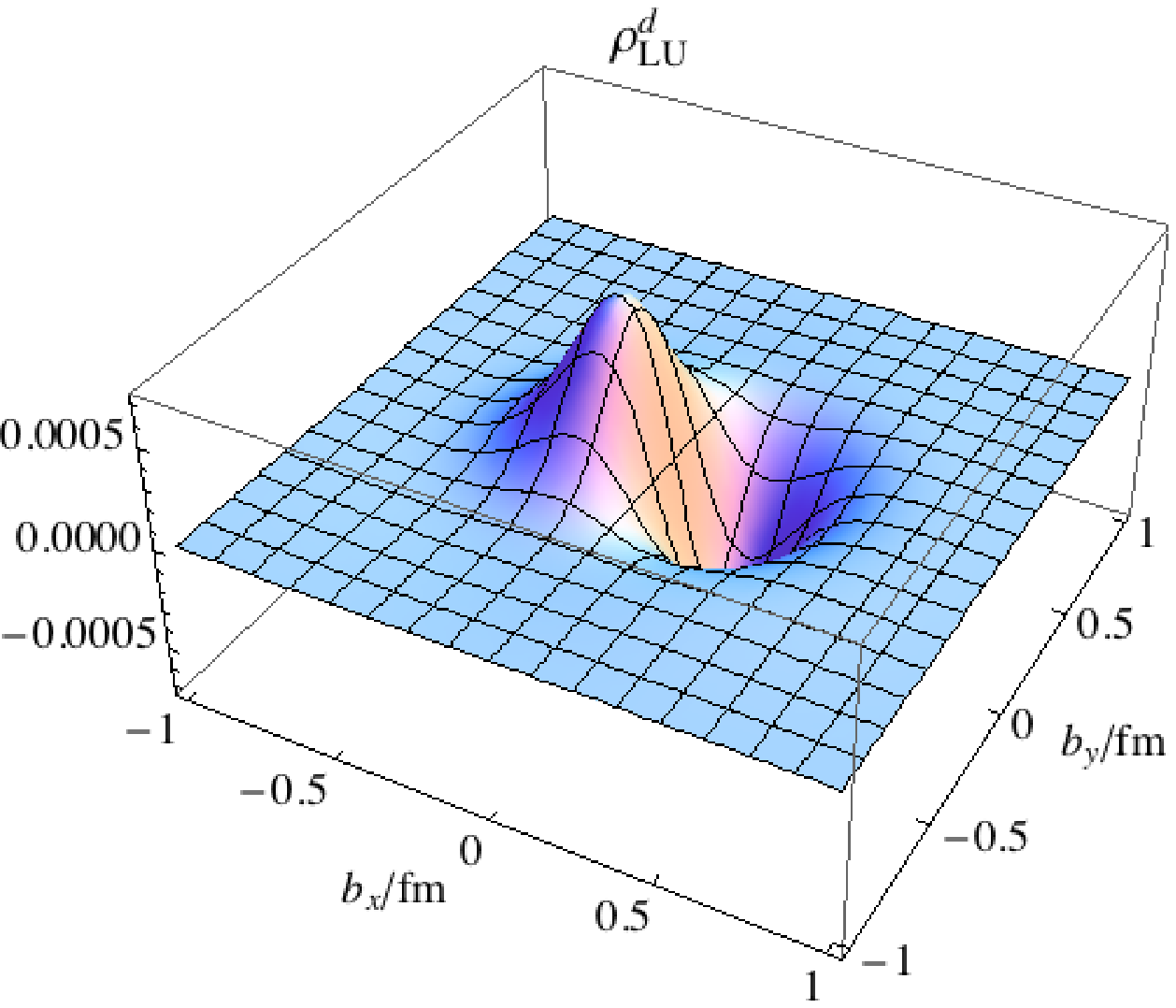}
\includegraphics[width=0.25\textwidth]{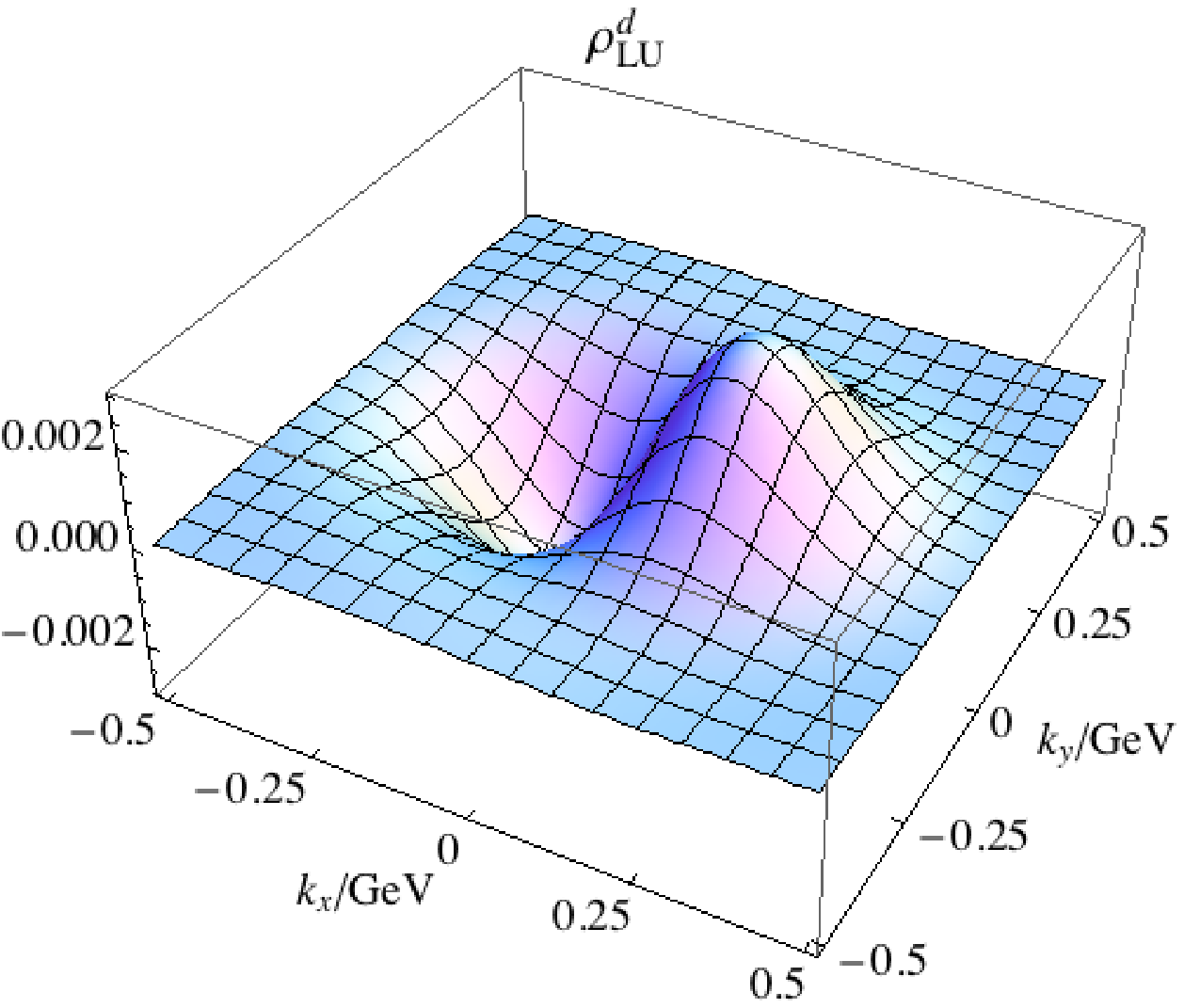}
\includegraphics[width=0.25\textwidth]{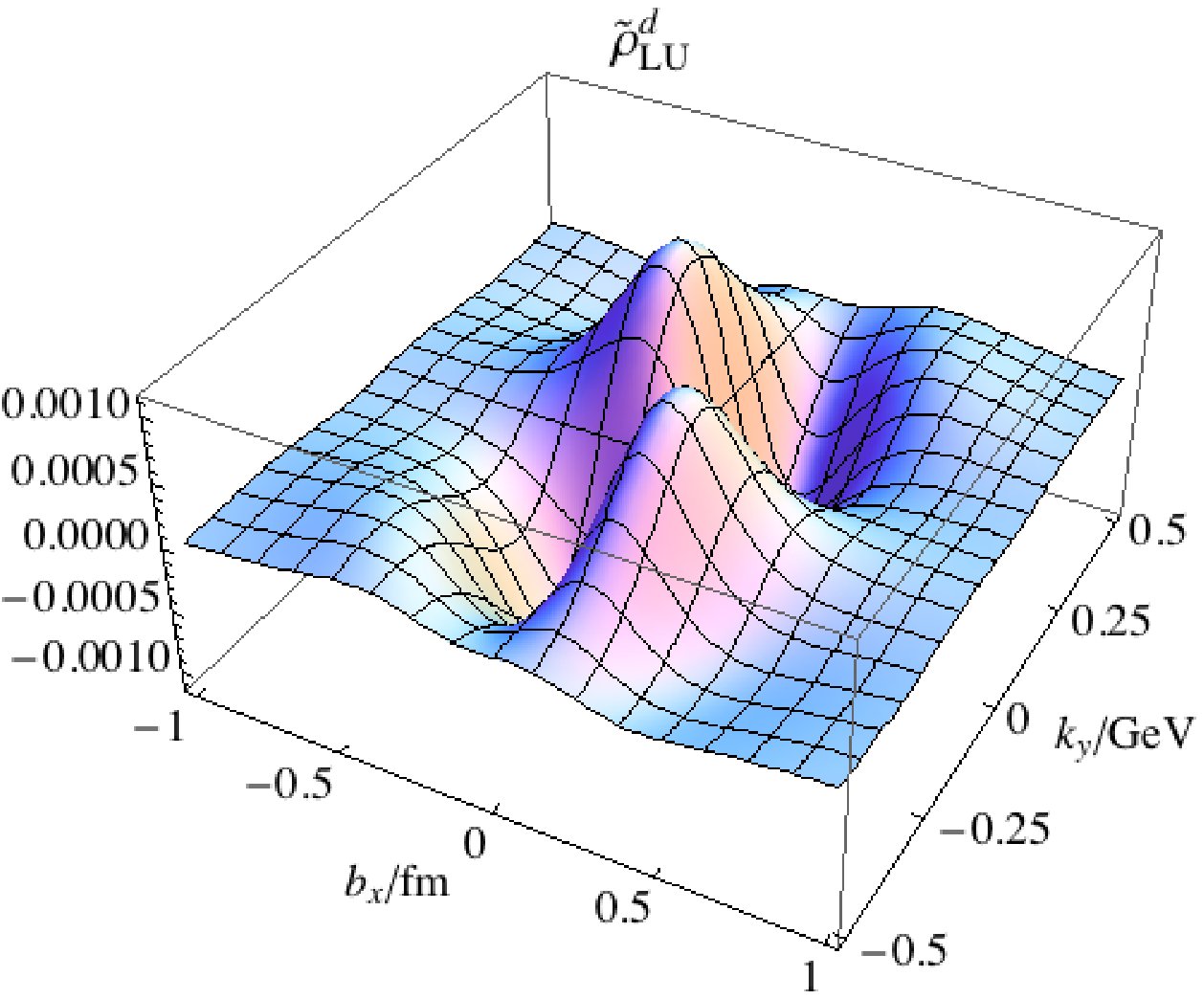}
\caption{(Color online). Longi-unpolarized Wigner distributions $\rho_{_\mathrm{LU}}$ and mixing distributions $\tilde{\rho}_{_\mathrm{LU}}$ for $u$ quark (upper) and $d$ quark (lower). The first column the Wigner distributions in transverse coordinate space with definite transverse momentum $\bm{k}_\perp=0.3\,\textrm{GeV}\,\hat{\bm{e}}_y$. The second column are the Wigner distributions in transverse momentum space with definite transverse coordinate $\bm{b}_\perp=0.4\,\textrm{fm}\,\hat{\bm{e}}_y$. The third column are the mixing distributions $\tilde{\rho}_{_\mathrm{LU}}$. \label{rholu}}
\end{figure}
\begin{figure}
\includegraphics[width=0.25\textwidth]{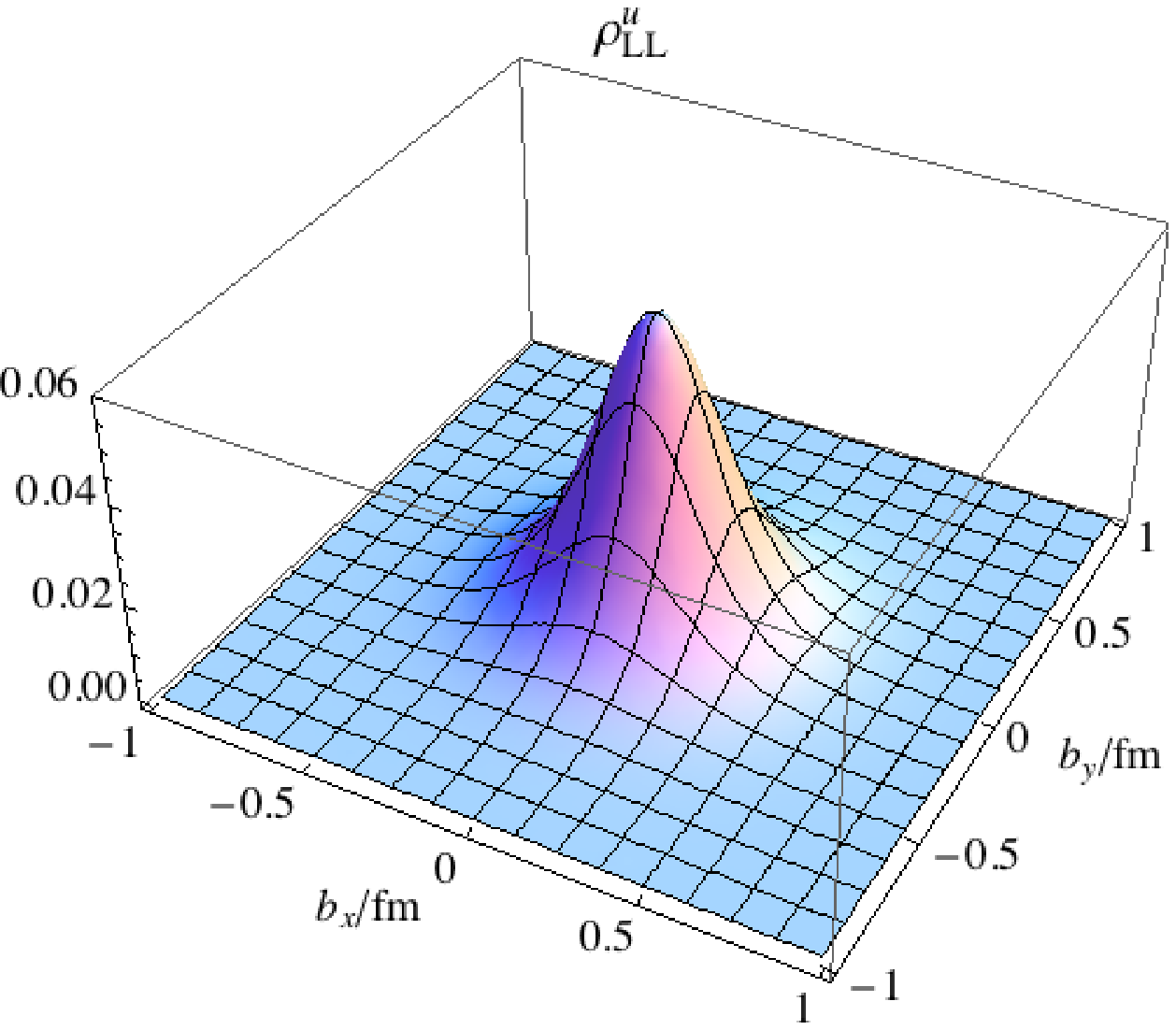}
\includegraphics[width=0.25\textwidth]{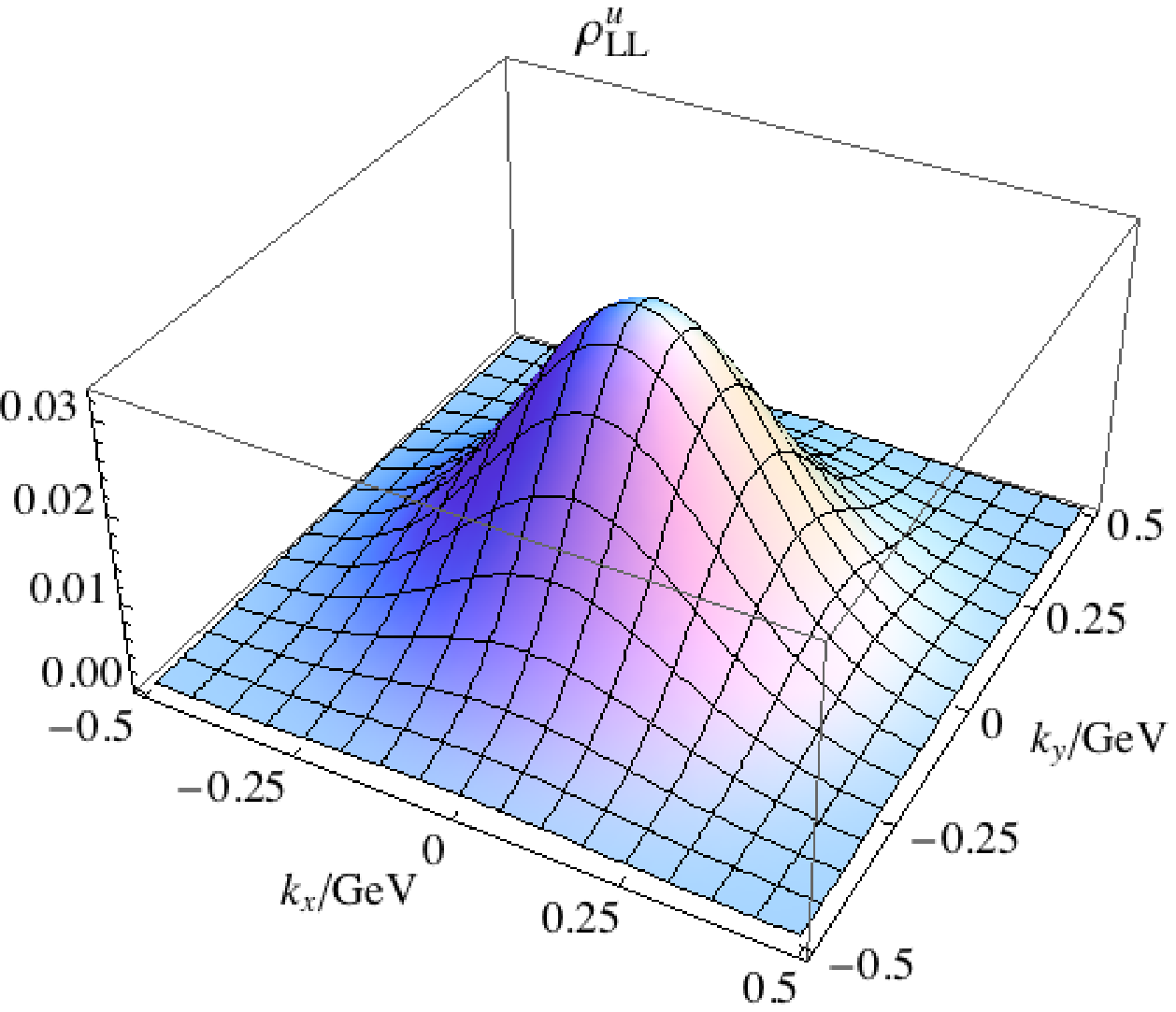}
\includegraphics[width=0.25\textwidth]{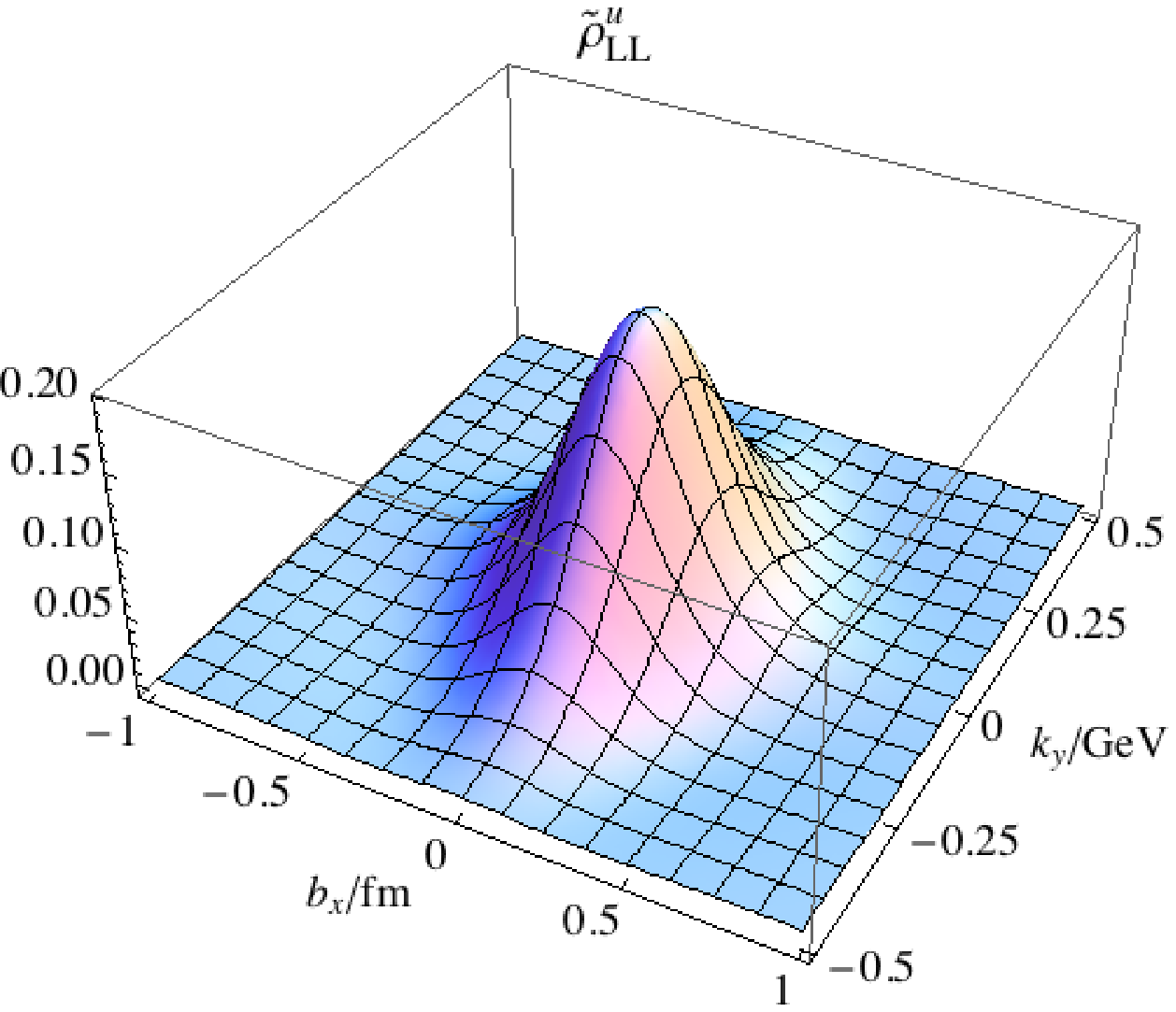}
\includegraphics[width=0.25\textwidth]{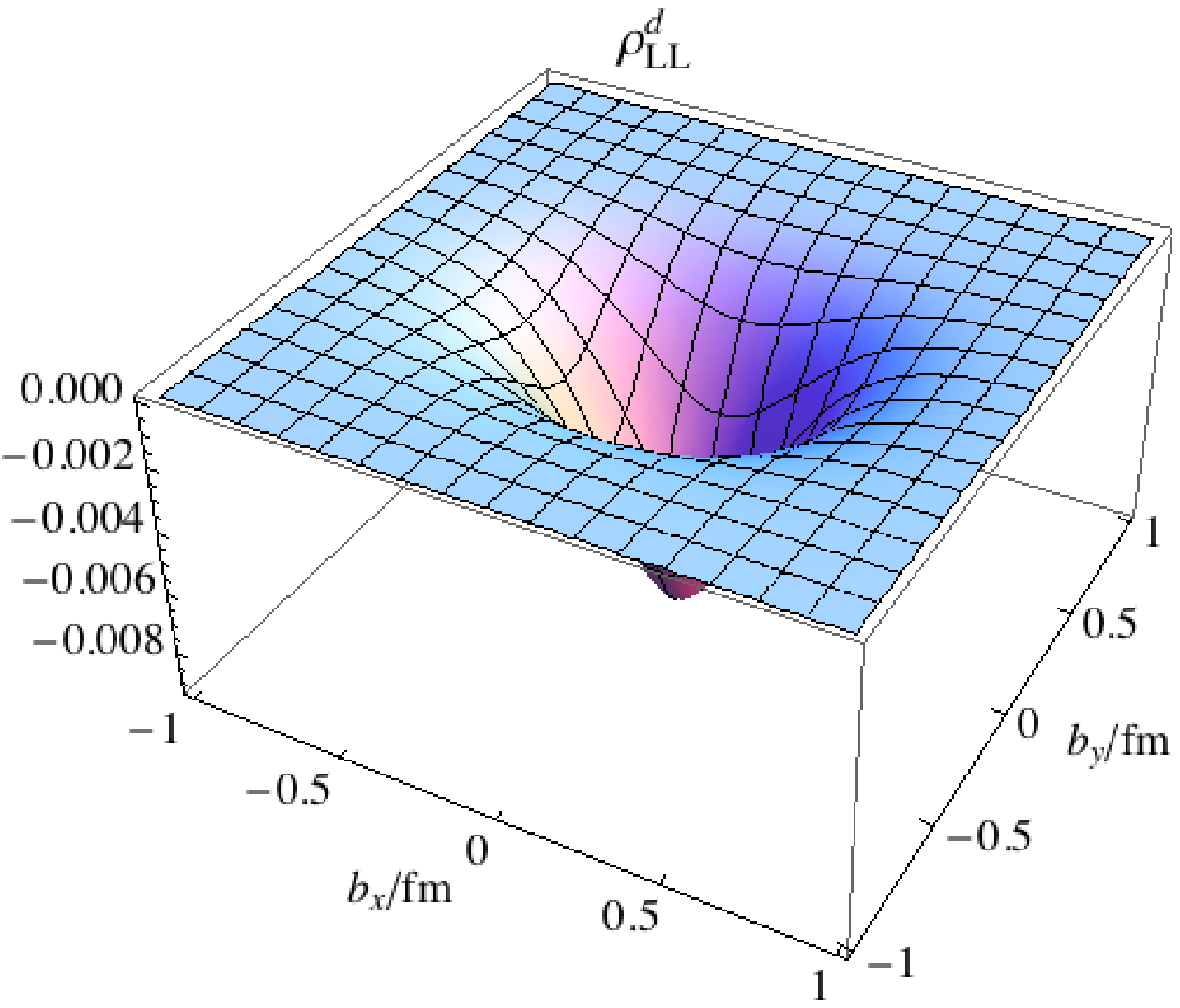}
\includegraphics[width=0.25\textwidth]{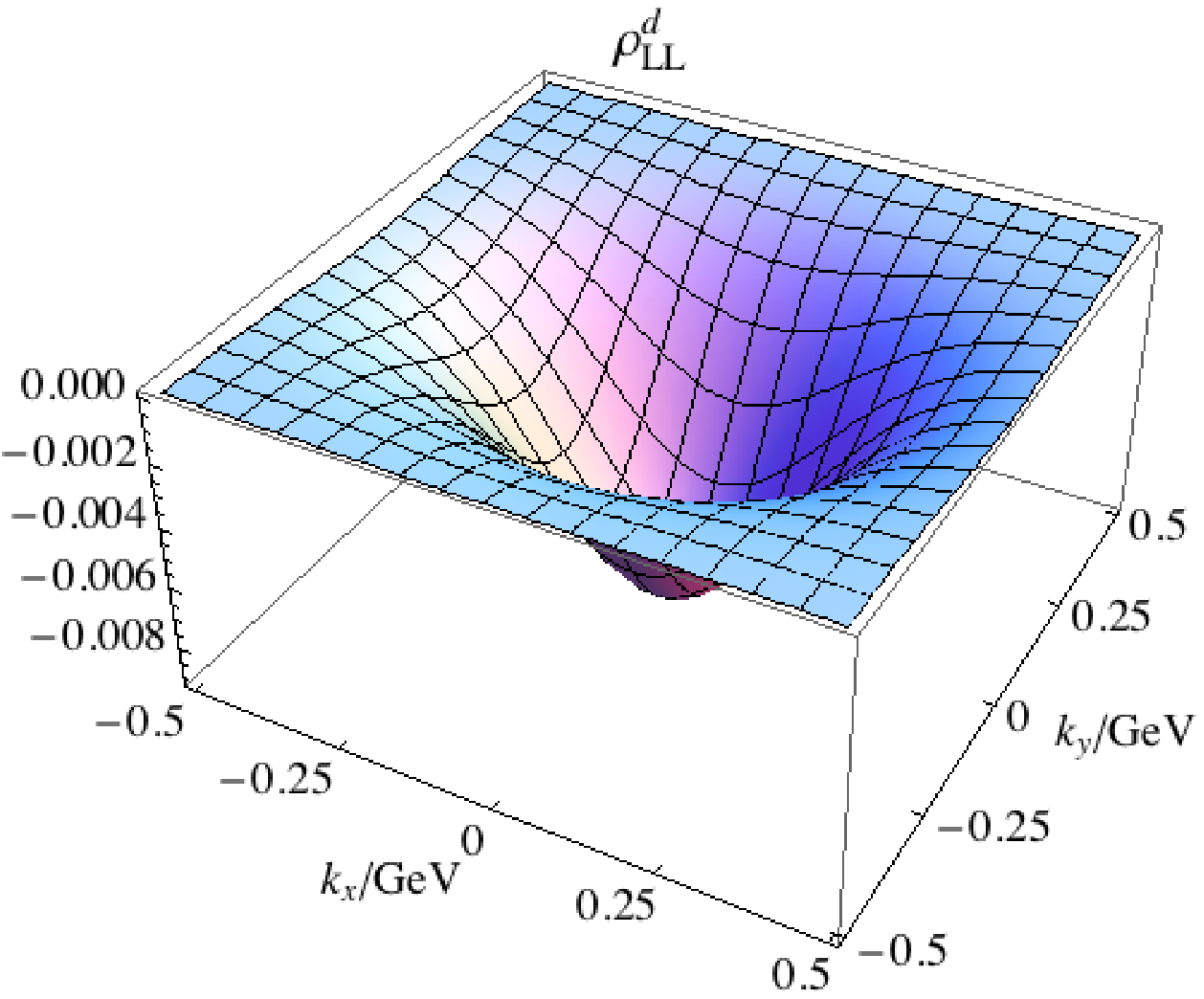}
\includegraphics[width=0.25\textwidth]{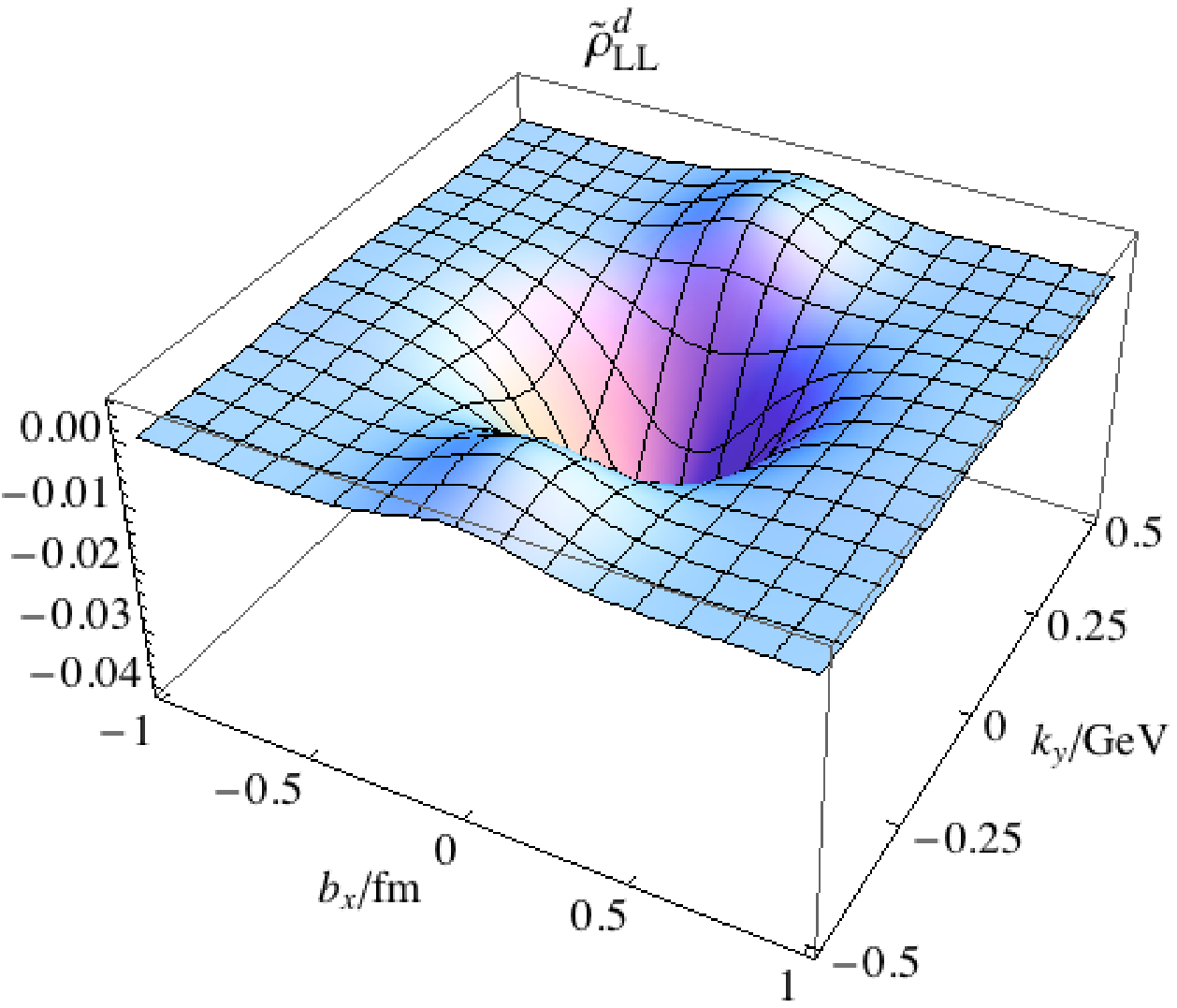}
\caption{(Color online). Longitudinal Wigner distributions $\rho_{_\mathrm{LL}}$ and mixing distributions $\tilde{\rho}_{_\mathrm{LL}}$ for the $u$ quark (upper) and the $d$ quark (lower). The first column the Wigner distributions in transverse coordinate space with definite transverse momentum $\bm{k}_\perp=0.3\,\textrm{GeV}\,\hat{\bm{e}}_y$. The second column are the Wigner distributions in transverse momentum space with definite transverse coordinate $\bm{b}_\perp=0.4\,\textrm{fm}\,\hat{\bm{e}}_y$. The third column are the mixing distributions $\tilde{\rho}_{_\mathrm{LL}}$. \label{rholl}}
\end{figure}
\begin{figure}
\includegraphics[width=0.23\textwidth]{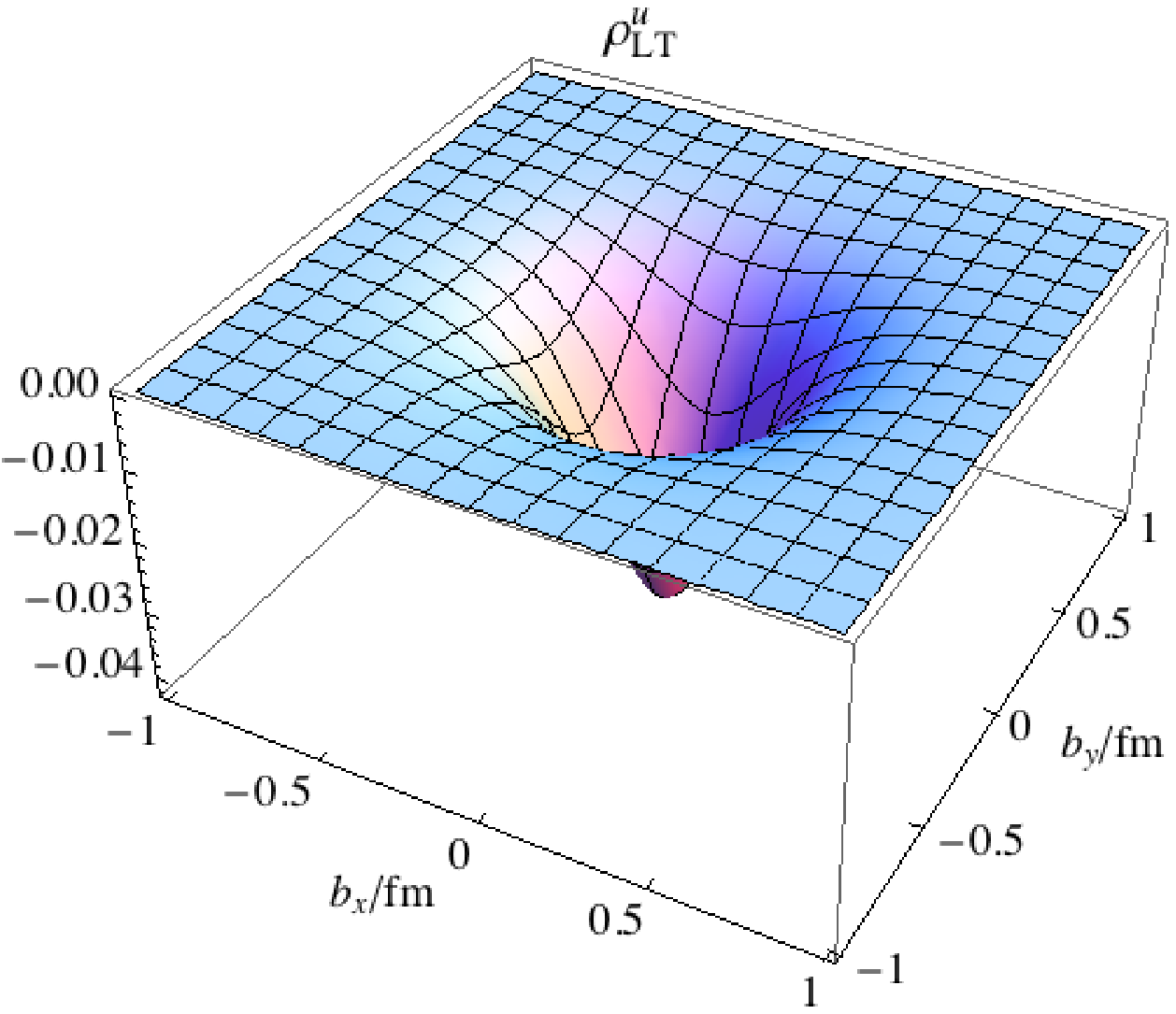}
\includegraphics[width=0.23\textwidth]{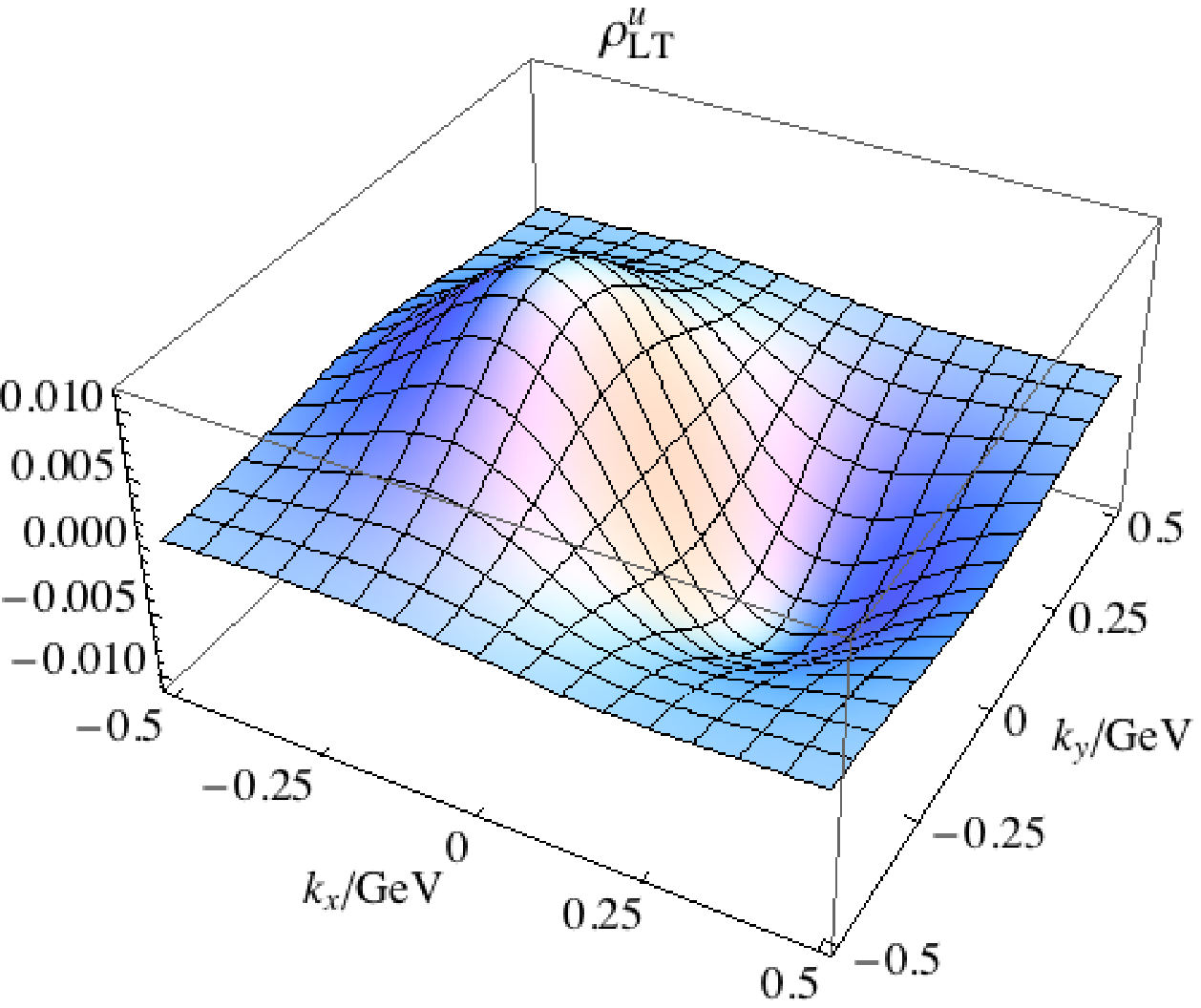}
\includegraphics[width=0.23\textwidth]{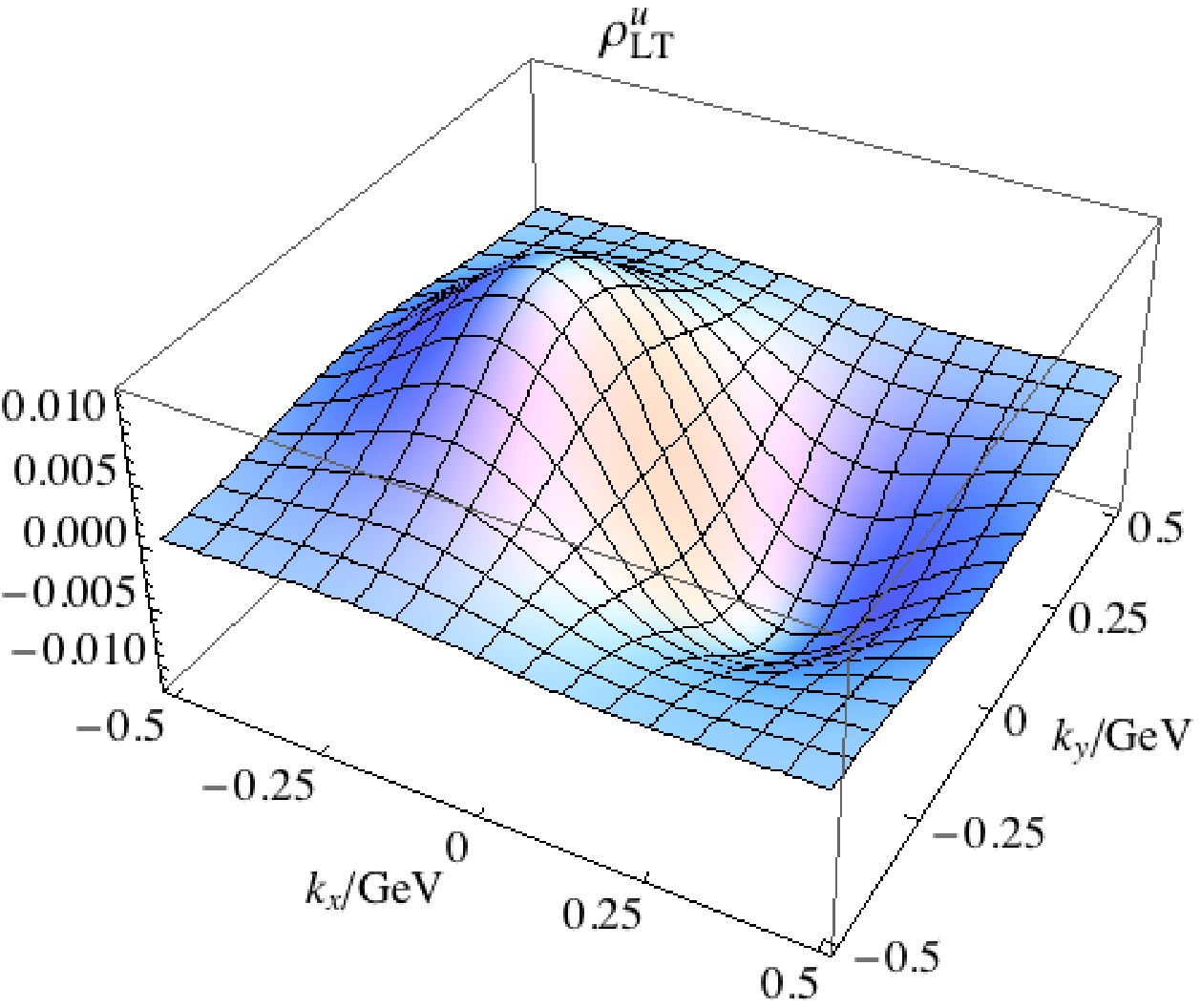}
\includegraphics[width=0.23\textwidth]{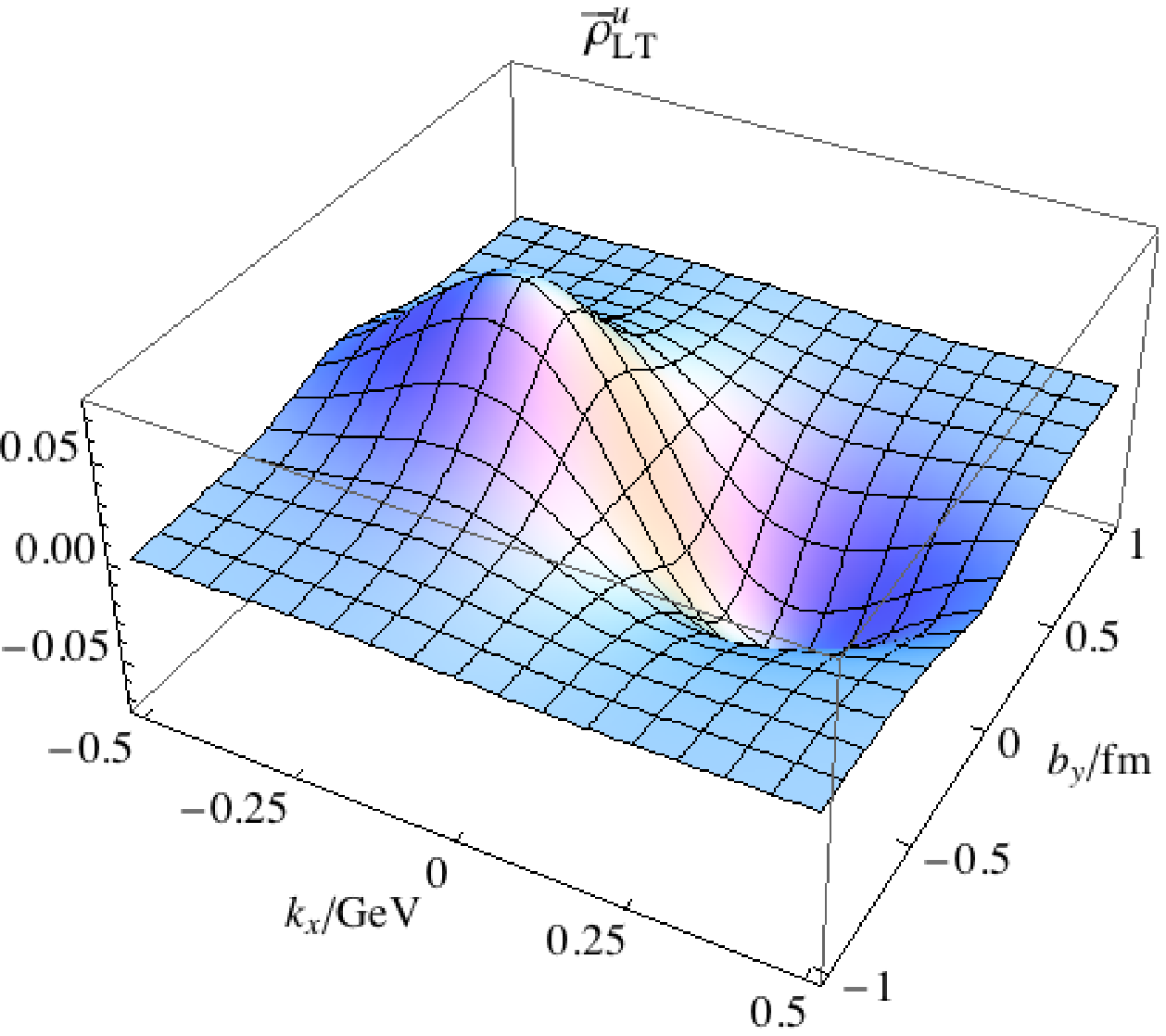}
\includegraphics[width=0.23\textwidth]{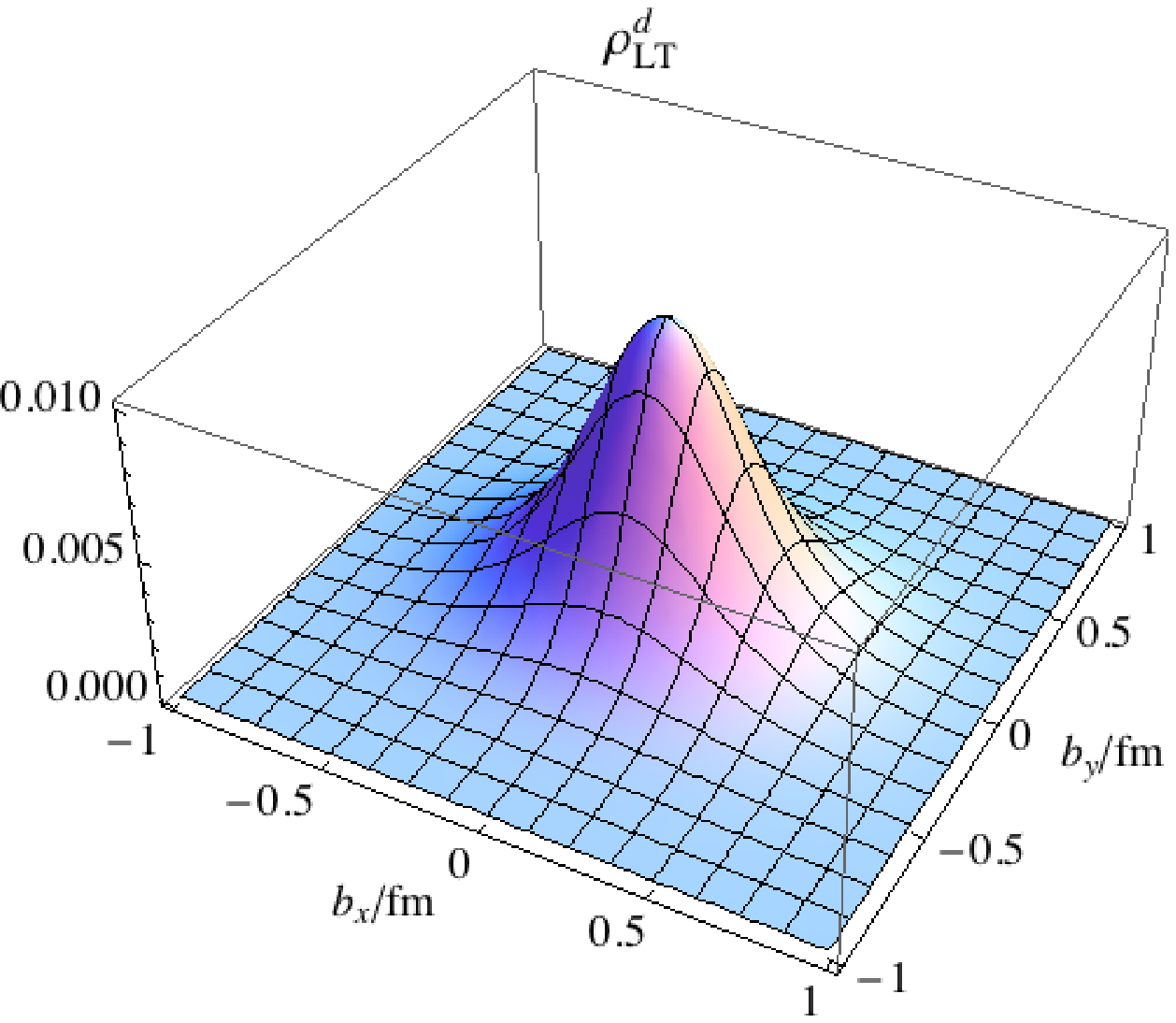}
\includegraphics[width=0.23\textwidth]{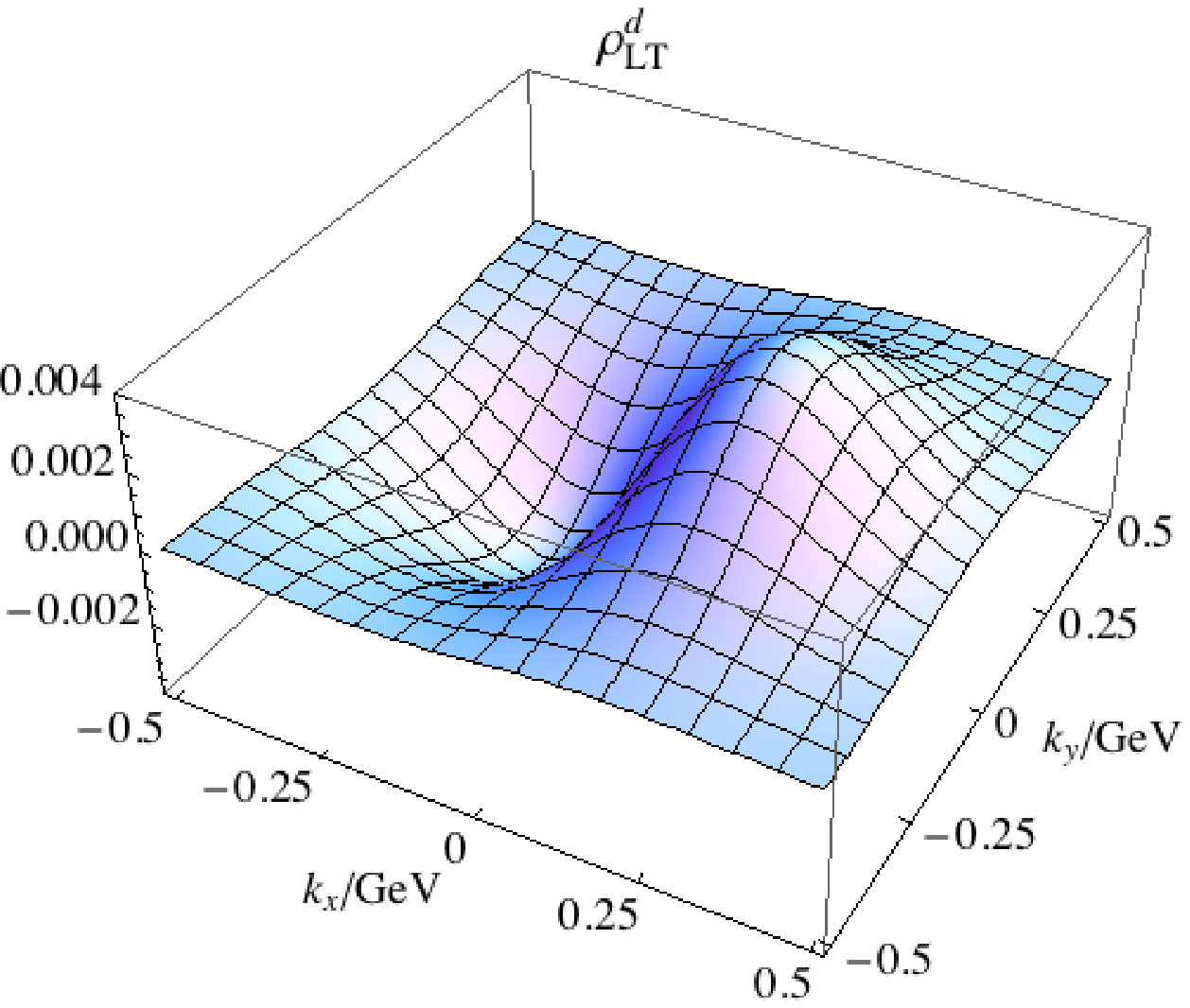}
\includegraphics[width=0.23\textwidth]{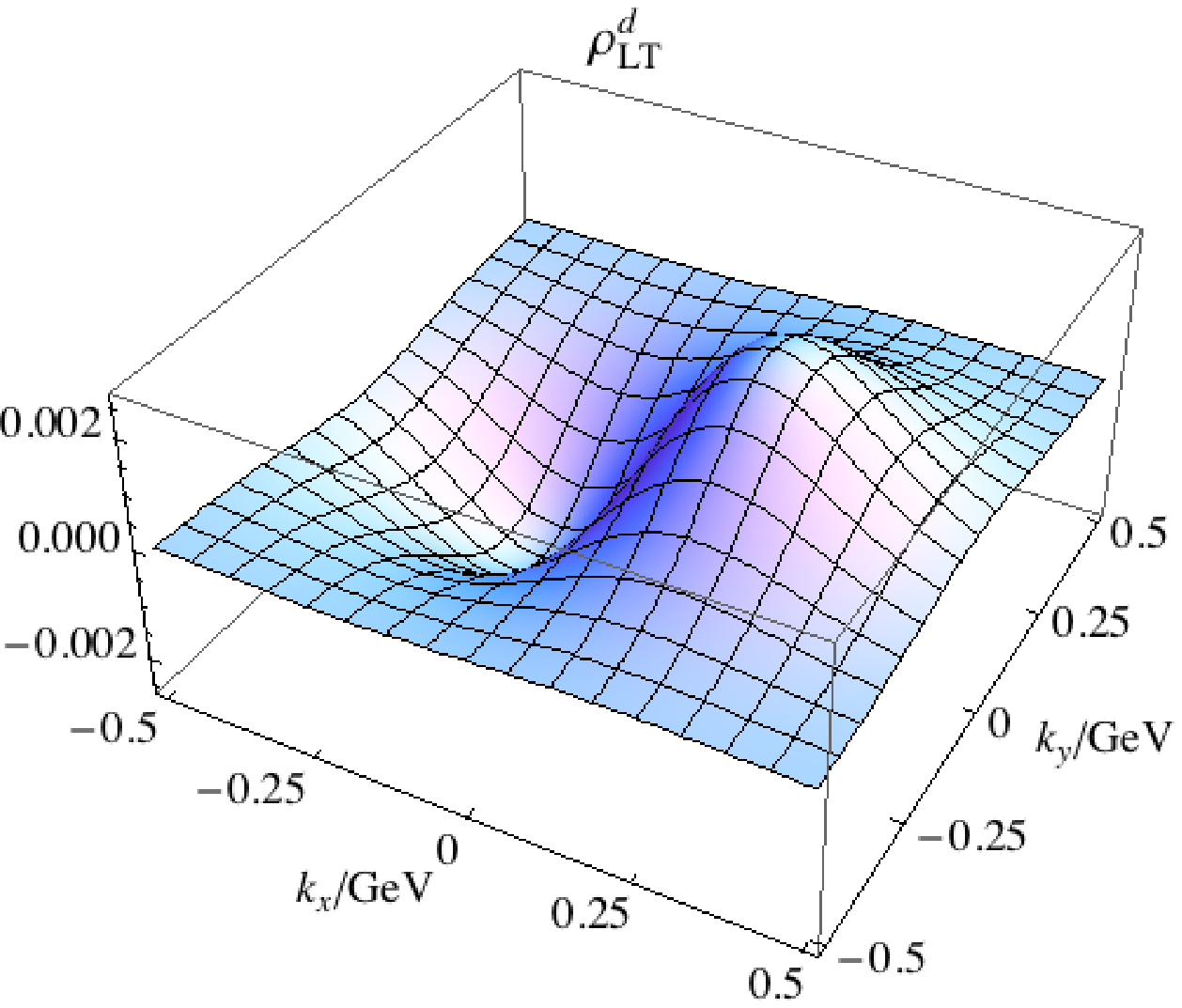}
\includegraphics[width=0.23\textwidth]{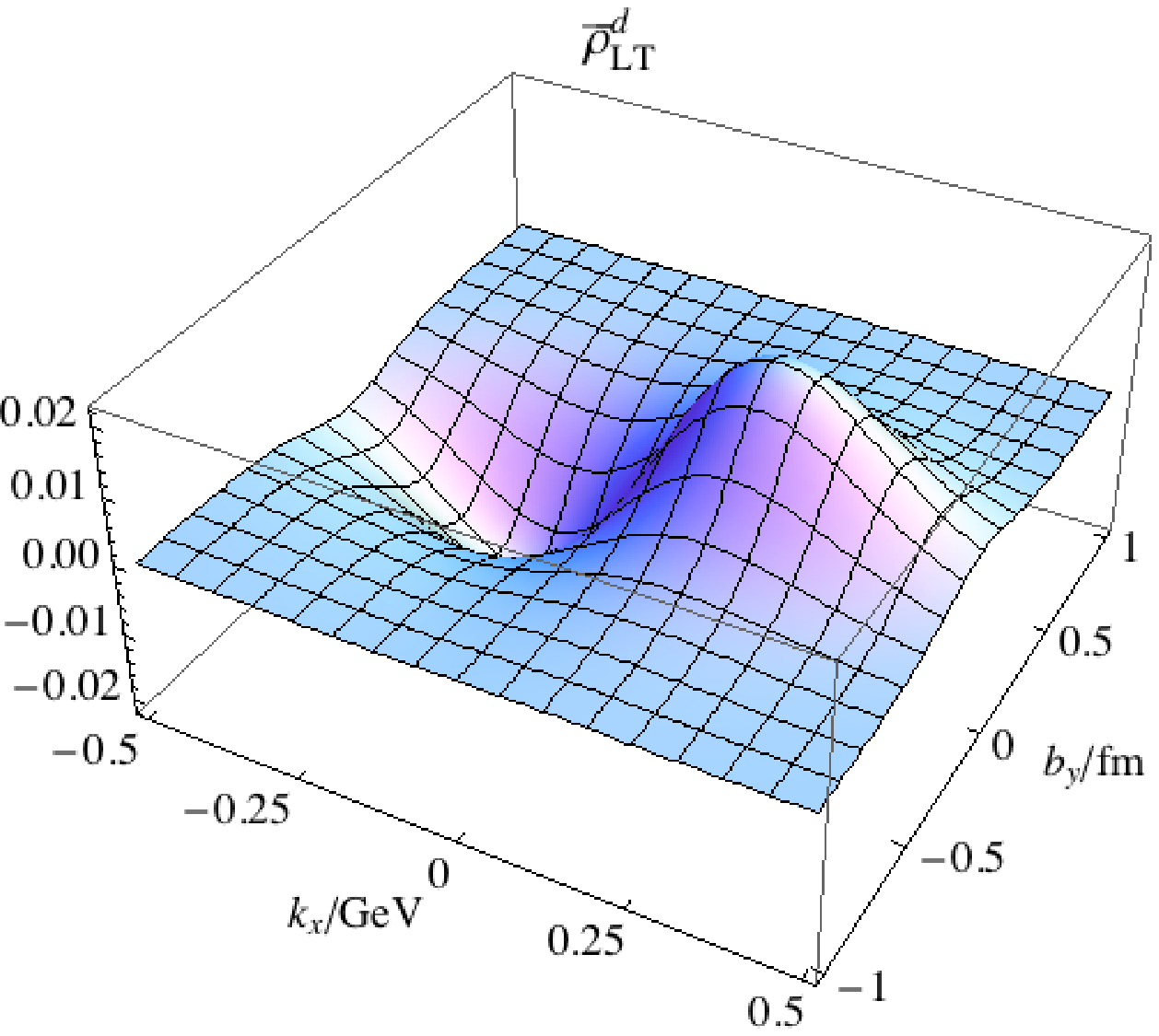}
\caption{(Color online). Longi-transversal Wigner distributions $\rho_{_\mathrm{LT}}$ with fixed $\bm{k}_\perp$ for the $u$ quark (upper) and the $d$ quark (lower). The first column are the distributions in transverse coordinate space with fixed transverse momentum $\bm{k}_\perp=0.3\,\textrm{GeV}\,\hat{\bm{e}}_x$ parallel to the quark polarization. The second column are the distributions in transverse momentrum space with fixed transverse coordinate $\bm{b}_\perp=0.4\,\textrm{fm}\,\hat{\bm{e}}_x$ parallel to the quark polarization, and the third column are those with fixed transverse coordinate $\bm{b}_\perp=0.4\,\textrm{fm}\,\hat{\bm{e}}_y$ perpendicular to the quark polarization. The fourth column are the mixing distributions $\bar{\rho}_{_\mathrm{LT}}$. \label{rholt}}
\end{figure}

In Fig. \ref{rholu}, we plot the longi-unpolarized Wigner distributions $\rho_{_\mathrm{LU}}$ and mixing distributions $\tilde{\rho}_{_\mathrm{LU}}$. They describe quark phase-space distributions in a longitudinal polarized proton without any information of quark spins. Thus they are suitable to study quark orbital angular momentum related issues~\cite{Lorce:2011ni}. Similar to the unpol-longitudinal Wigner distribution, no twist-two TMDs or IPDs are related to the longi-unpolarized Wigner distribution. Therefore it will vanish at TMD or IPD limit, and the phase-space behavior in this distribution cannot generally be extracted from leading twist TMDs or IPDs.

The longi-unpolarized Wigner distributions have dipole structures, and correspondingly the mixing distributions have quadrupole structures. The preference of quark orbital motions, clockwise or anticlockwise, are clearly observed from them. We find in our calculations that the $u$ quark has positive orbital angular momentum while the $d$ quark has negative orbital angular momentum, but this is model dependent and the choice of the wave functions may also change this behavior. In addition, the sign change of the $d$ quark distribution in large coordinate or momentum region is found, as also observed in some constituent quark model. Comparing the longi-unpolarized distributions with the unpol-longitudinal distributions, we find in this model that quark orbital motions have stronger correlation to quark spins than to nucleon spins. However, nor does this property is general. Therefore more careful investigation on this distribution is required.

In Fig. \ref{rholl}, we plot the longitudinal Wigner distributions $\rho_{_\mathrm{LL}}$ and mixing distributions $\tilde{\rho}_{_\mathrm{LL}}$. They describe the longitudinal polarized quark in a longitudinal polarized proton, and correspond to the helicity distributions after integrating over transverse variables.

In the simple model, we find the $u$ quark is positive polarized, while the $d$ quark is negative polarized. This is qualitatively consistant with our knowledge from the axial charge. In the large transverse momentum region, a sign change is clearly observed in $d$ quark mixing distribution, as also found in some constituent quark model. This kind of sign change is also found in large longitudinal momentum region of $d$ quark helicity distributions in similar models~\cite{Bacchetta:2008af}. Besides, it is nature to find that the quark helicity distributions concentrate on the center in the phase-space, though it is a result of a simplistic model.

In Fig. \ref{rholt}, we plot the longi-transverse Wigner distributions $\rho_{_\mathrm{LT}}$ and mixing distributions $\bar{\rho}_{_\mathrm{LT}}$. They discribe the correlation of the phase-space distribution with quark transverse spin and proton longitudinal spin. At the TMD limit, the longi-transverse Wigner distribution will reduce to one worm-gear function, the longi-transversity $h_{1L}^\perp$. At the IPD limit, it is related to the IPDs $H_T$ and $\tilde{H}_T$ together with other distributions. Unlike the unpol-transverse distributions, both the TMD and the IPDs are related to the T-even part of the longi-transverse Wigner distribution at certain limits. Thus it is possible to build some relation between them, though no general relations have been found yet.

In the longi-transverse Wigner distribution, the transverse SO(2) symmetry is explicitly broken by the transverse polarization direction which we refer to as the $x$-direction.  With a trivial Wilson line as adopted in this paper, the Wigner distribution as well as the corresponding mixing distribution is vanishing when quark transverse momentum is perpendicular to the transverse polarization. Careful examinations including a nontrivial Wilson line may lead to non-zero values in this situation. In general, this behavior reflects a strong correlation to the direction of quark intrinsic transverse momentum, and in contrast the correlation to the direction of transverse coordinate is weak in this model.

\subsection{Transverse polarized proton}

\begin{figure}
\includegraphics[width=0.23\textwidth]{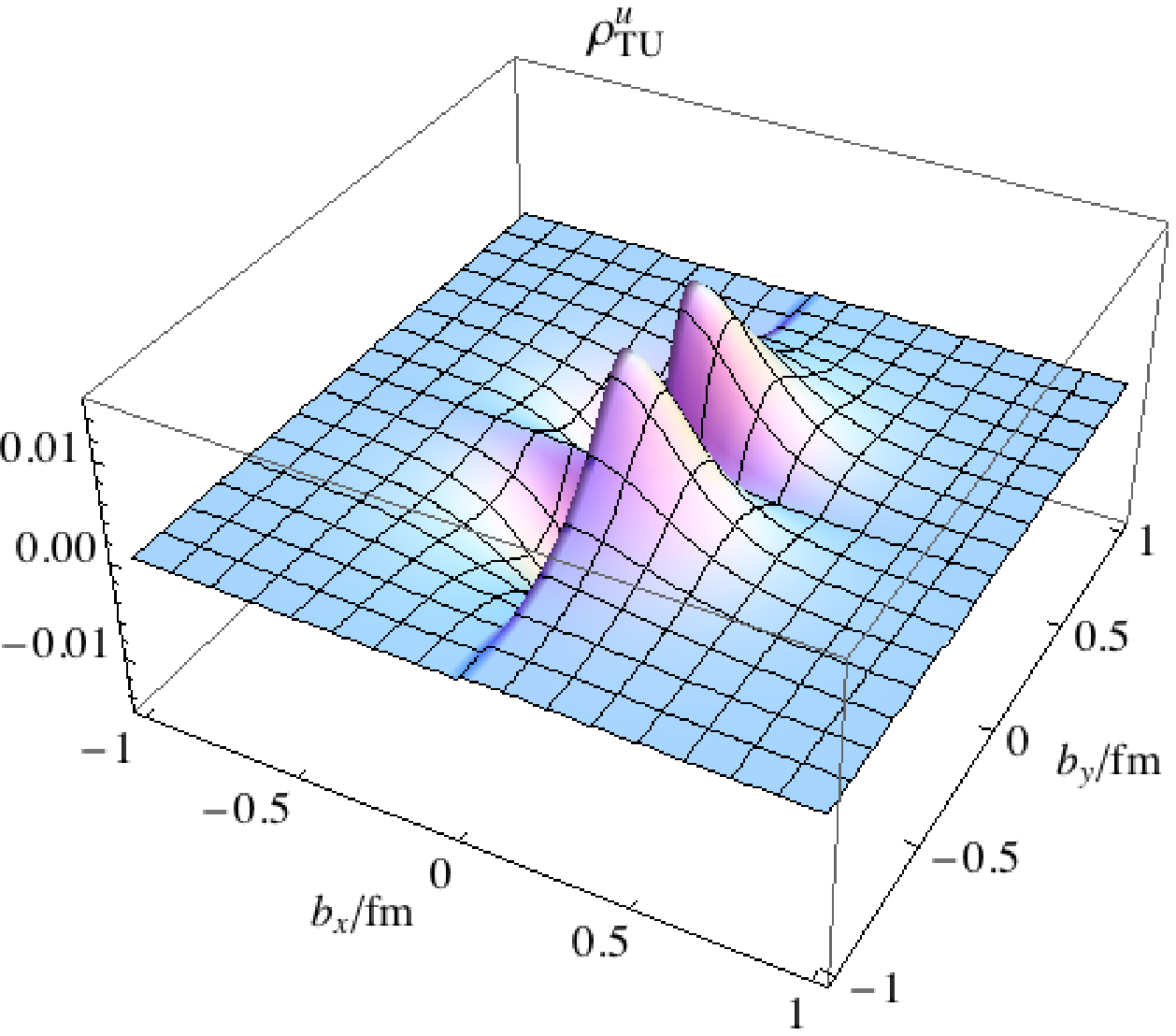}
\includegraphics[width=0.23\textwidth]{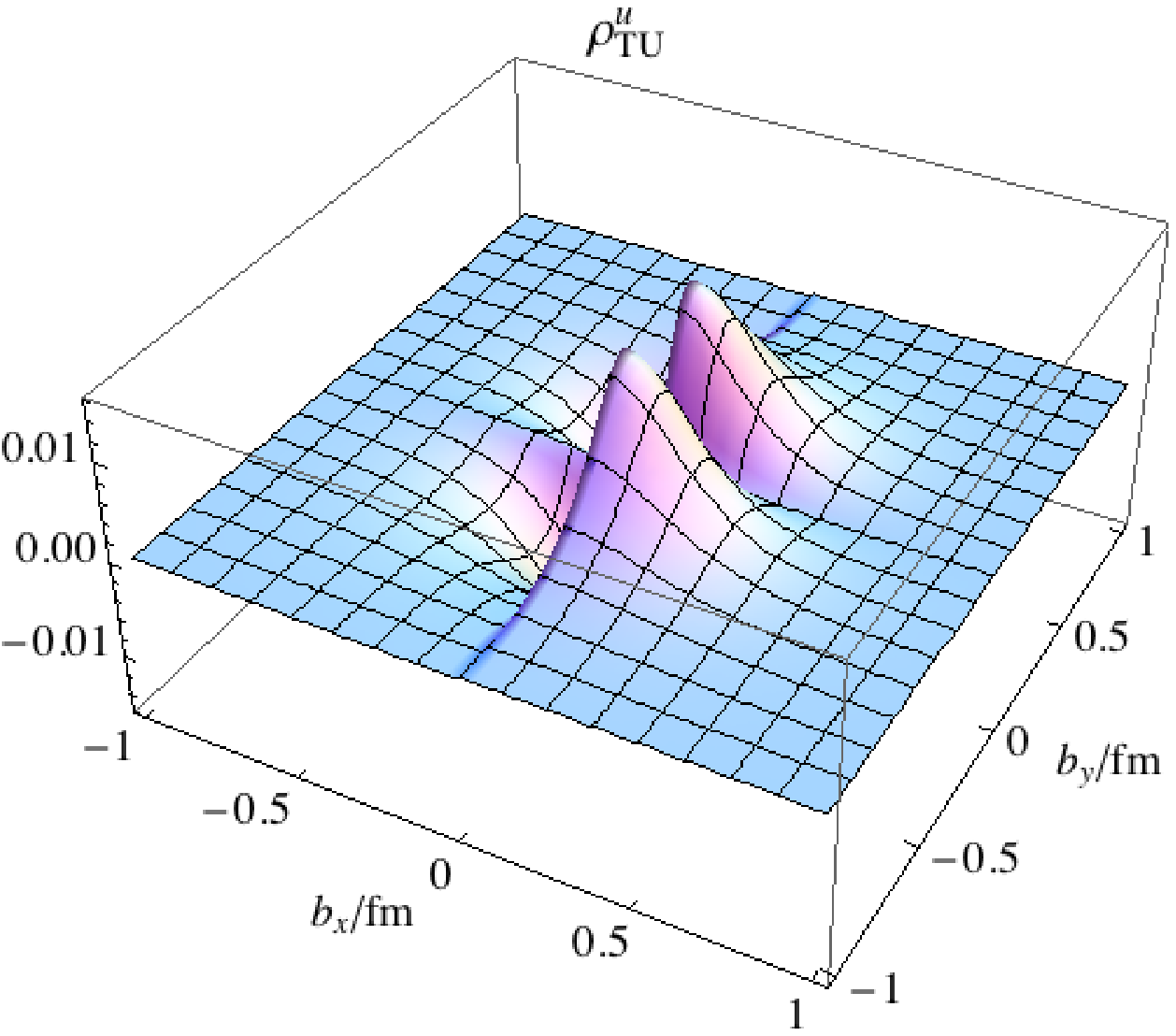}
\includegraphics[width=0.23\textwidth]{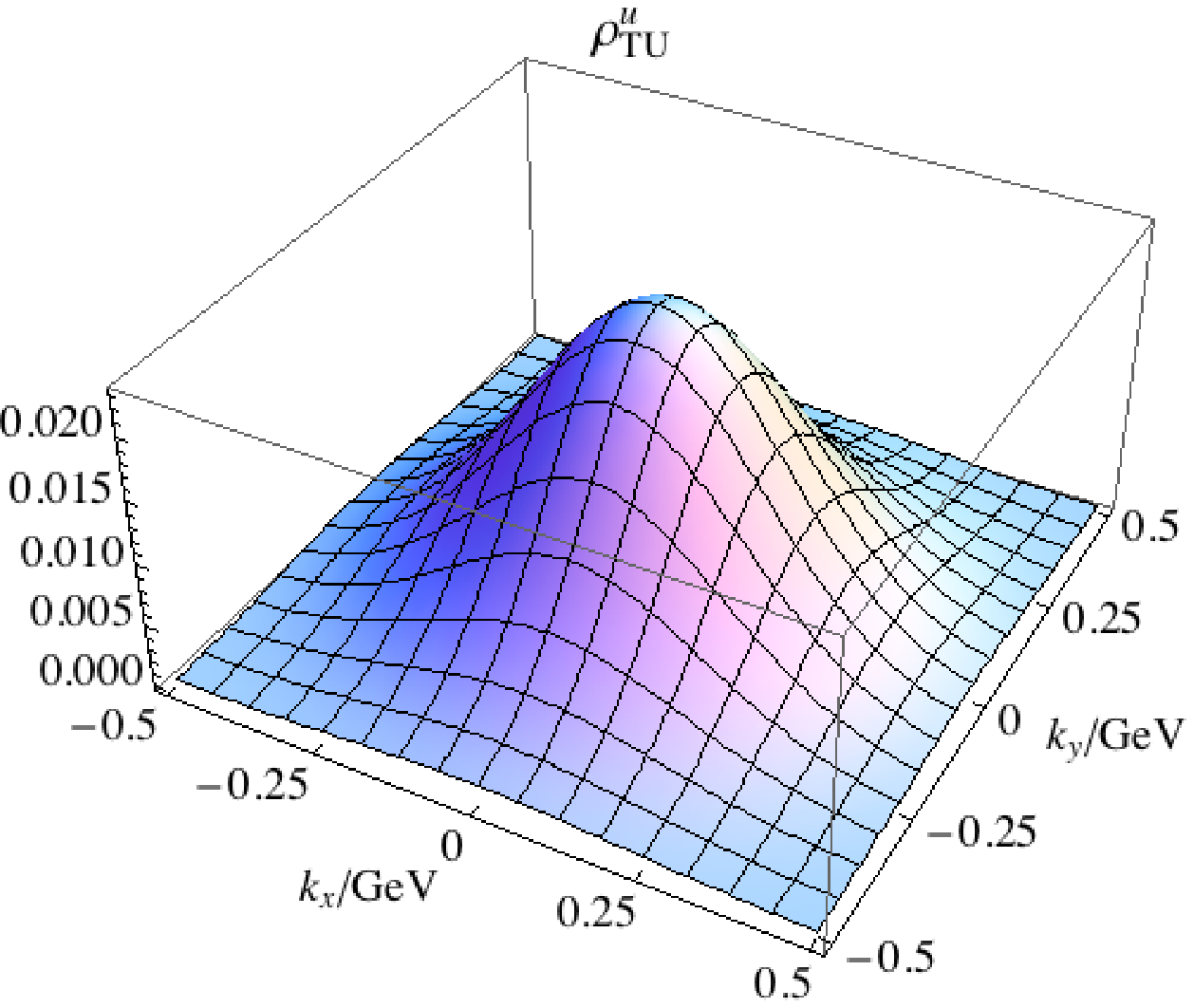}
\includegraphics[width=0.23\textwidth]{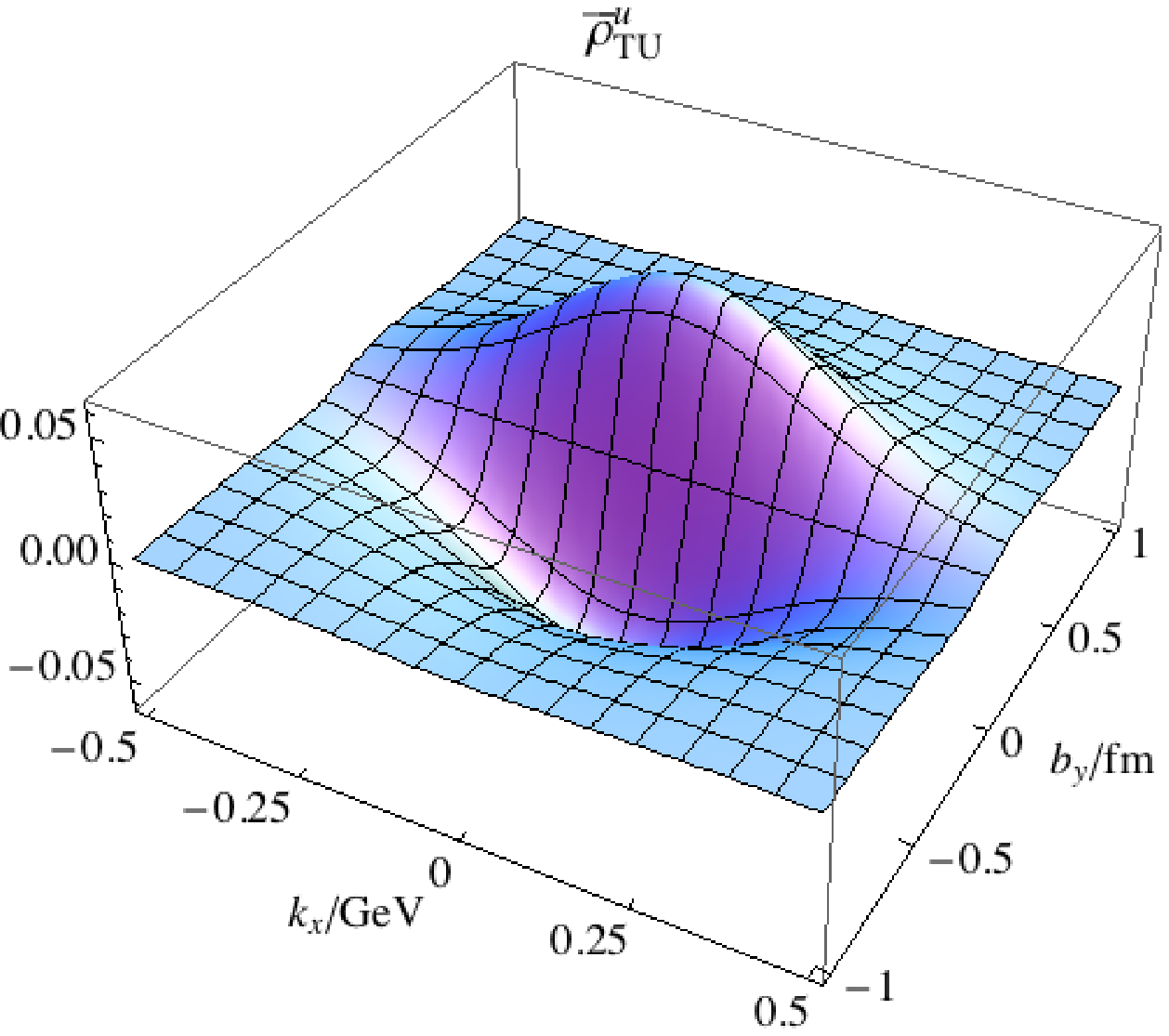}
\includegraphics[width=0.23\textwidth]{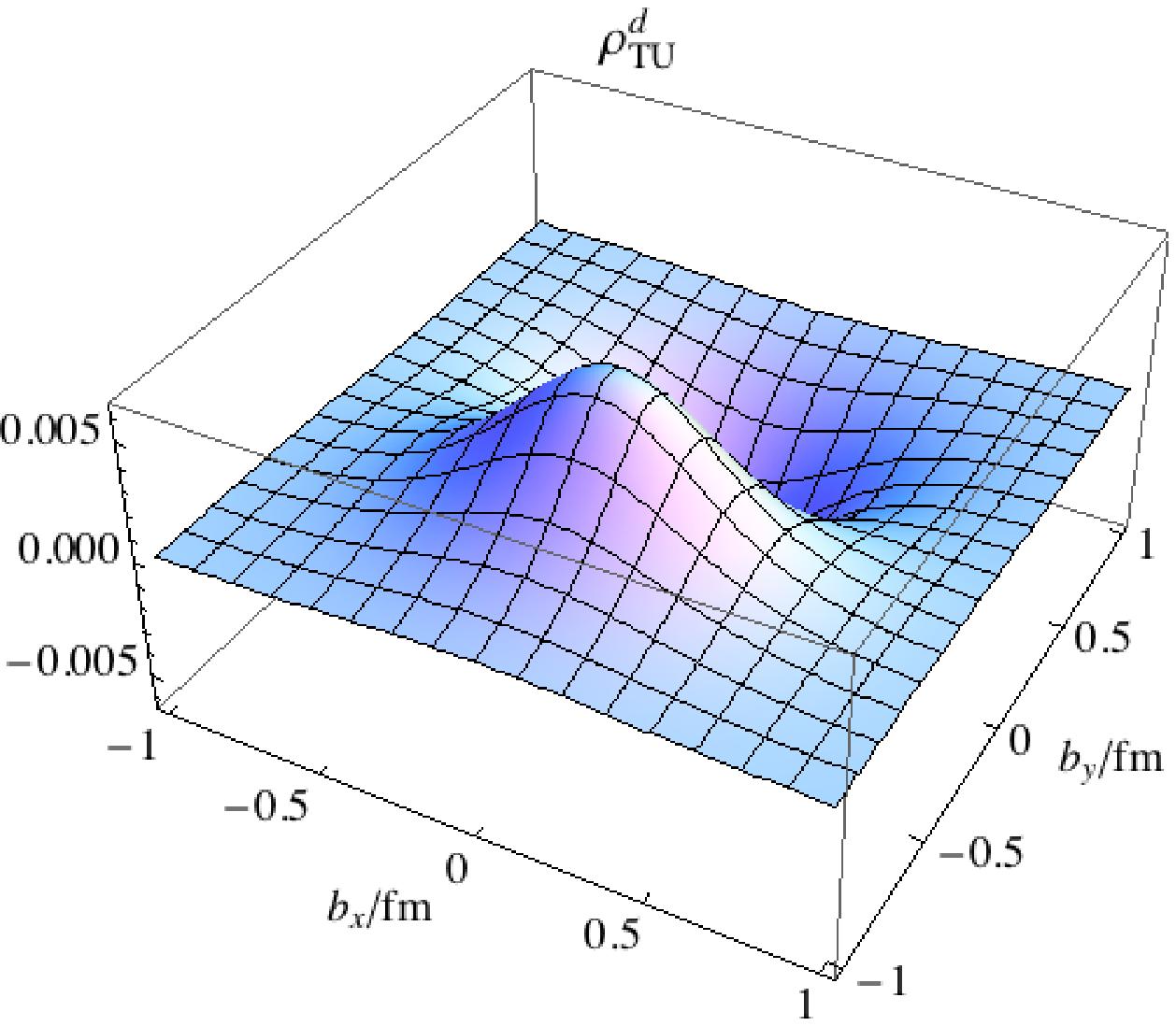}
\includegraphics[width=0.23\textwidth]{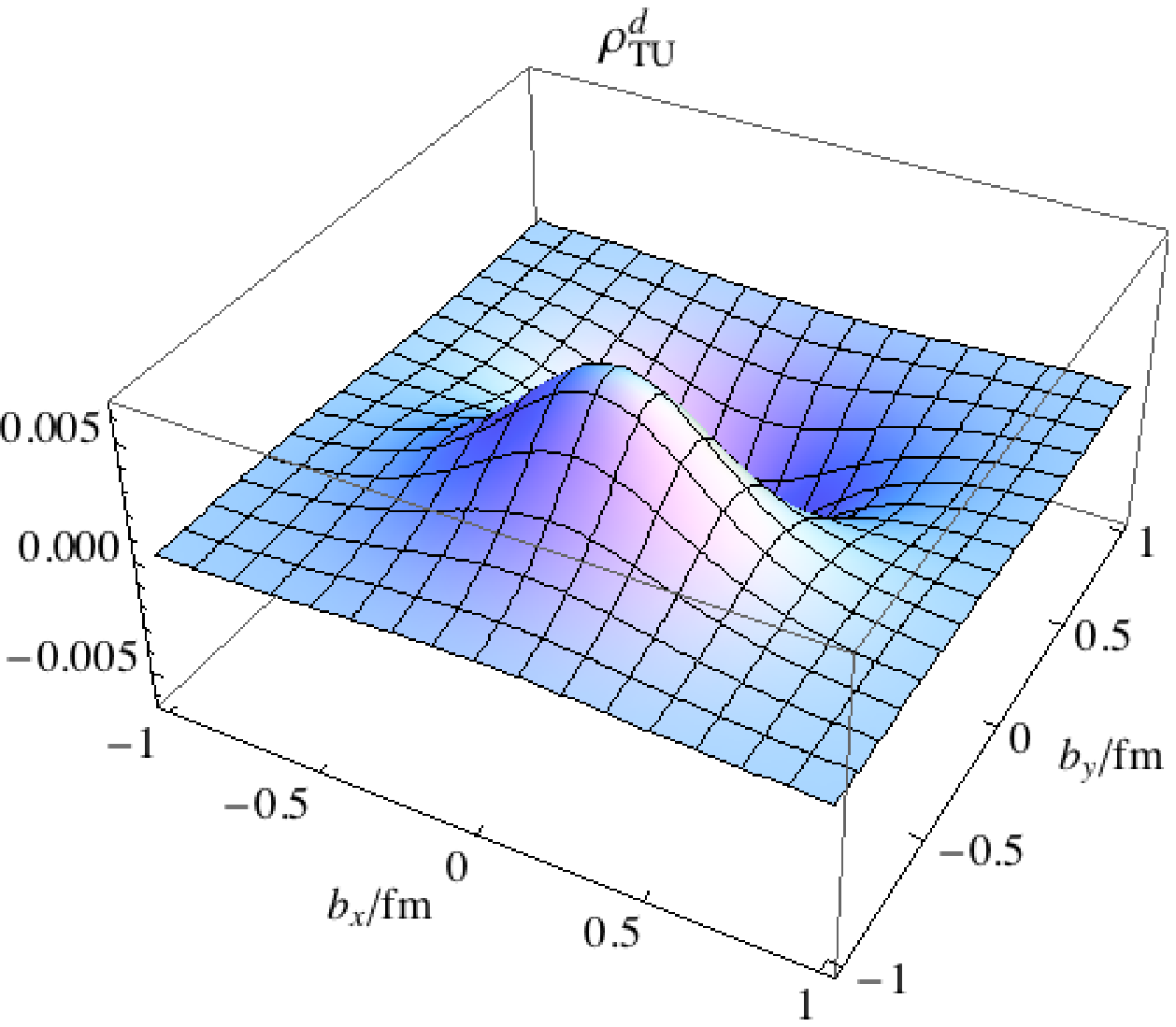}
\includegraphics[width=0.23\textwidth]{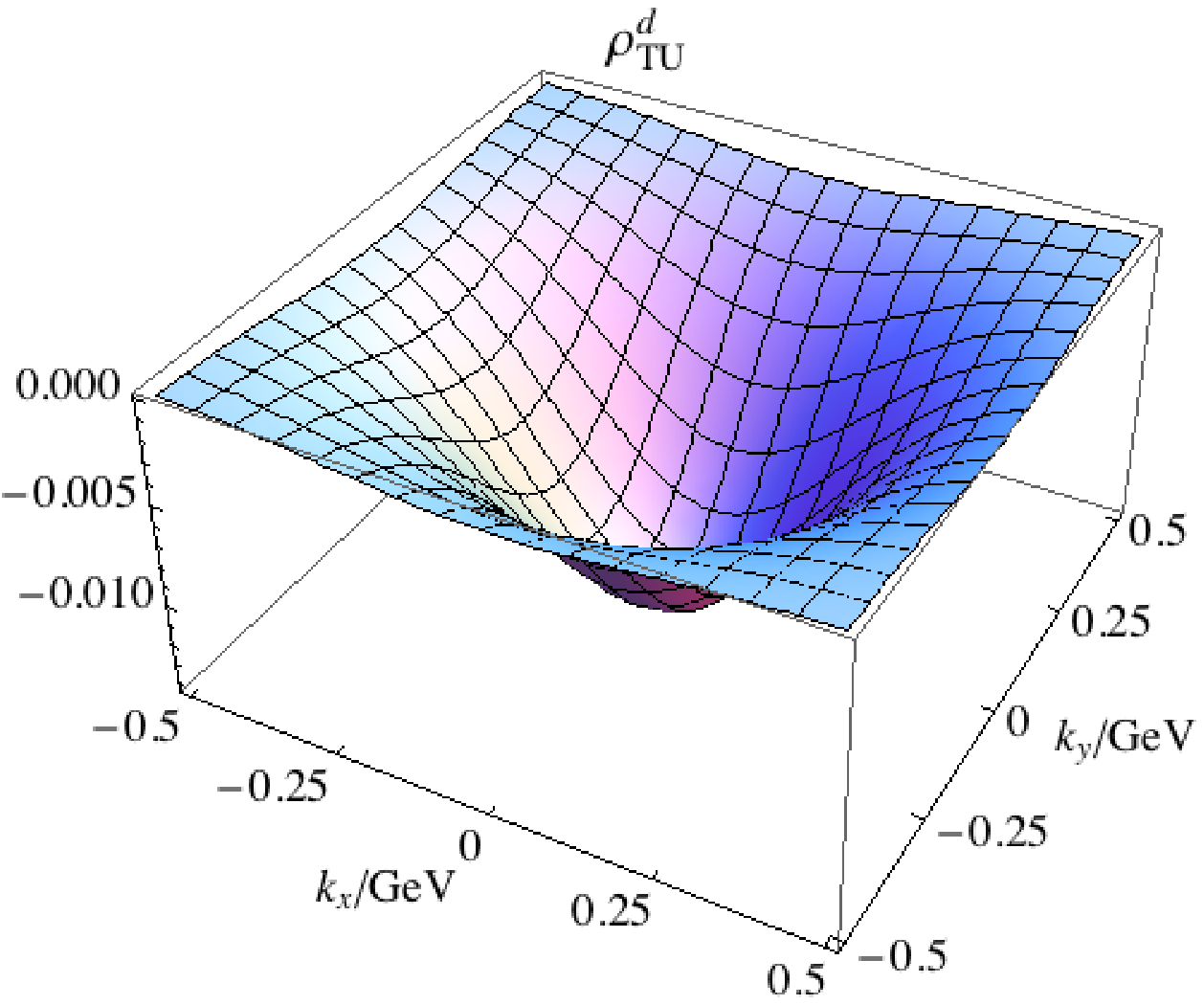}
\includegraphics[width=0.23\textwidth]{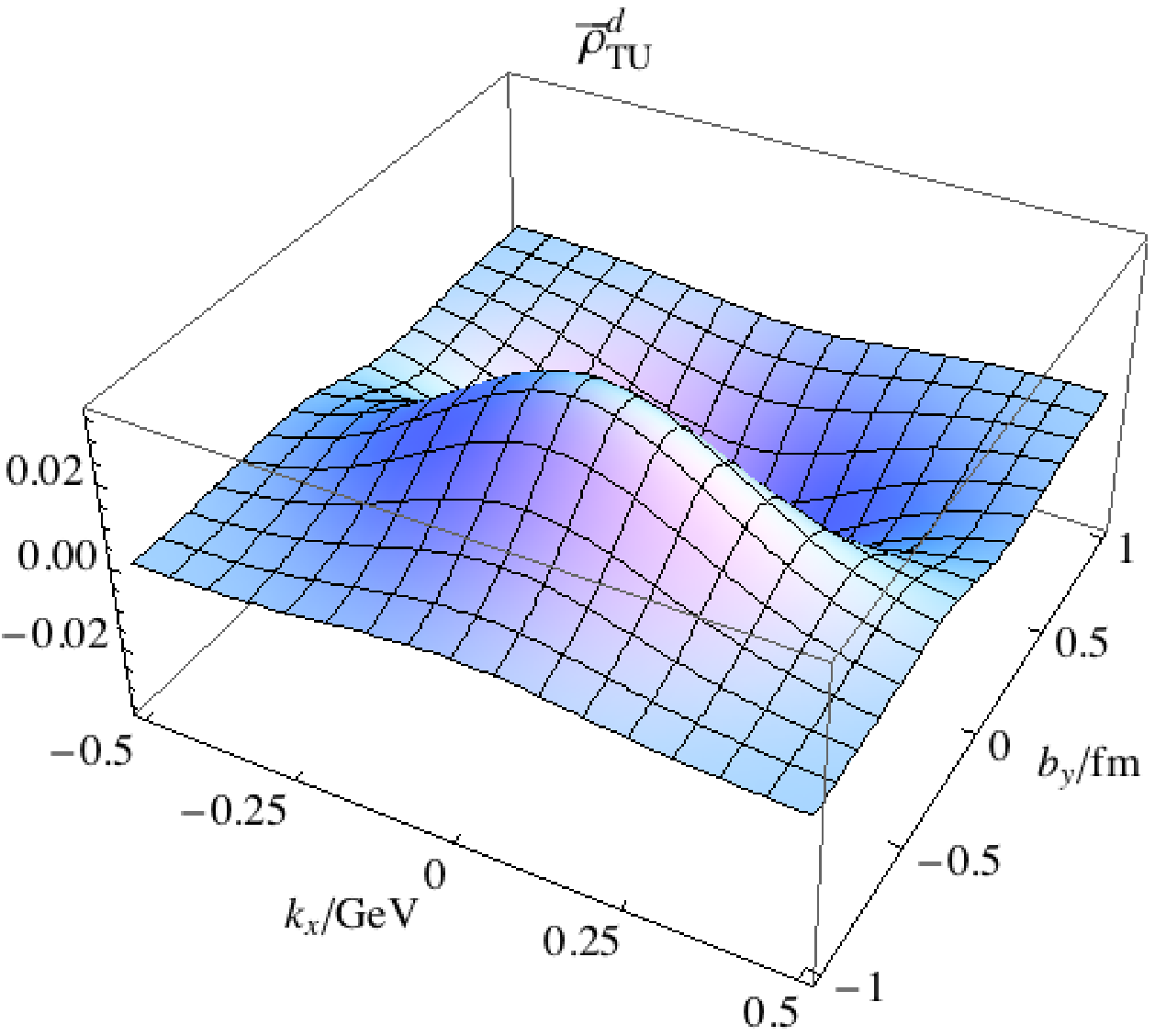}
\caption{(Color online). Trans-unpolarized Wigner distributions $\rho_{_\mathrm{TU}}$ and mixing distributions $\bar{\rho}_{_\mathrm{TU}}$ for the $u$ quark (upper) and the $d$ quark (lower). The first column are the distributions in transverse coordinate space with fixed transverse momentum $\bm{k}_\perp=0.3\,\textrm{GeV}\,\hat{\bm{e}}_x$ parallel to the proton polarization, and the second column are those with fixed transverse momentum $\bm{k}_\perp=0.3\,\textrm{GeV}\,\hat{\bm{e}}_y$ perpendicular to the proton polarization. The third column are the distributions in transverse momentrum space with fixed transverse coordinate $\bm{b}_\perp=0.4\,\textrm{fm}\,\hat{\bm{e}}_y$ perpendicular to the proton polarization. The fourth column are the mixting distributions. \label{rhotu}}
\end{figure}
\begin{figure}
\includegraphics[width=0.23\textwidth]{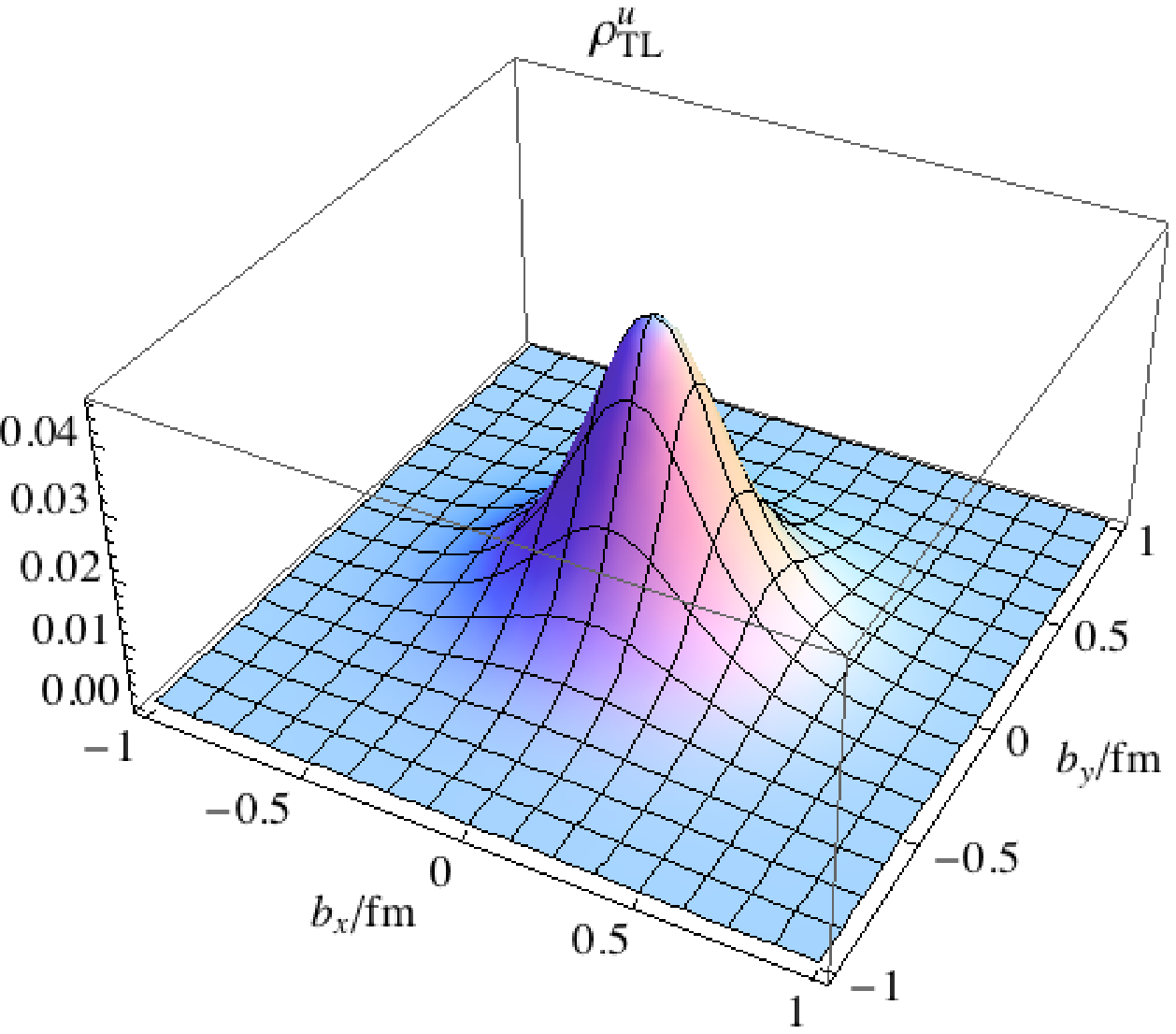}
\includegraphics[width=0.23\textwidth]{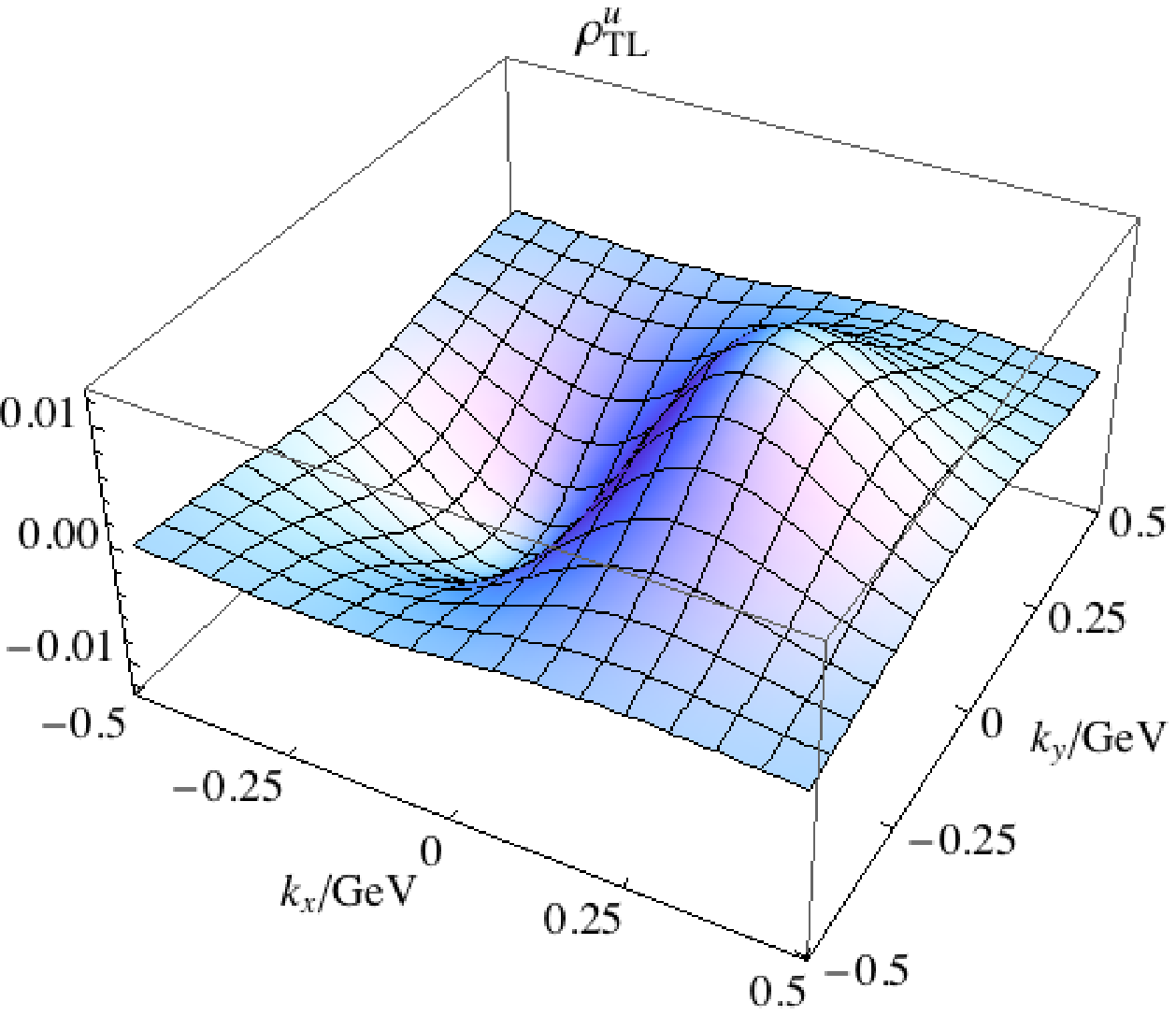}
\includegraphics[width=0.23\textwidth]{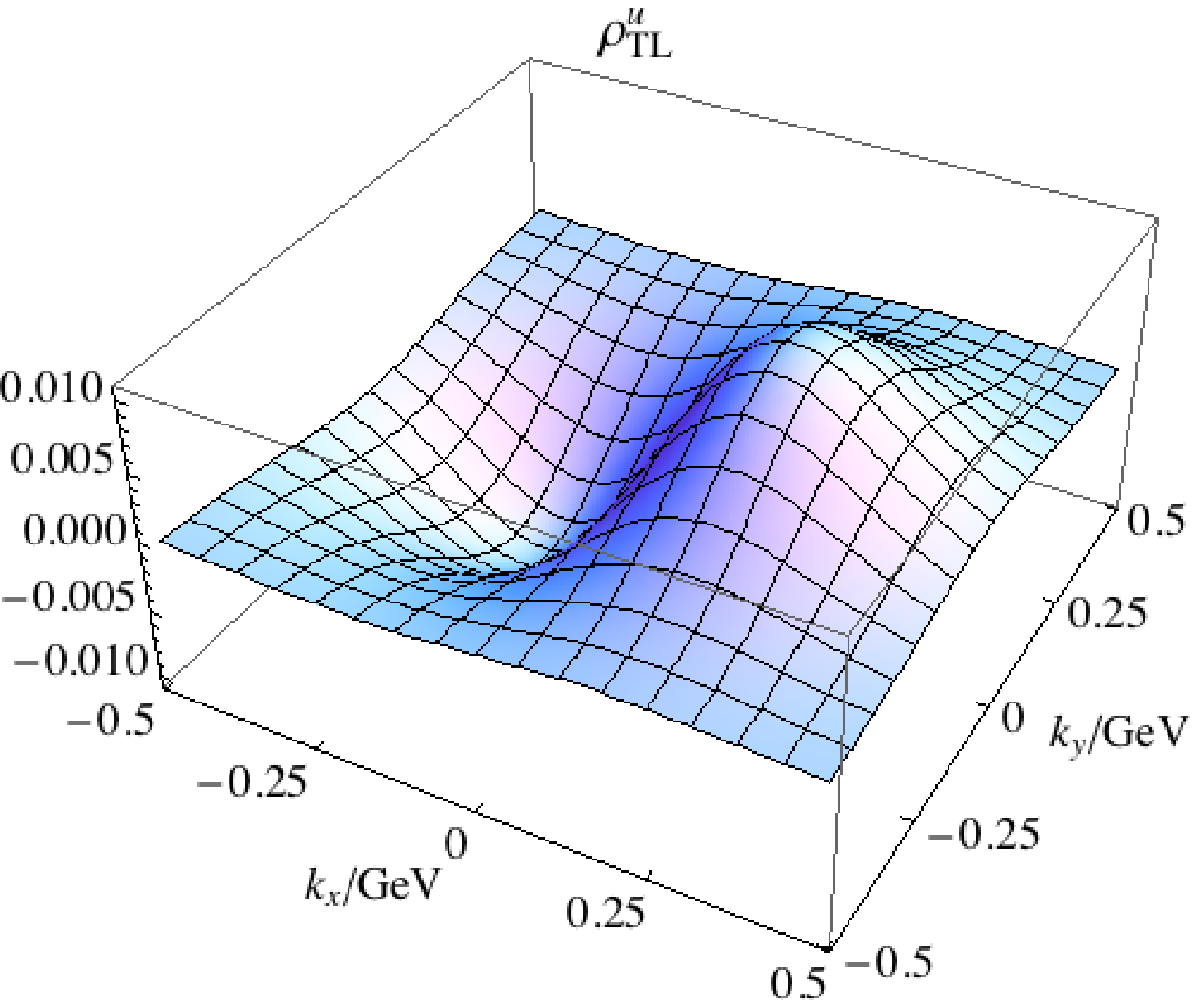}
\includegraphics[width=0.23\textwidth]{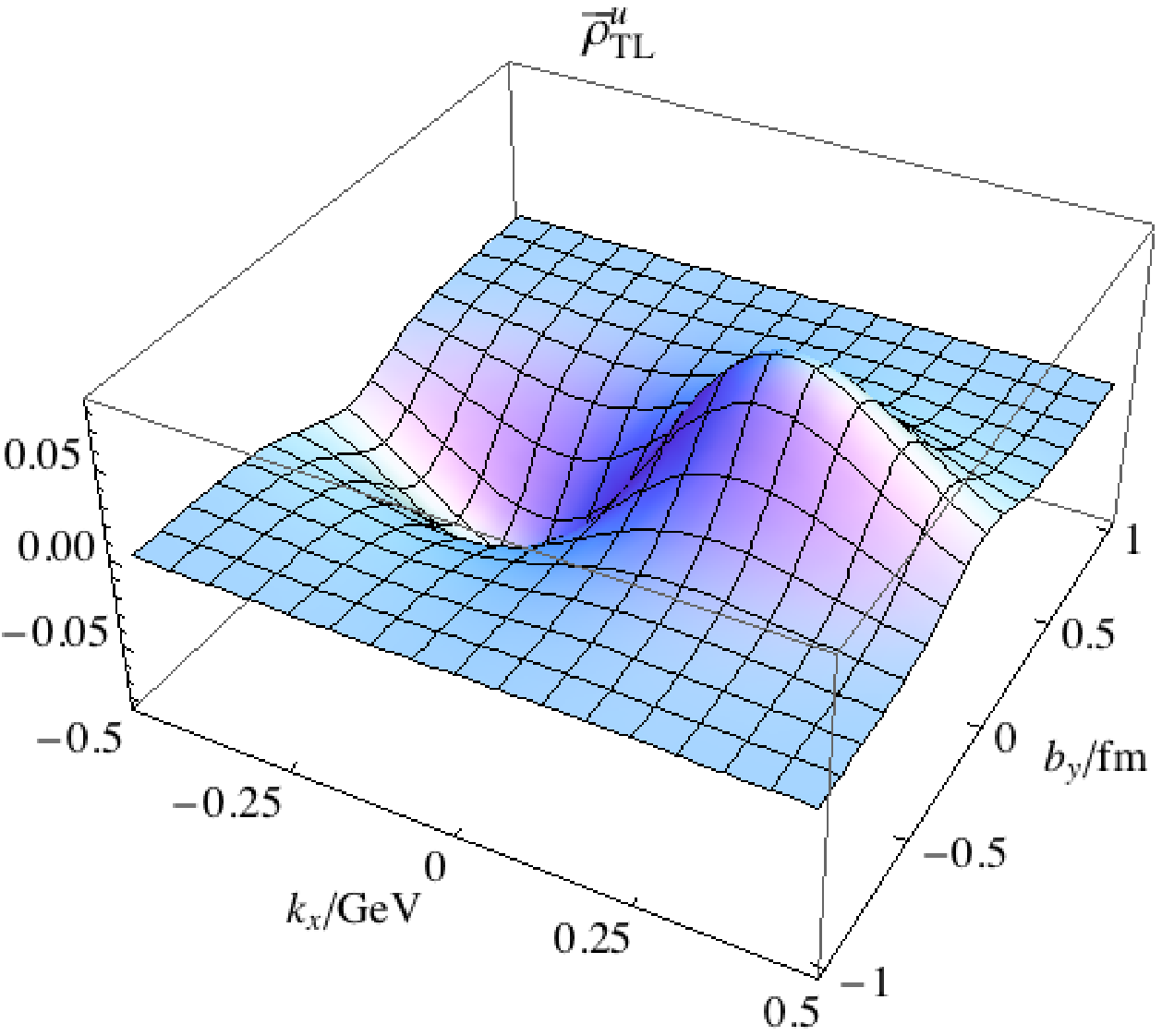}
\includegraphics[width=0.23\textwidth]{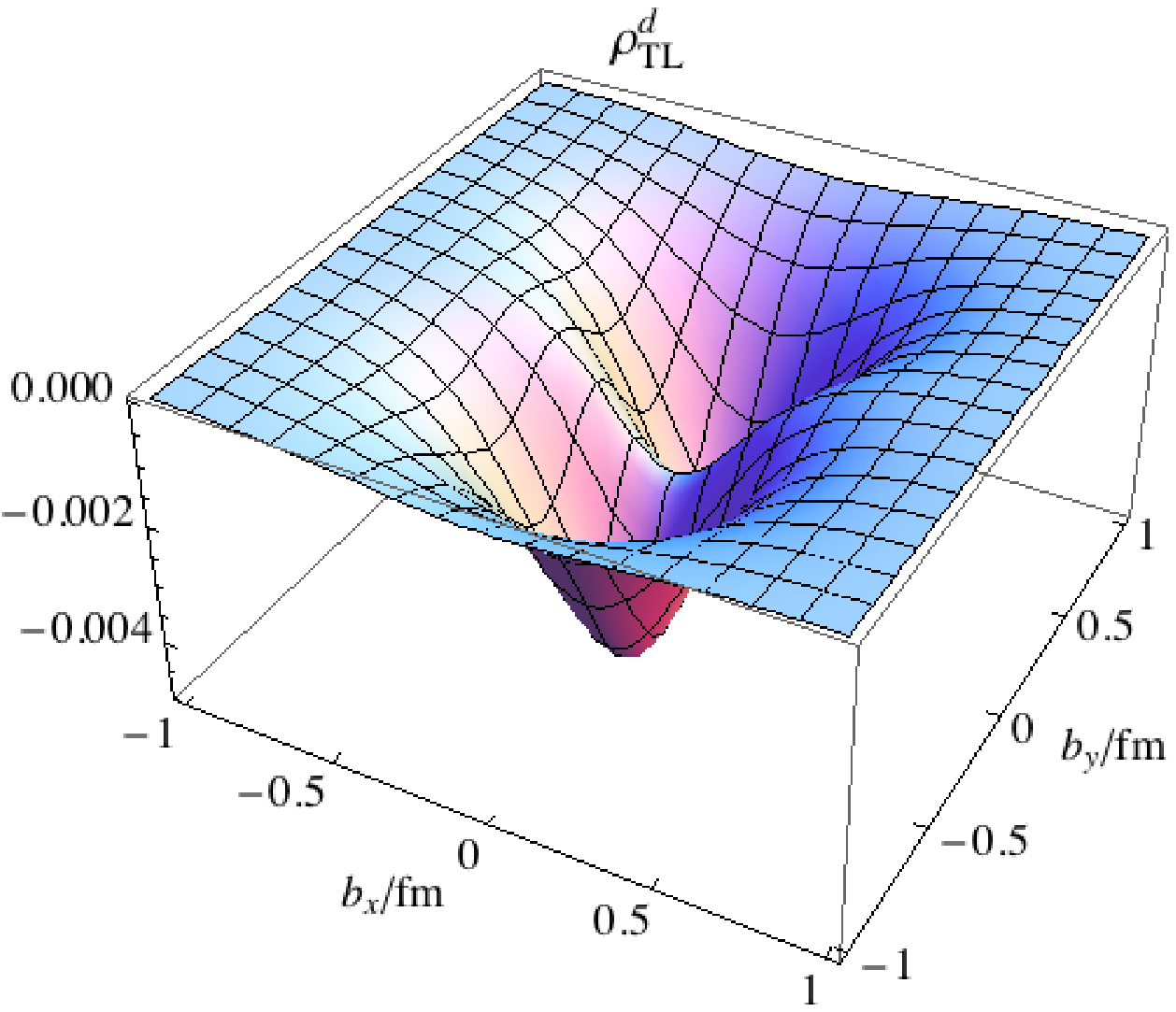}
\includegraphics[width=0.23\textwidth]{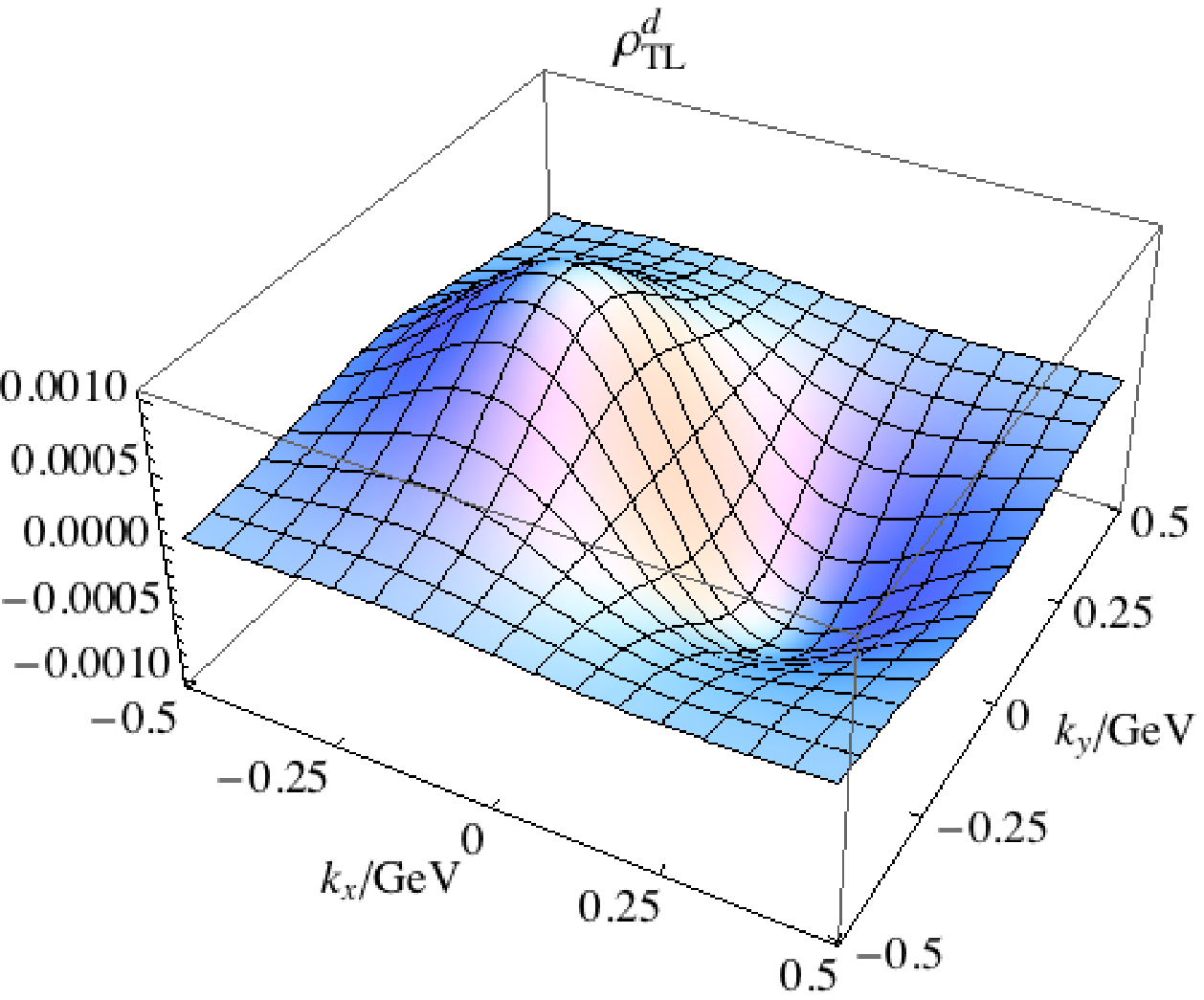}
\includegraphics[width=0.23\textwidth]{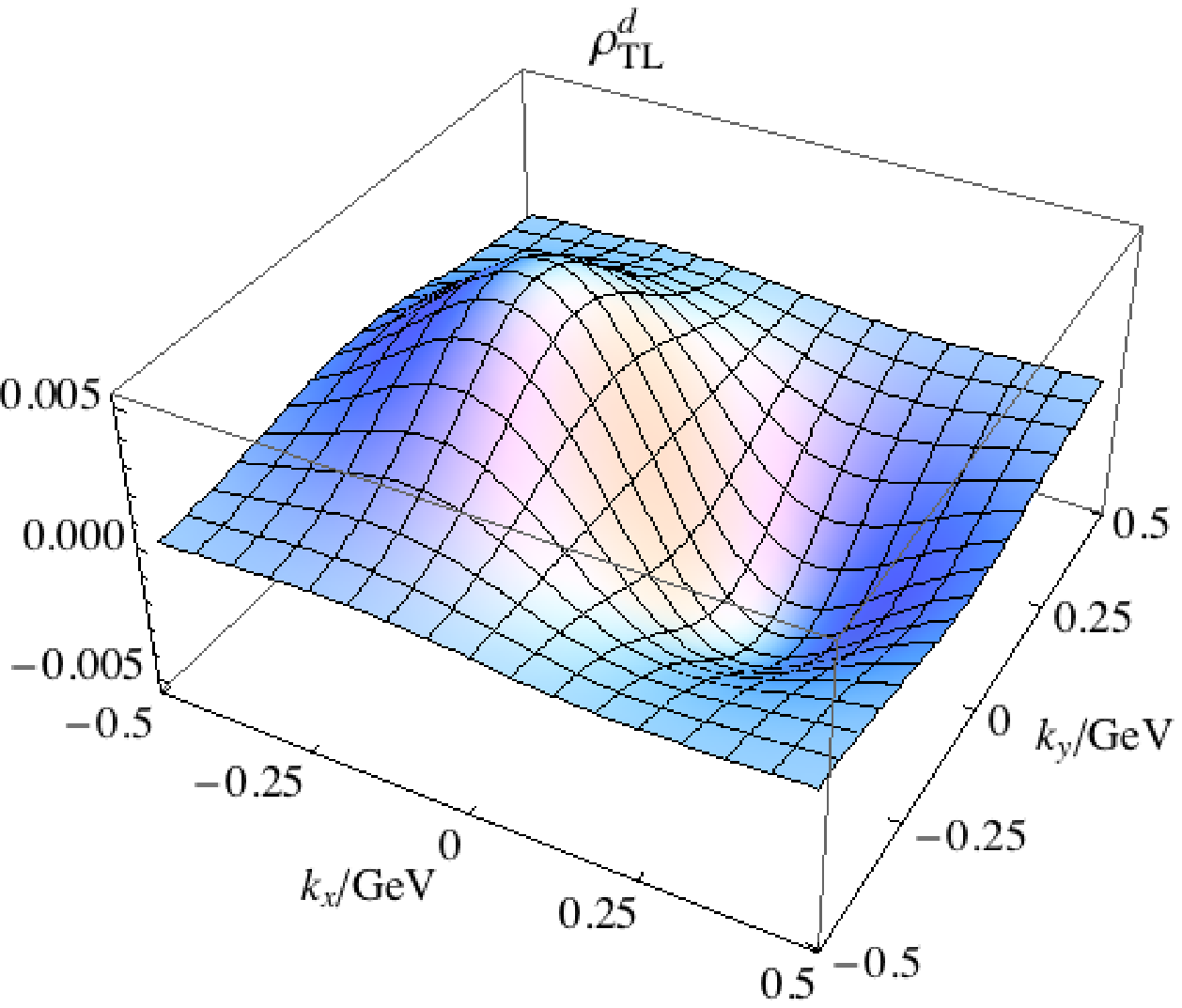}
\includegraphics[width=0.23\textwidth]{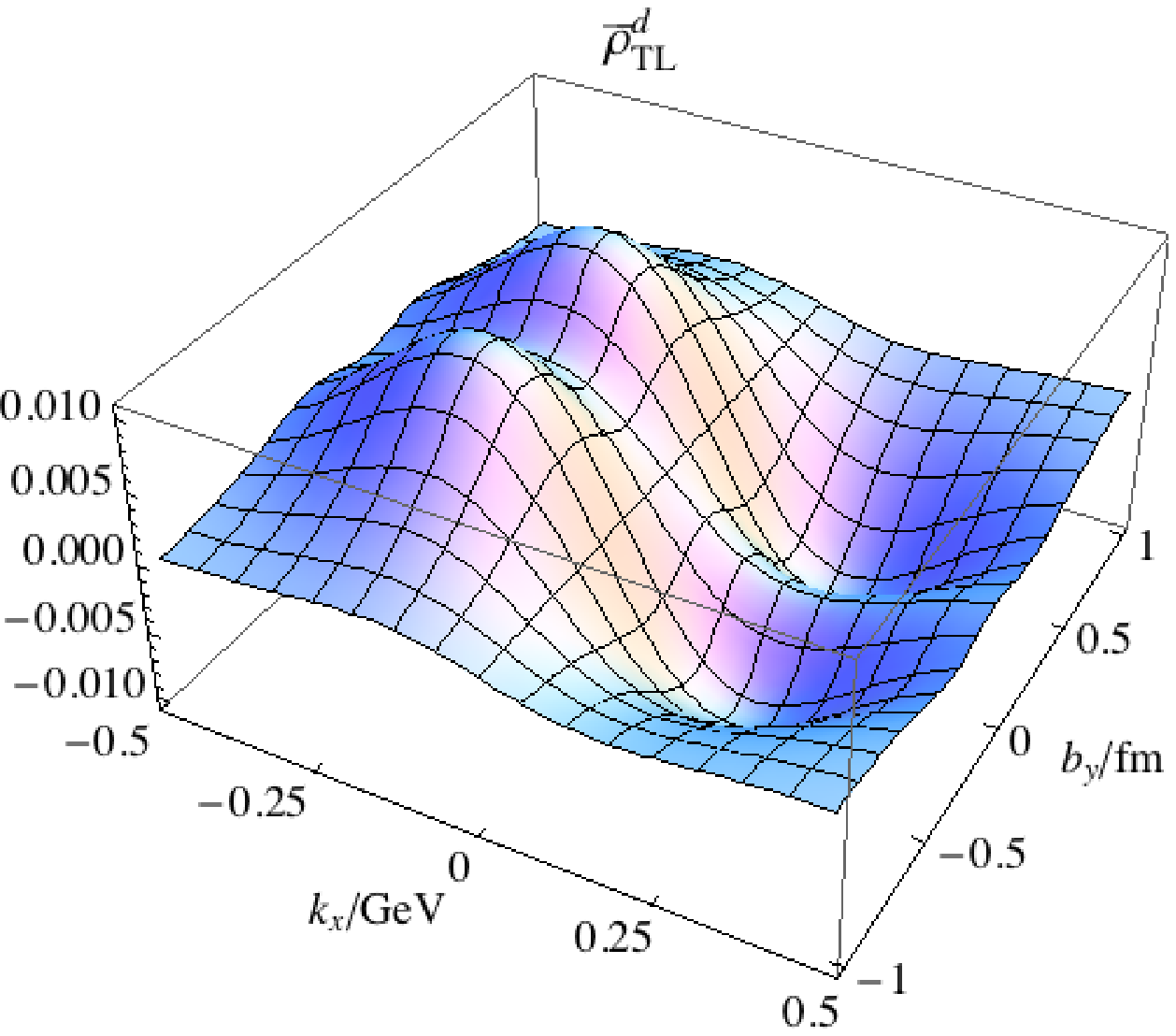}
\caption{(Color online). Trans-longitudinal Wigner distributions $\rho_{_\mathrm{TL}}$ and mixing distributions $\bar{\rho}_{_\mathrm{TL}}$ for the $u$ quark (upper) and the $d$ quark (lower). The first column are the distributions in transverse coordinate space with fixed transverse momentum $\bm{k}_\perp=0.3\,\textrm{GeV}\,\hat{\bm{e}}_x$ parallel to the proton polarization. The second column are the distributions in transverse momentrum space with fixed transverse coordinate $\bm{b}_\perp=0.4\,\textrm{fm}\,\hat{\bm{e}}_x$ parallel to the proton polarization, and the third column are those with fixed transverse coordinate $\bm{b}_\perp=0.4\,\textrm{fm}\,\hat{\bm{e}}_y$ perpendicular to the proton polarization. The fourth column are the mixing distributions. \label{rhotl}}
\end{figure}
\begin{figure}
\includegraphics[width=0.23\textwidth]{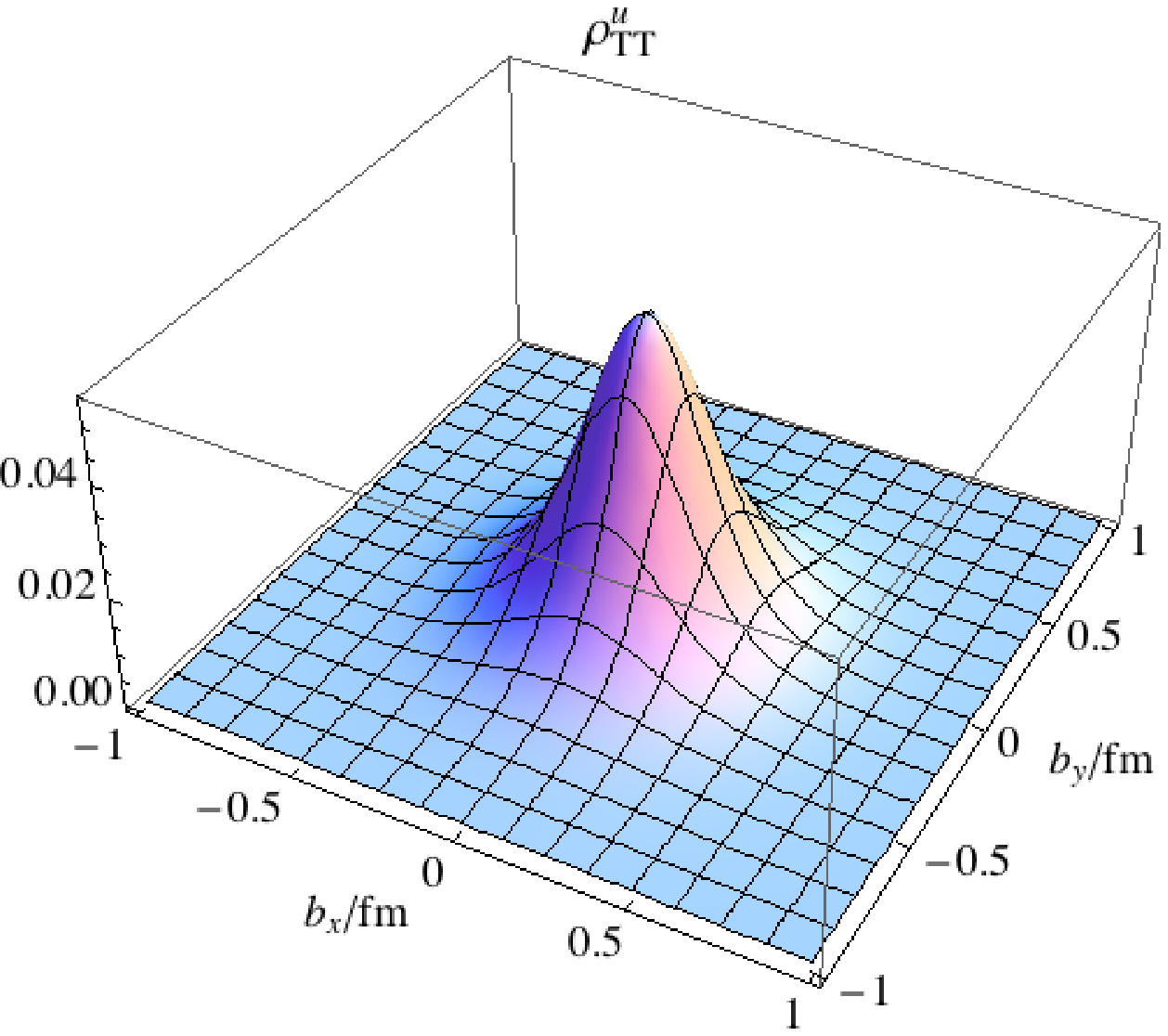}
\includegraphics[width=0.23\textwidth]{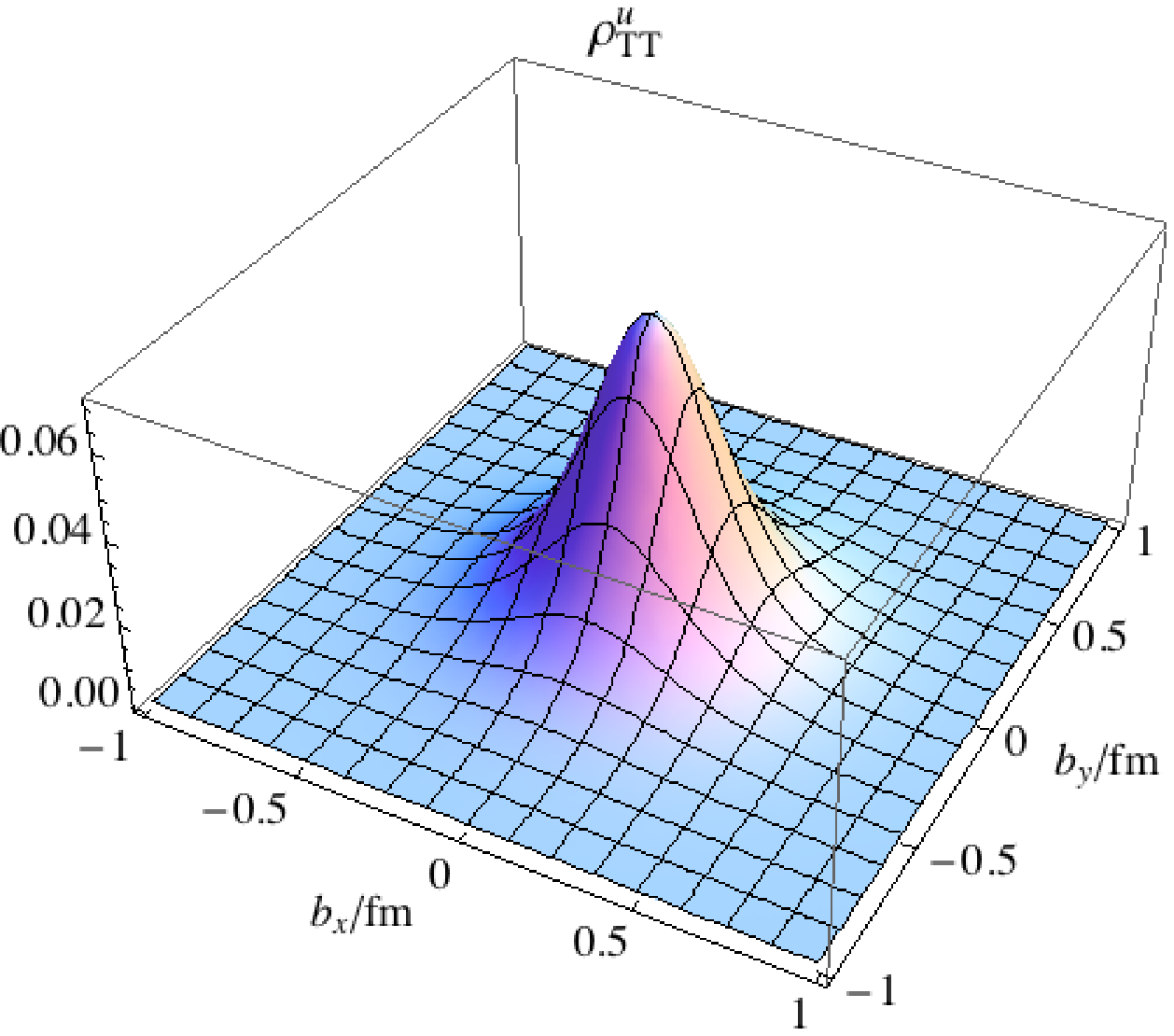}
\includegraphics[width=0.23\textwidth]{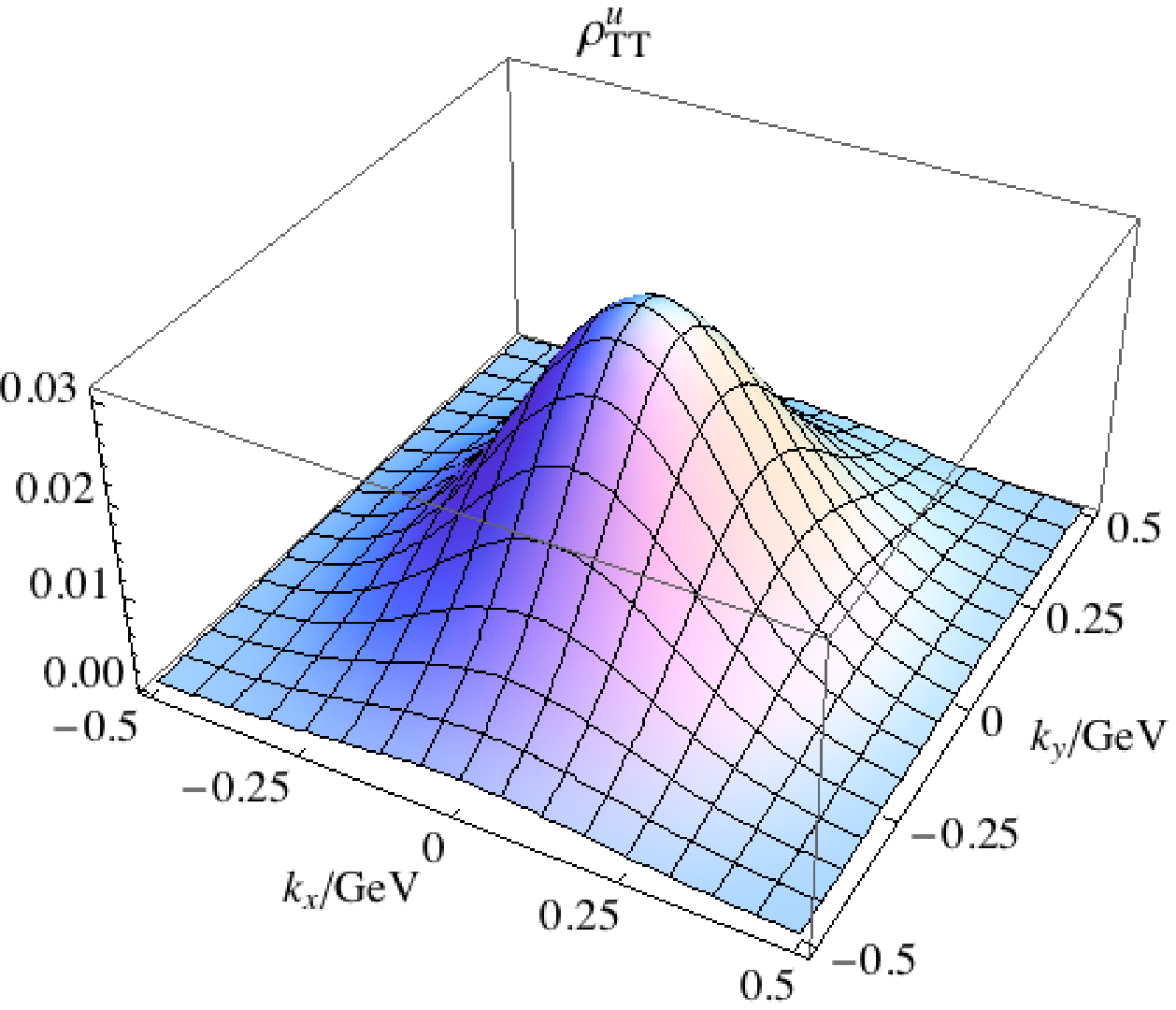}
\includegraphics[width=0.23\textwidth]{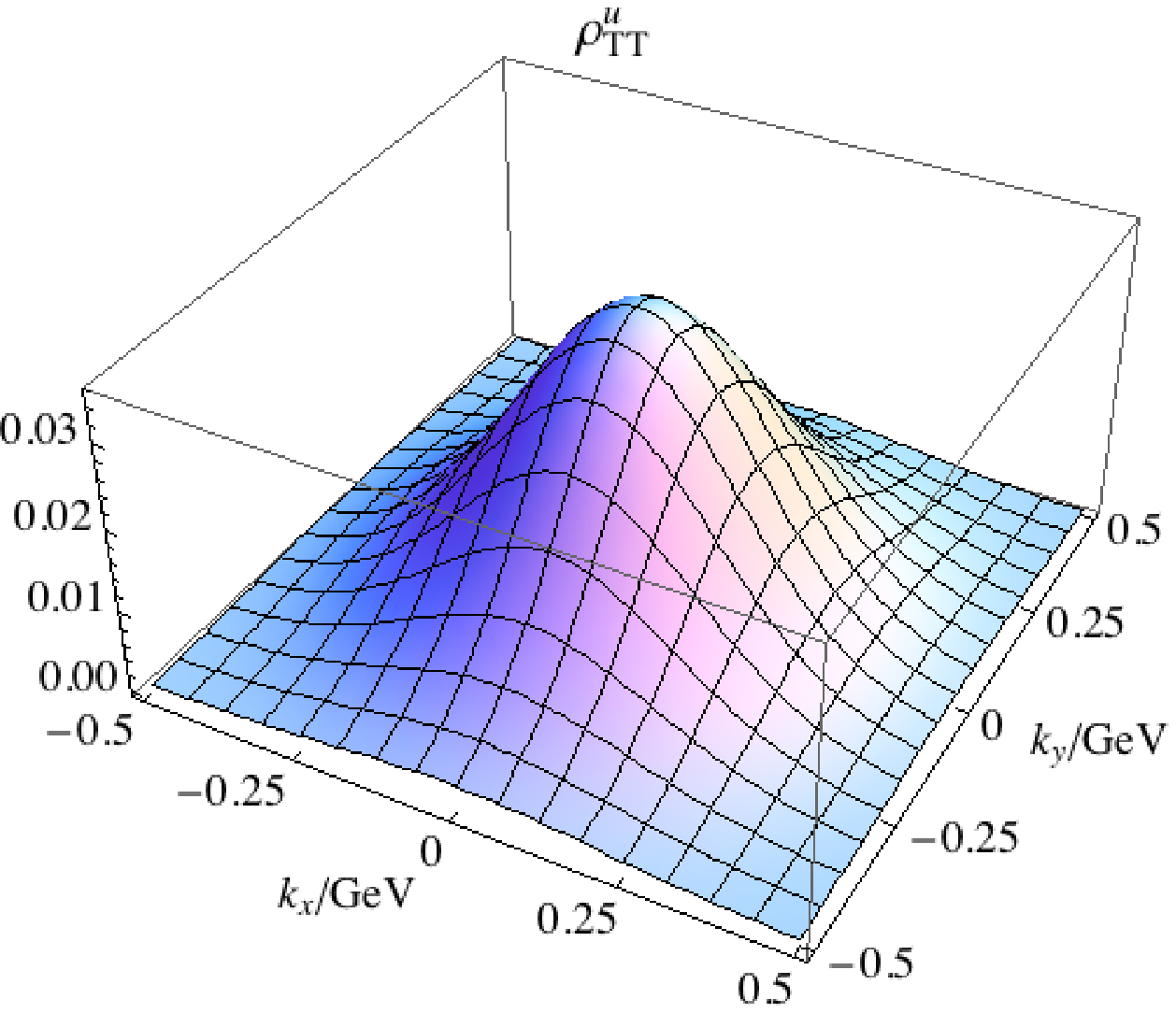}
\includegraphics[width=0.23\textwidth]{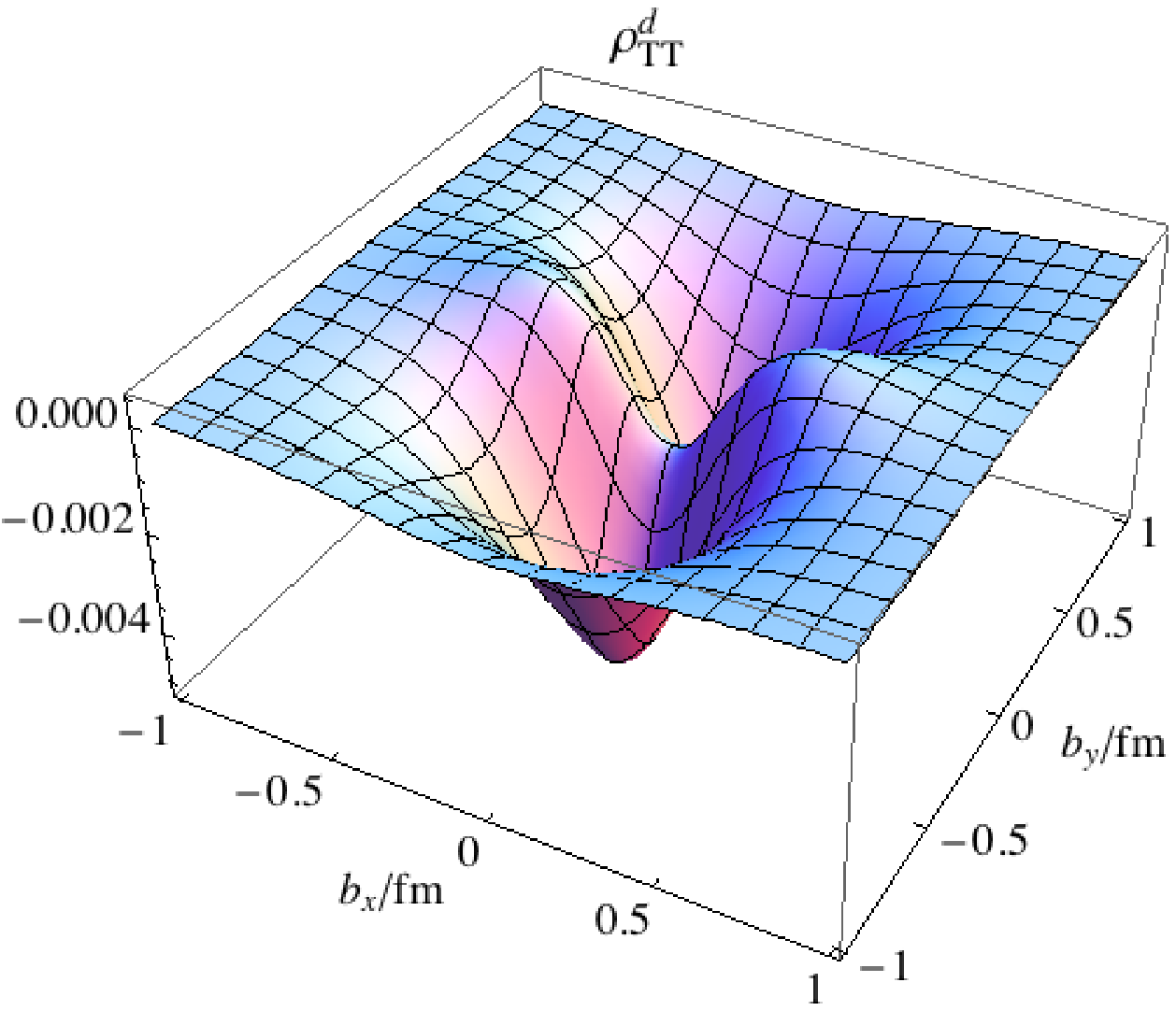}
\includegraphics[width=0.23\textwidth]{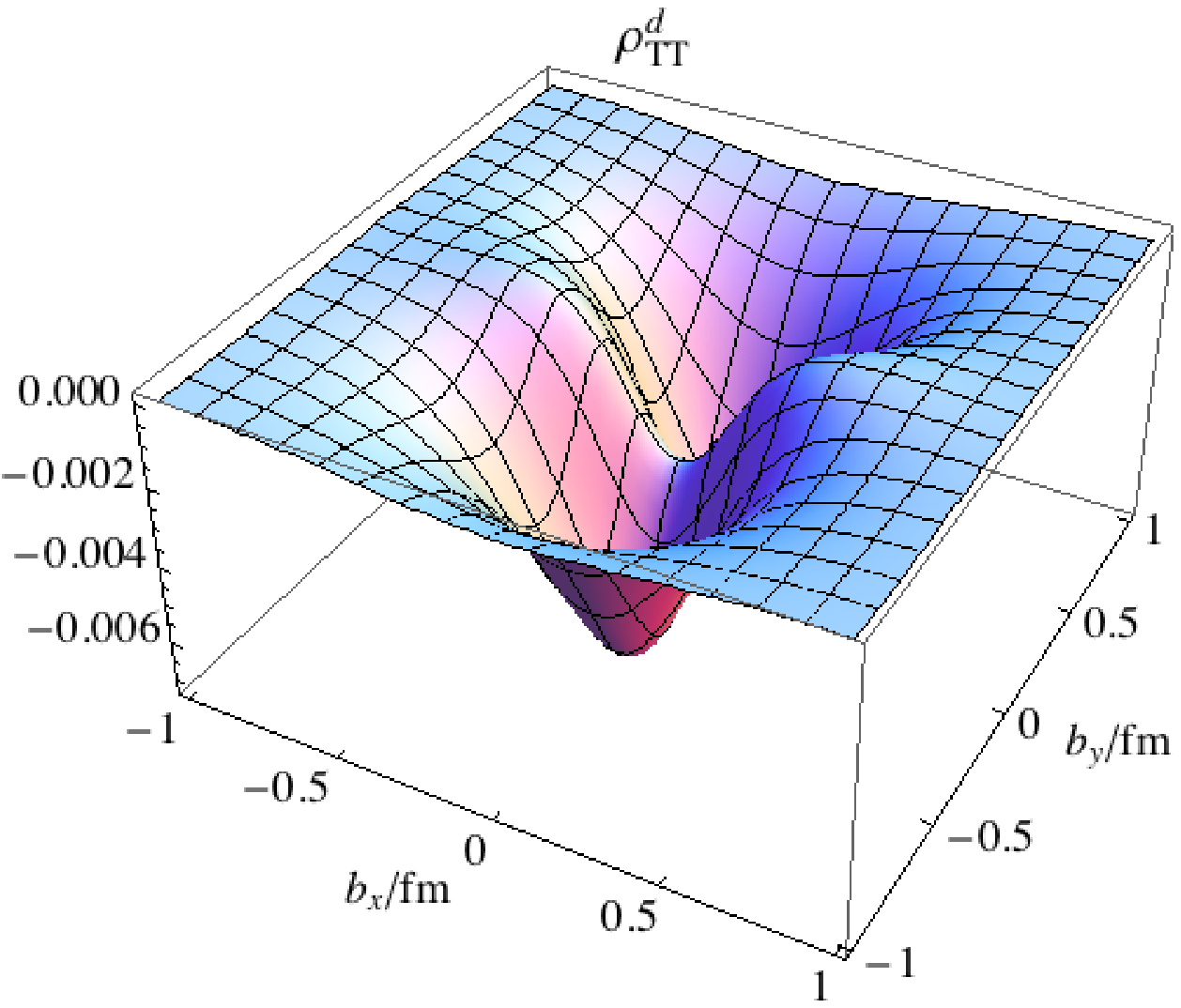}
\includegraphics[width=0.23\textwidth]{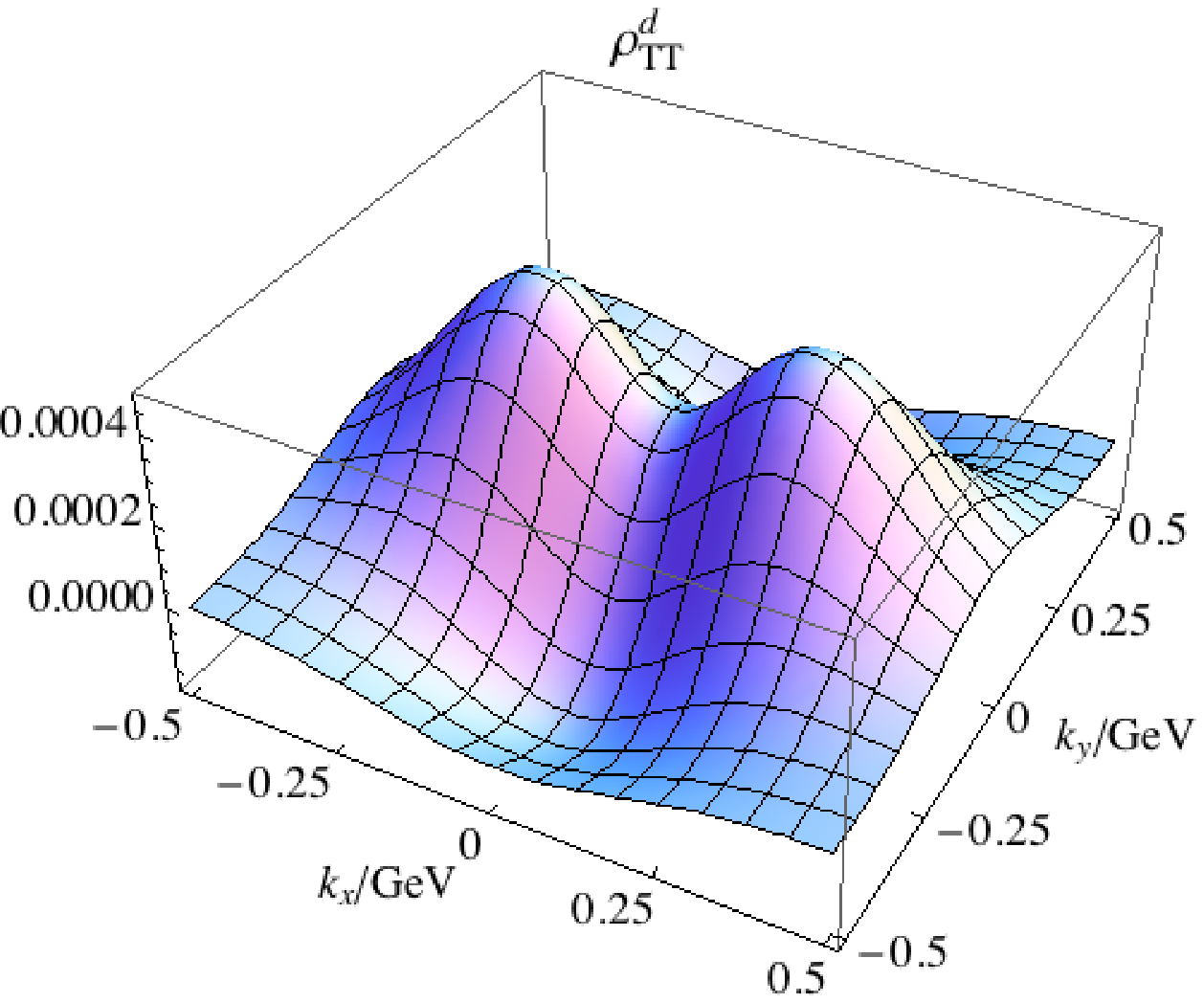}
\includegraphics[width=0.23\textwidth]{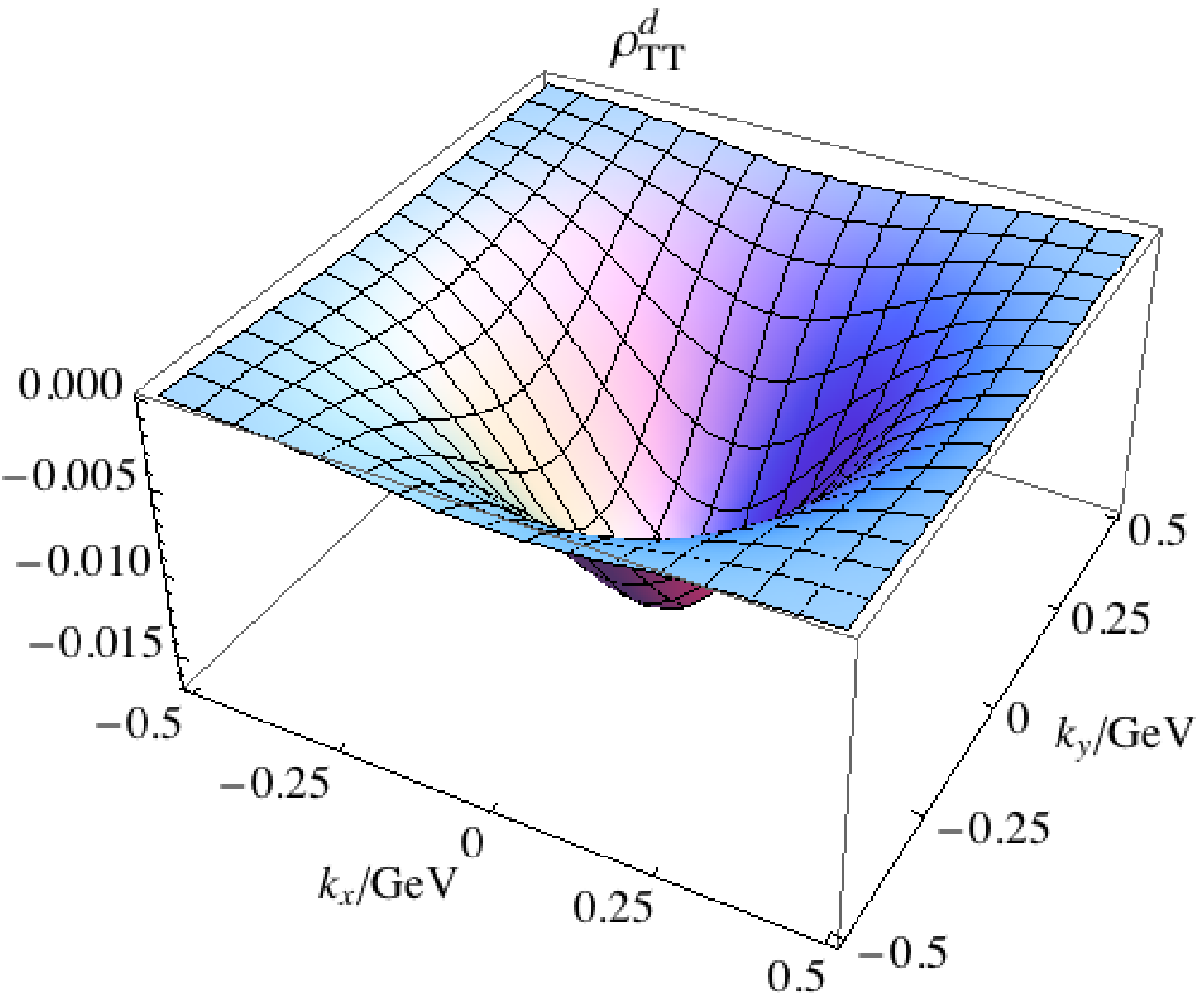}
\caption{(Color online). Transverse Wigner distributions $\rho_{_\mathrm{TT}}$ for the $u$ quark (upper) and the $d$ quark (lower). The first column are the distributions in transverse coordinate space with fixed transverse momentum $\bm{k}_\perp=0.3\,\textrm{GeV}\,\hat{\bm{e}}_x$ parallel to the quark polarization, and the second column are those with fixed transverse momentum $\bm{k}_\perp=0.3\,\textrm{GeV}\,\hat{\bm{e}}_y$ perpendicular to the quark polarization. The third column are the distributions in transverse momentrum space with fixed transverse coordinate $\bm{b}_\perp=0.4\,\textrm{fm}\,\hat{\bm{e}}_x$ parallel to the quark polarization, and the fourth column are those with fixed transverse coordinate $\bm{b}_\perp=0.4\,\textrm{fm}\,\hat{\bm{e}}_y$ perpendicular to the quark polarization. \label{rhott}}
\end{figure}
\begin{figure}
\includegraphics[width=0.25\textwidth]{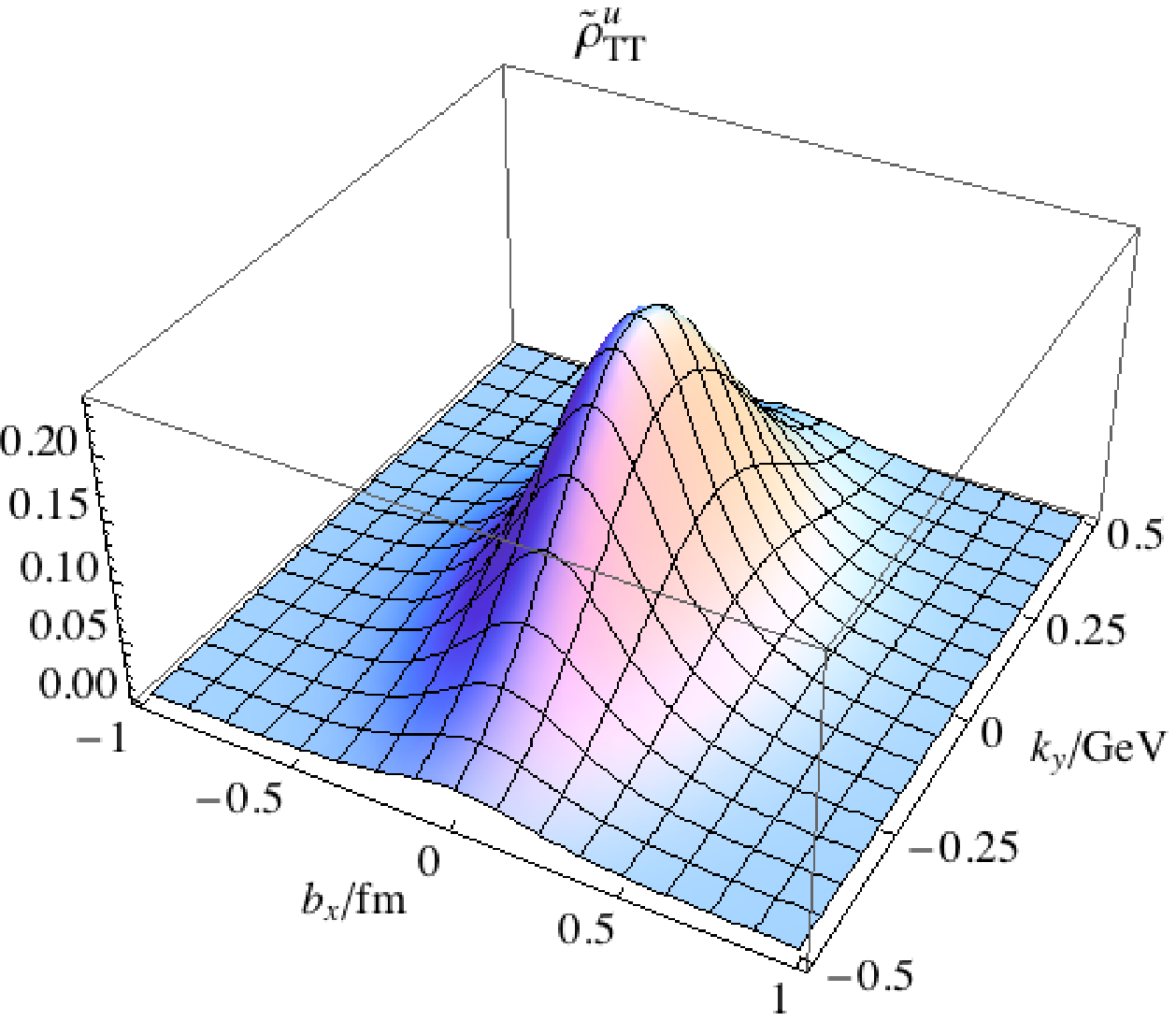}
\includegraphics[width=0.25\textwidth]{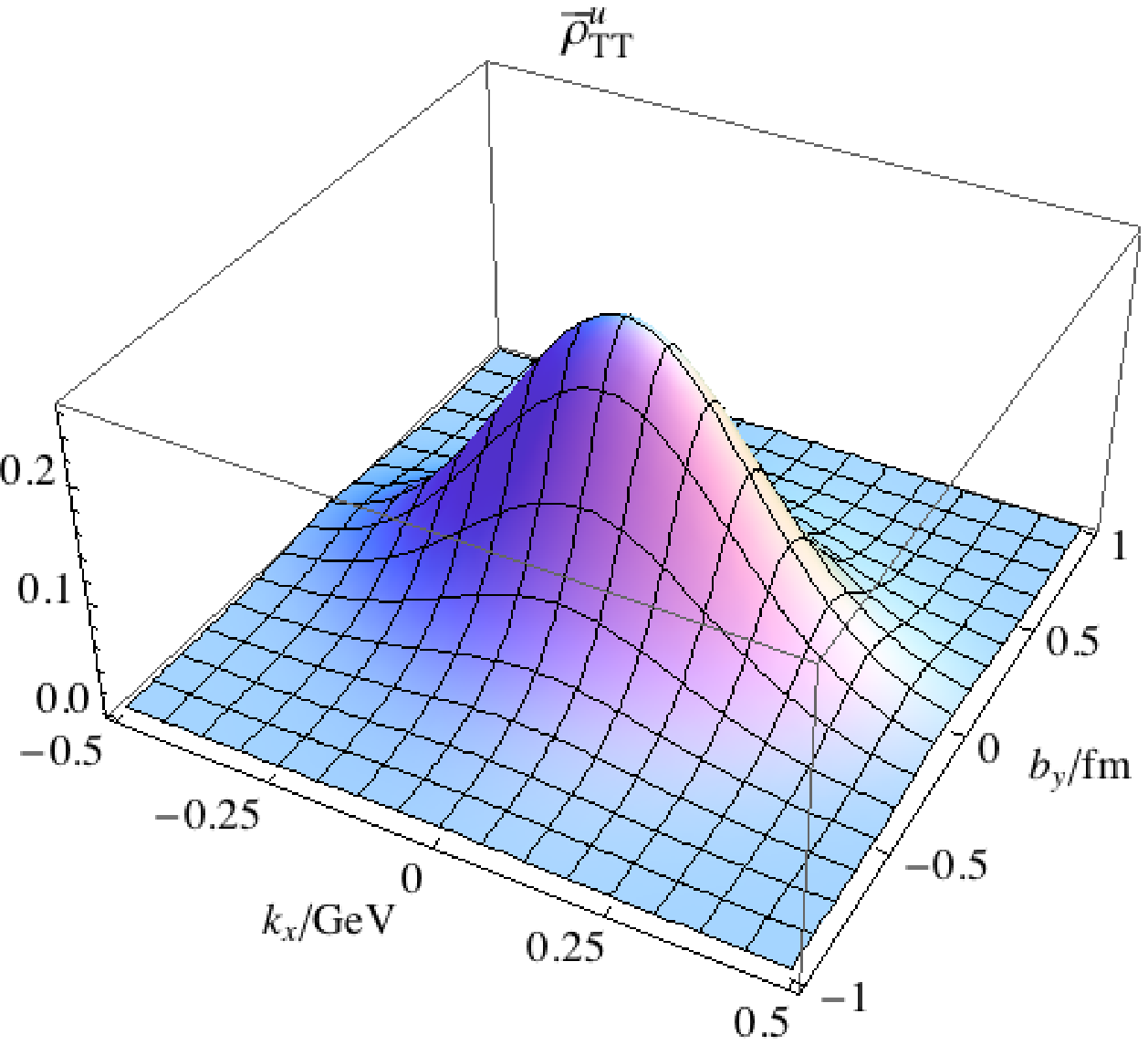}\\
\includegraphics[width=0.25\textwidth]{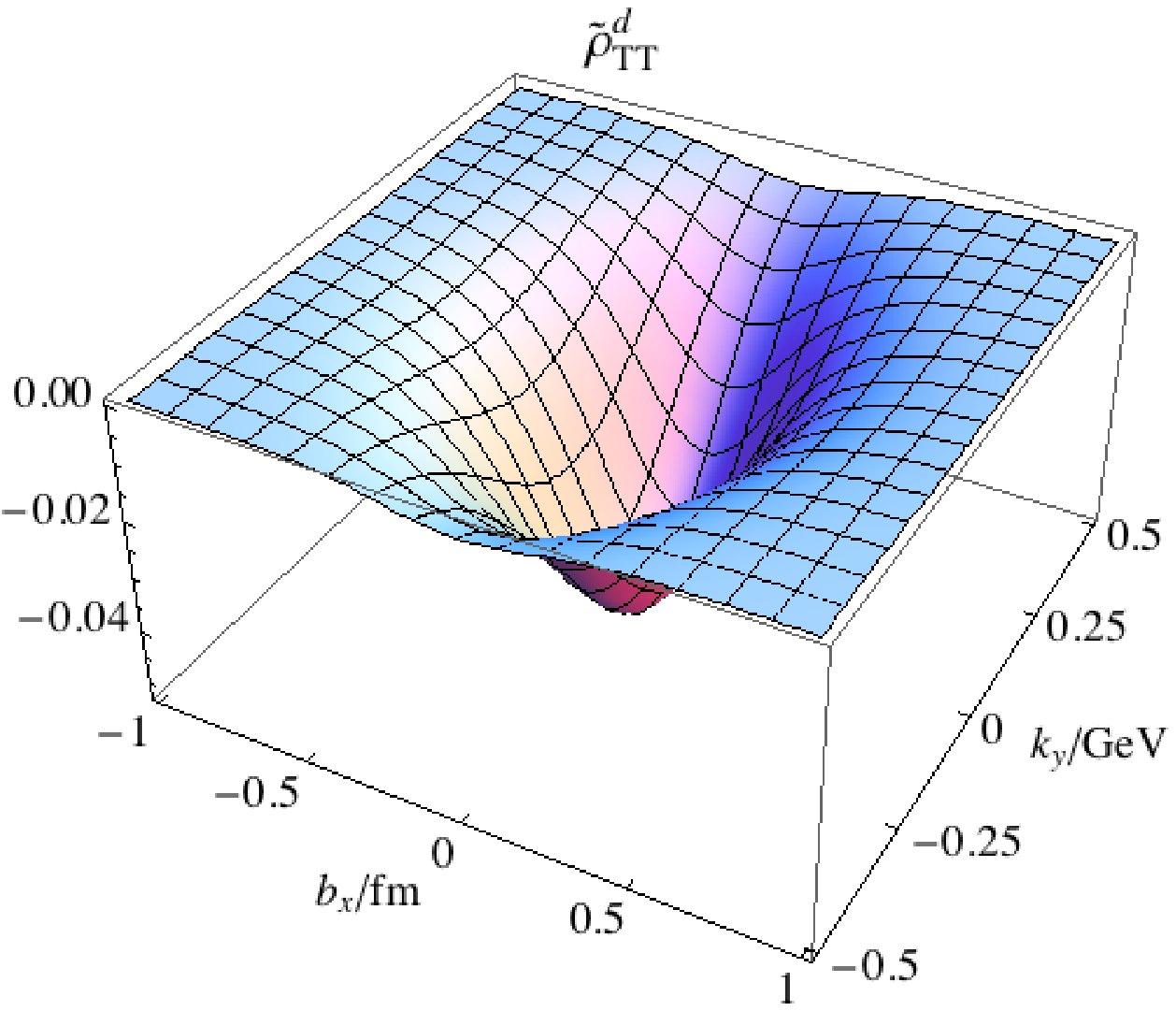}
\includegraphics[width=0.25\textwidth]{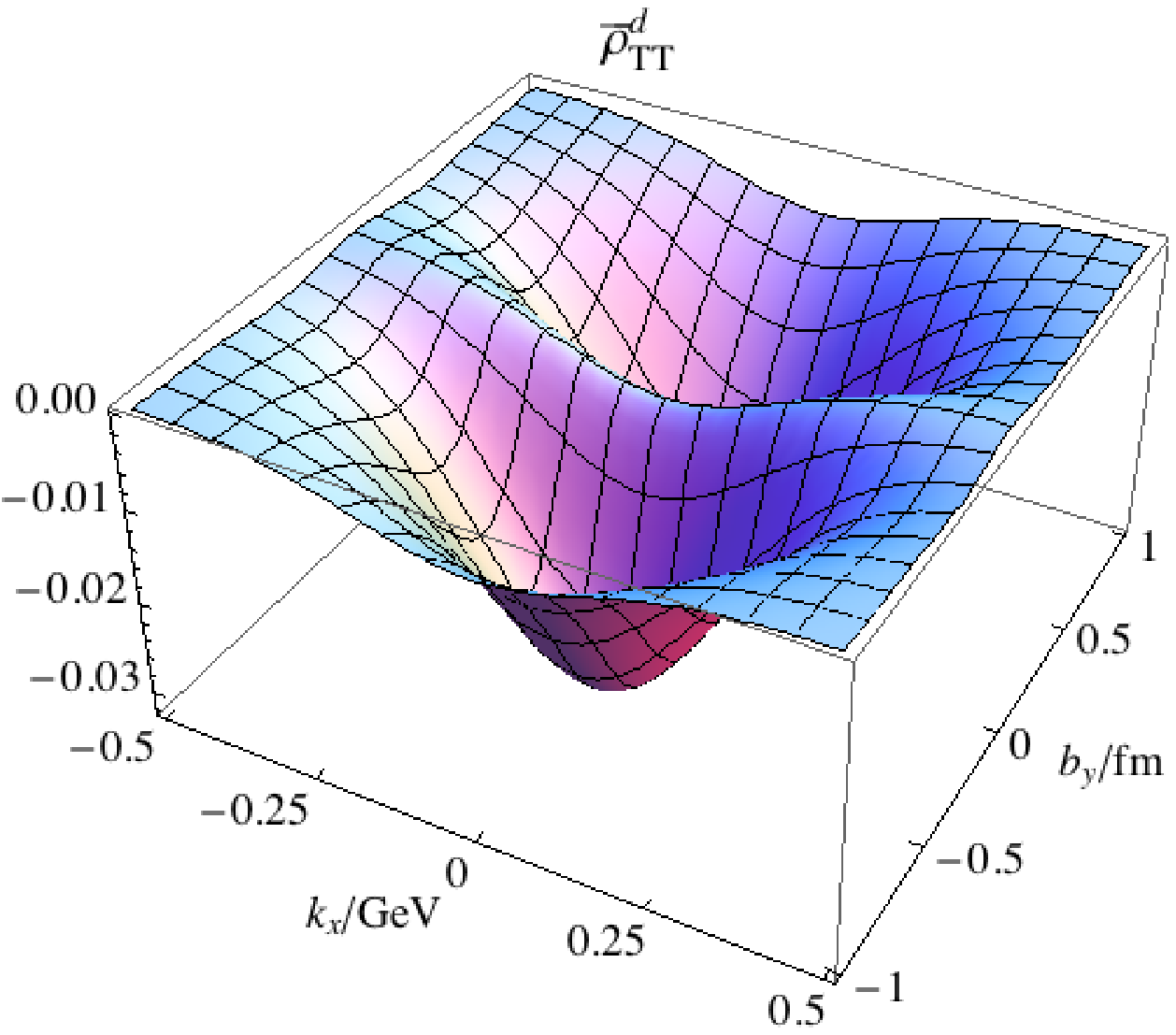}
\caption{(Color online). Transverse mixing distributions $\tilde{\rho}_{_\mathrm{TT}}$ (left) and $\bar{\rho}_{_\mathrm{TT}}$ (right) for $u$ quark (upper) and $d$ quark (lower) with quark polarized along the $x$-direction.\label{rhott2}}
\end{figure}
\begin{figure}
\includegraphics[width=0.23\textwidth]{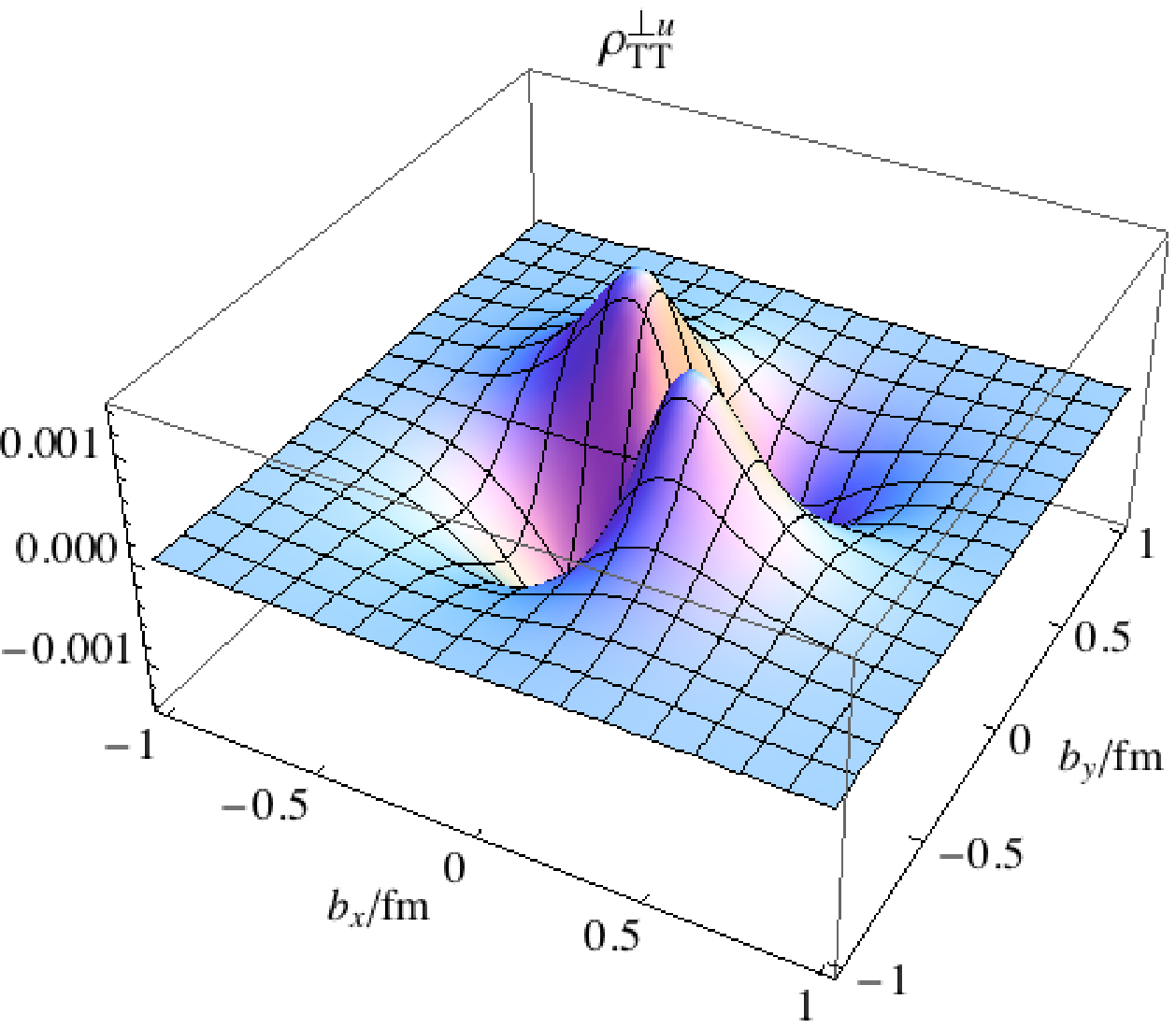}
\includegraphics[width=0.23\textwidth]{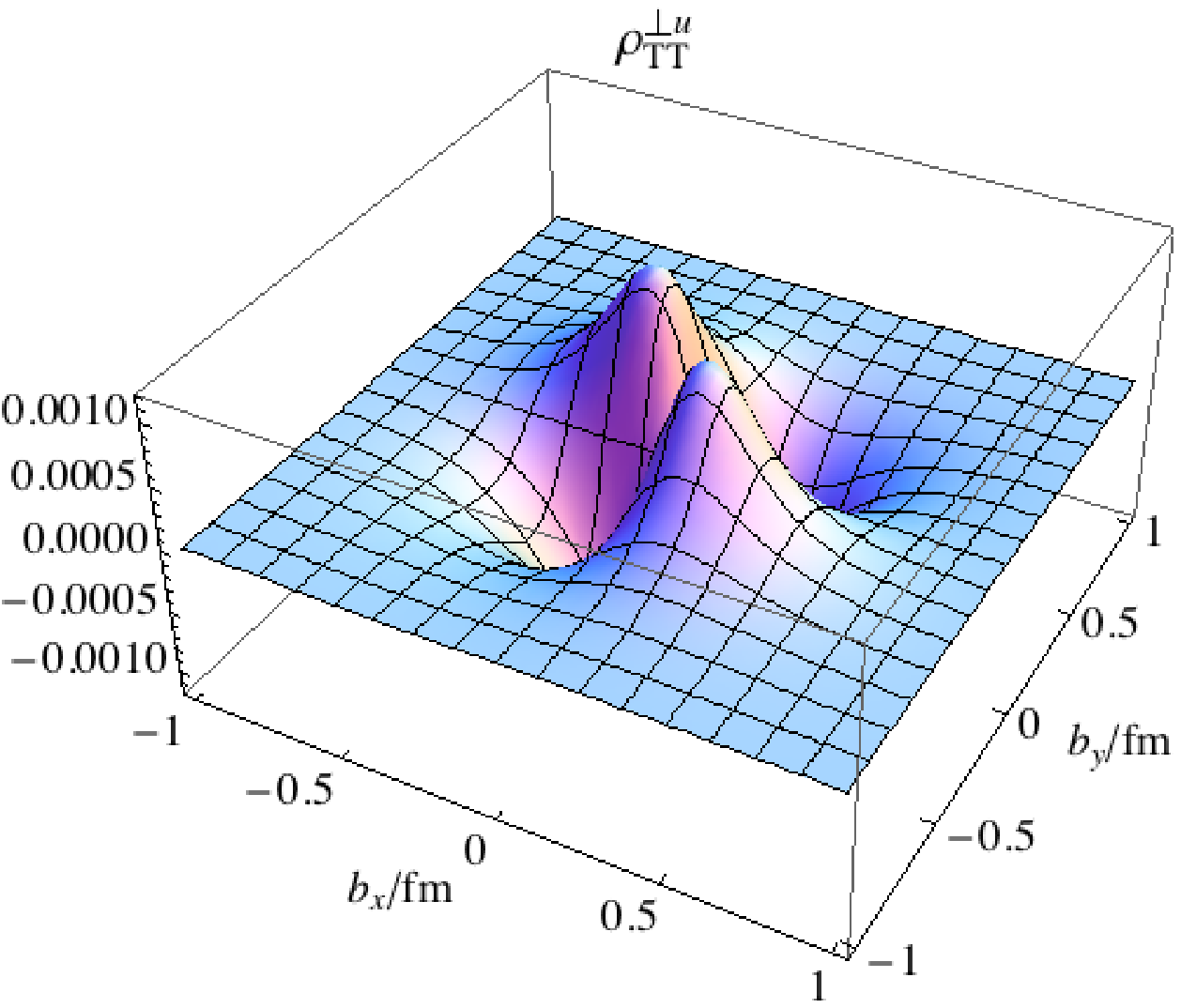}
\includegraphics[width=0.23\textwidth]{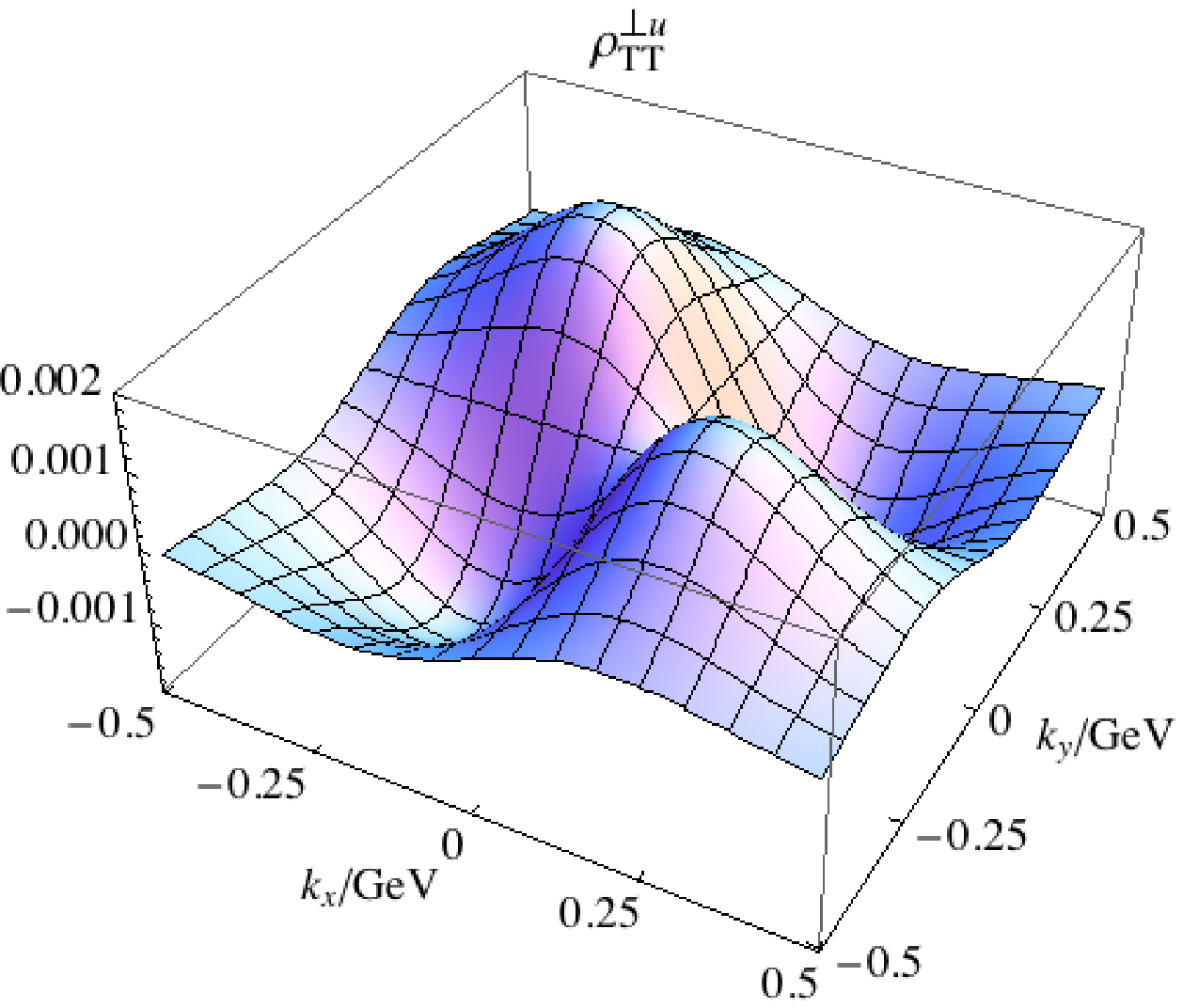}
\includegraphics[width=0.23\textwidth]{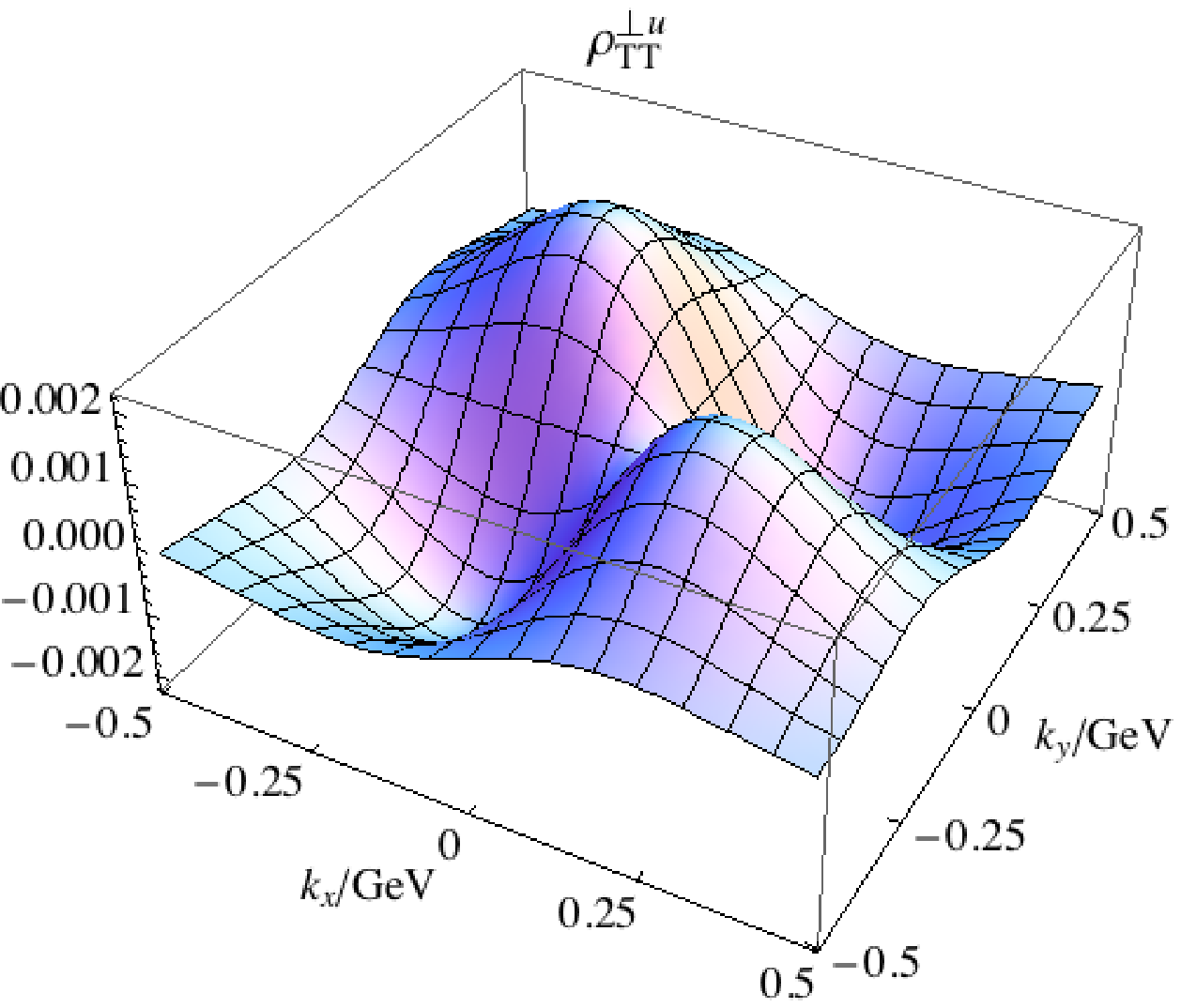}
\includegraphics[width=0.23\textwidth]{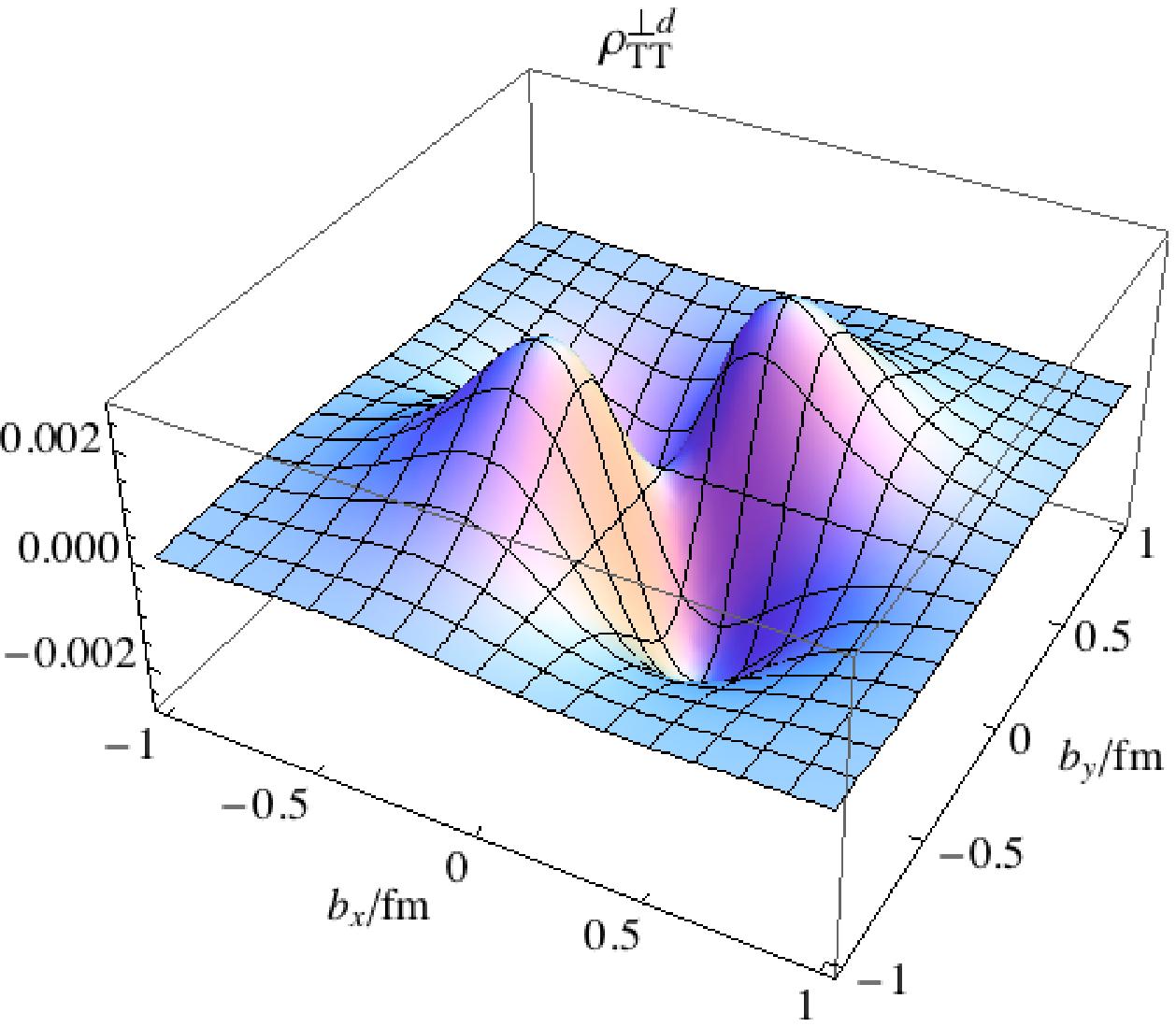}
\includegraphics[width=0.23\textwidth]{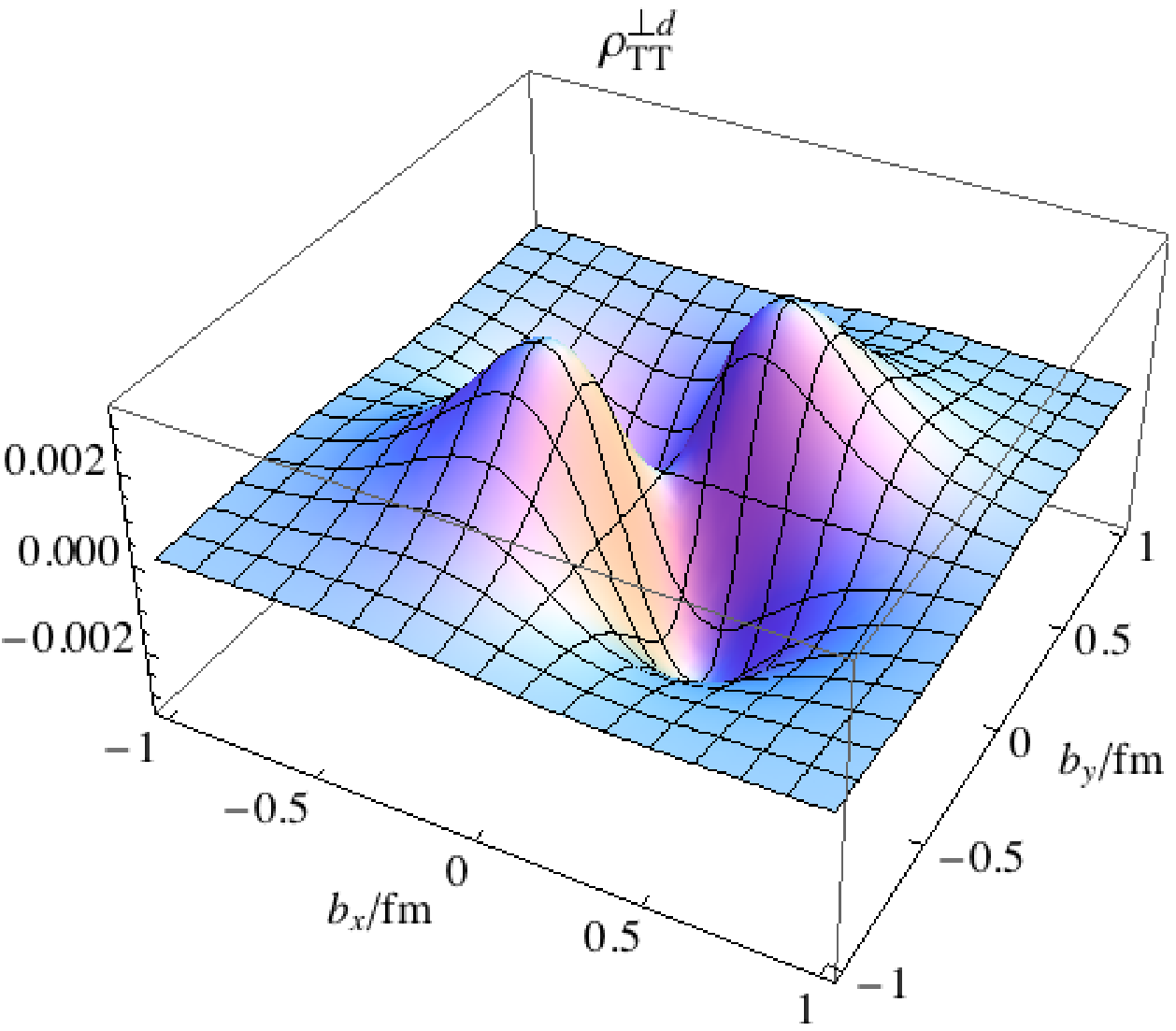}
\includegraphics[width=0.23\textwidth]{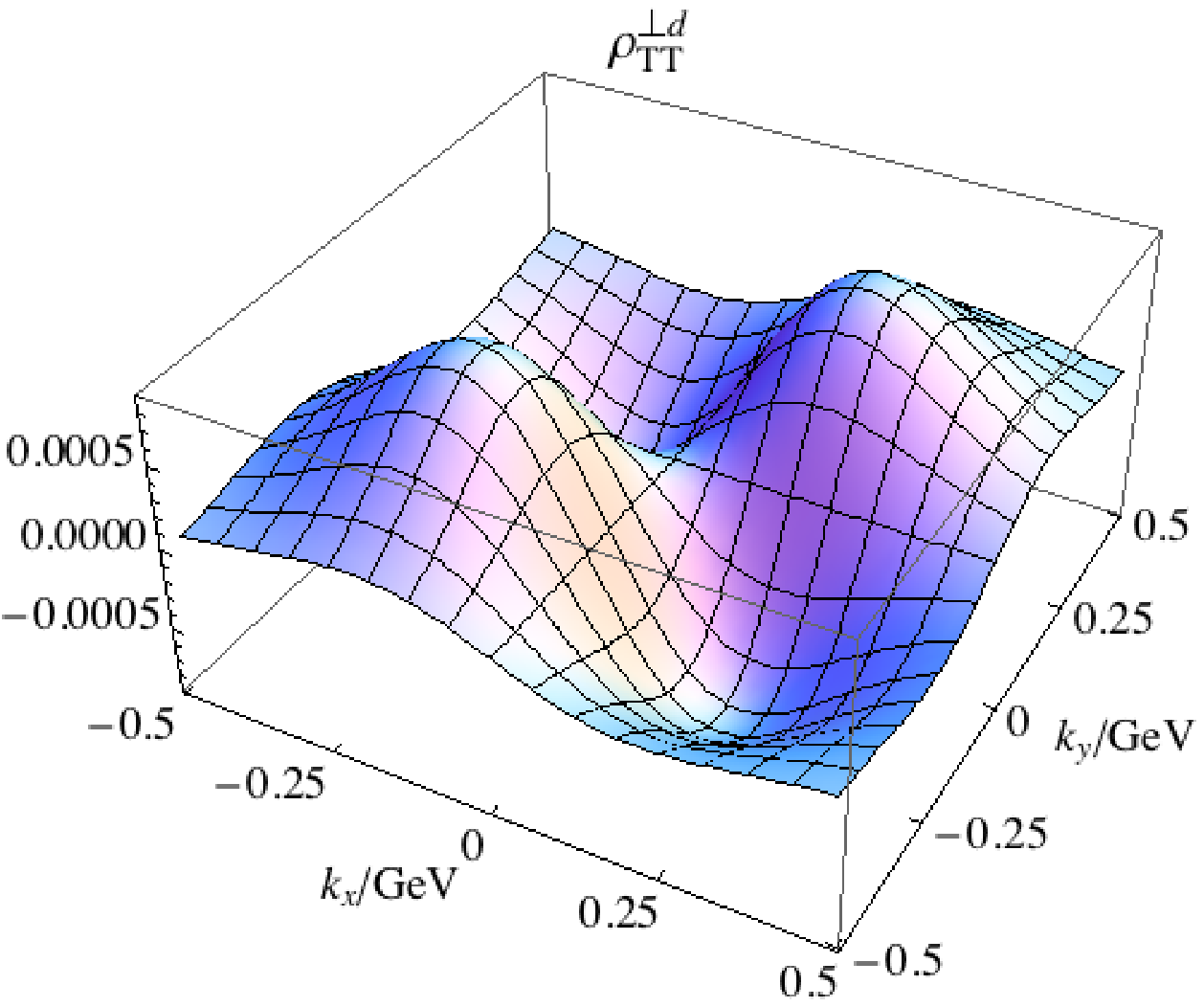}
\includegraphics[width=0.23\textwidth]{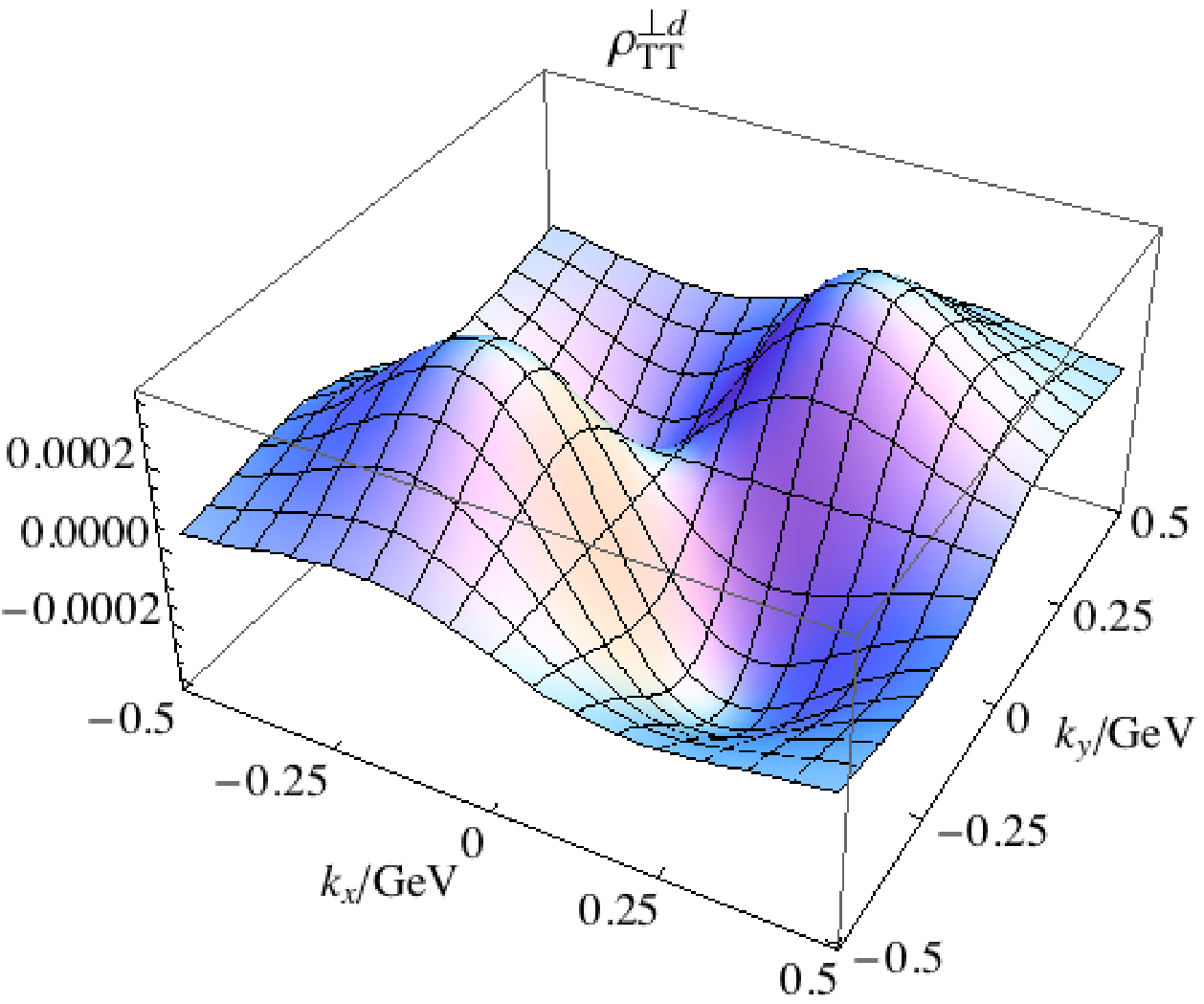}
\caption{(Color online). Pretzelous Wigner distributions $\rho_{_\mathrm{TT}}^\perp$ for $u$ quark (upper) and $d$ quark (lower). The first column are the distributions in transverse coordinate space with fixed transverse momentum $\bm{k}_\perp=0.3\,\textrm{GeV}\,\hat{\bm{e}}_x$ parallel to the quark polarization, and the second column are those with fixed transverse momentum $\bm{k}_\perp=0.3\,\textrm{GeV}\,\hat{\bm{e}}_y$ parallel to the proton polarization. The third column are the distributions in transverse momentrum space with fixed transverse coordinate $\bm{b}_\perp=0.4\,\textrm{fm}\,\hat{\bm{e}}_x$ parallel to the quark polarization, and the fourth column are those with fixed transverse coordinate $\bm{b}_\perp=0.4\,\textrm{fm}\,\hat{\bm{e}}_y$ parallel to the proton polarization. \label{rhottp}}
\end{figure}

In Fig. \ref{rhotu}, we plot the trans-unpolarized Wigner distributions $\rho_{_\mathrm{TU}}$ and mixing distributions $\bar{\rho}_{_\mathrm{TU}}$. They describe the unpolarized quark in a transverse polarized proton, and thus represent the correlation of quark transverse distributions and proton transverse spin. A priviledged direction in the transverse plane is determined by the proton polarization which is referred to as the $x$-direction.

At the TMD limit, the trans-unpolarized Wigner distribution corresponds to a naive T-odd distribution, the Sivers function $f_{1T}^\perp$. At the IPD limit, it is related to the IPDs $H$ and $E$ together with other distributions. However, these two limits select two different part of the Wigner distribution. The Sivers TMD corresponds to the T-odd part, while the IPDs $H$ and $E$ only depends on the T-even part. Thus there is no general relation between them if no further assumptions are applied to connect the T-odd and T-even parts. However some relation between the Sivers function $f_{1T}^\perp$ and the GPD $E$ is suggected by some models calculations~\cite{Burkardt:2003je,Bacchetta:2011gx} and also implicated in some experiment~\cite{Allada:2013nsw}. Thus this kind of connection is not impossible. In our calculation, since the effect from the Wilson line is neglected, vanishing result is obtained at the TMD limit. Therefore the nontrivial gauge link plays an important role in establishing the possible relation between the Sivers TMD and the GPDs.

Similar to the situation in the unpol-transverse Wigner distribution, the trans-unpolarized Wigner distribution is vanishing if quark transverse coordinate is parallel to the proton polarization. Together with the behavior of the unpol-transverse Wigner distribution, quark transverse coordinate has no correlations to either quark parallel transverse spin or proton parallel transverse spin. Besides, its behavior has no dependence on the direction of quark transverse momentum, but a nontrivial Wilson line may change this property.

In Figs. \ref{rhotl}, we plot the trans-longitudinal Wigner distributions $\rho_{_\mathrm{TL}}$ and mixing distributions $\bar{\rho}_{_\mathrm{TL}}$. They describe the longitudinal polarized quark in a transverse polarized proton. At the TMD limit, the trans-longitudinal Wigner distribution will reduce to the other worm-gear function, the trans-helicity $g_{1L}$. At the IPD limit, it is related to the IPDs $\tilde{H}$ and $\tilde{E}$ together with other distributions. Both of the trans-helicity TMD $g_{1L}$ and the IPDs $\tilde{H}$ and $\tilde{E}$ only depend on the T-even part of the trans-longitudinal distribution. Thus it is possible to establish some relations between them. Besides, the T-odd part, which is neglected in our model calculation, will provide us information beyond TMDs and IPDs.

Similar to the situation of the longi-transverse Wigner distribution, the isotropy in the transverse plane is explicitly violated by the transverse polarization, which we refer to as the $x$-direction, in trans-longitudinal Wigner distributions. Determined by the combination of the polarization configurations, the phase-space distribution is zero when quark transverse momentum is perpendicular to the proton spin, but a nontrivial Wilson line may lead to non-zero values. This behavior reflects a strong correlation to the direction of quark transverse momentum. Compared to the longi-transverse Wigner distribution, the correlation to the direction of quark transverse coordinate is not weak in this case.

In Figs. \ref{rhott}-\ref{rhott2}, we plot the transverse Wigner distributions $\rho_{_\mathrm{TT}}$ and mixing distributions $\tilde{\rho}_{_\mathrm{TT}}$ and $\bar{\rho}_{_\mathrm{TT}}$. They describe the phase-space distribution in the case that both the quark and the proton are transverse polarized. Due to the two degrees of freedom of the choice of the transverse polarization direction, there are two independent combinations of the transverse polarizations of the quark and the proton. For the transverse Wigner distributions, the quark and proton are parallel polarized referred to as the $x$-direction. The situation that the quark and proton are transverse polarized along two orthogonal directions is described by the pretzelous Wigner distributions, as plotted in Fig. \ref{rhottp}.

The transverse Wigner distributions will reduce to the transversity distributions when the transverse phase-space is integrated. Unlike the longitudinal Wigner distributions where the quark helicity concentrates on the center of the phase-space, we find in the model that the maximum of $d$ quark transversity distribution in the phase-space is not on the center point. Although this behavior is not forbidden by any physical principles, it is beyond our intuition that usually adopted in parametrizations~\cite{Anselmino:2007fs}. Though this is a model dependent result, it provides us the possibility that may be taken into account in realistic analyses.

For the pretzelous Wigner distribution in Fig. \ref{rhottp}, a quadrupole structure is observed. This behavior is essentially determined by the spin structures. In this model, we find opposite correlation in the phase-space for $u$ quark and $d$ quark in this situation.

\section{Summary}

In this paper, we investigate quark Wigner distributions in a lignt-cone spectator model. The Wigner distribution, as a phase-space distributions, contains full one-parton information in the proton. All TMDs and IPDs can be obtained from the Wigner distributions at certain limits. Therefore, it is a possible bridge to build the relation between the TMDs and GPDs. Besides, the Wigner distributions do contain information that cannot in general be extracted from the TMDs and GPDs. Thus the study on Wigner distributions will provide us new knowledge on understanding the nucleon structures.

We perform the calculations of all twist-two quark Wigner distributions in a spectator model, which has been applied to study many physical observables in high energy scattering experiments. In our calculation, we include both the scalar and the axial vector spectators to have flavor separation. For the Wilson line, we take a crude truncation in this paper, and this means the neglection of the naive T-odd part in Wigner distributions. Though the calculations are performed in a simplistic model, we can still find some general properties of quark Wigner distributions from the results. Thus the plots displayed here can be view as qualitative results. More realistic analyses in QCD is required in the future.

The dipole and quadrupole structures of some Wigner distributions and mixing distributions are determined by the spin structures of the quark and proton, and should be general properties. Although the Wilson line is neglected in this study, some possible effects from nontrivial Wilson line are discussed. Some Wigner distributions that have no correponding TMDs or IPDs are also interesting, such as the unpol-longitudinal distribution $\rho_{_\mathrm{UL}}$ and the longi-unpolarized distribution $\rho_{_\mathrm{LU}}$. They are related to the issues on spin-orbit correlation and orbit angular momentum, and variant results can be obtained from different models. Therefore more careful investigation on these Wigner distributions is important to clarify the physical pictures. In addition, we have some discussion on the possibility to establish the relations between TMDs and GPDs. Though no general relations have been found, it is suggested by some model calculations and implicated by some experiments to relate the Sivers function with the GPD $E$. Since they are related to the same Wigner distribution but different parts, it is an opportunity to find the relation at Wigner distribution level. Therefore the investigation on Wigner distributions can improve our understandings on the nucleon structure, and more realistic analyses are deserved.

\acknowledgments{This work is supported by National Natural Science Foundation of China (Grants No.~11035003 and No.~11120101004).}

\appendix

\section{Light-cone wave function overlap representation for the Wigner distribution}

In this section, we take the $\rho^{[\gamma^+]}(\bm{b}_\perp,\bm{k}_\perp,x,\uparrow)$ as an example to demonstrate how the twist-two Wigner distributions are expressed in terms of the overlap of light-cone wave functions in the spectator model.

Interpolating the Wigner operator in \eqref{wigneru} between two proton state with a transverse momentum $\bm{\Delta}_\perp$ transferred, we can obtain the Wigner distribution with \eqref{wignerdistr} as
{\allowdisplaybreaks
\begin{align*}
&\rho^{[\gamma^+]}(\bm{b}_\perp,\bm{k}_\perp,x,\uparrow)\\
=&\int\frac{d^2\bm{\Delta}_\perp}{(2\pi)^2}\sum_{\sigma',s'}\int\frac{dx_q'd^2\bm{k}'_{q\perp}}{2\sqrt{x'_q}(2\pi)^3}\int\frac{dx_D'd^2\bm{k}'_{D\perp}}{2\sqrt{x'_D}(2\pi)^3}16\pi^3\delta(1-x_q'-x_D')\delta^{(2)}(\bm{k}'_{q\perp}+\bm{k}'_{D\perp})\\
&\sum_{\sigma,s}\int\frac{dx_qd^2\bm{k}_{q\perp}}{2\sqrt{x_q}(2\pi)^3}\int\frac{dx_Dd^2\bm{k}_{D\perp}}{2\sqrt{x_D}(2\pi)^3}16\pi^3\delta(1-x_q-x_D)\delta^{(2)}(\bm{k}_{q\perp}+\bm{k}_{D\perp})\\
&\frac{1}{2}\int\frac{d\ell'^+d^2\bm{\ell}'_\perp}{(2\pi)^3}\int\frac{d\ell^+d^2\bm{\ell}_\perp}{(2\pi)^3}e^{-i(\bm{\ell}'_\perp-\bm{\ell}_\perp)\cdot\bm{b}_\perp}\delta(k^+-\frac{\ell'^++\ell^+}{2})\delta^{(2)}(\bm{k}_\perp-\frac{\bm{\ell}'_\perp+\bm{\ell}_\perp}{2})\\
&(2P^+)^2\sqrt{x'_qx'_D}\sqrt{x_qx_D}\psi_{\sigma's'}^{\uparrow *}(x'_q,\bm{k}'_{q\perp})\psi_{\sigma s}^{\uparrow}(x_q,\bm{k}_{q\perp})\langle0|a_{s'}(x'_DP^+,\bm{k}'_{D\perp}+x'_D\frac{\bm{\Delta}_\perp}{2})b_{\sigma'}(x'_qP^+,\bm{k}'_{q\perp}+x'_q\frac{\bm{\Delta}_\perp}{2})\\
&\sum_\lambda b_\lambda^\dagger(\ell'^+,\bm{\ell}'_\perp)b_\lambda(\ell^+,\bm{\ell}_\perp)b_\sigma^\dagger(x_qP^+,\bm{k}_{q\perp}-x_q\frac{\bm{\Delta}_\perp}{2})a_s^\dagger(x_DP^+,\bm{k}_{D\perp}-x_D\frac{\bm{\Delta}_\perp}{2})|0\rangle\\
=&\sum_{\sigma',s'}\sum_{\sigma,s}\sum_\lambda\frac{1}{2}\int\frac{d^2\bm{\Delta}_\perp}{(2\pi)^2}\int\frac{dx_q'd^2\bm{k}'_{q\perp}}{(2\pi)^3}\int\frac{dx_qd^2\bm{k}_{q\perp}}{(2\pi)^3}\int\frac{d\ell^+d^2\bm{\ell}'_\perp}{(2\pi)^3}\int\frac{d\ell^+d^2\bm{\ell}_\perp}{(2\pi)^3}\\
&(P^+)^2\psi_{\sigma's'}^{\uparrow *}(x'_q,\bm{k}'_{q\perp})\psi_{\sigma s}^{\uparrow}(x_q,\bm{k}_{q\perp})e^{-i(\bm{\ell}'_\perp-\bm{\ell}_\perp)\cdot\bm{b}_\perp}\delta(k^+-\frac{\ell'^++\ell^+}{2})\delta^{(2)}(\bm{k}_\perp-\frac{\bm{\ell}'_\perp+\bm{\ell}_\perp}{2})\\
&(2\pi)^3\delta(x'_qP^+-x_qP^+)\delta^{(2)}(-\bm{k}'_{q\perp}+\bm{k}_{q\perp}+(1-\frac{x_q'+x_q}{2})\bm{\Delta}_\perp)\delta_{s's}\\
&(2\pi)^3\delta(\ell'^+-x_q'P^+)\delta^{(2)}(\bm{\ell}'_\perp-\bm{k}'_{q\perp}-x'_q\frac{\bm{\Delta}_\perp}{2})\delta_{\sigma'\lambda}
(2\pi)^3\delta(\ell^+-x_qP^+)\delta^{(2)}(\bm{\ell}_\perp-\bm{k}_{q\perp}-x_q\frac{\bm{\Delta}_\perp}{2})\delta_{\sigma\lambda}\\
=&\sum_{\lambda,s}\frac{1}{2}\int\frac{d^2\bm{\Delta}_\perp}{(2\pi)^2}\int\frac{dx'_qd^2\bm{k}'_{q\perp}}{(2\pi)^3}\int\frac{dx_qd^2\bm{k}_{q\perp}}{(2\pi)^3}(P^+)^2\psi_{\lambda s}^{\uparrow *}(x'_q,\bm{k}'_{q\perp})\psi_{\lambda s}^{\uparrow}(x_q,\bm{k}_{q\perp})\\
&e^{-i(\bm{k}'_{q\perp}-\bm{k}_{q\perp}+\frac{x_q'+x_q}{2}\bm{\Delta}_\perp)\cdot\bm{b}_\perp}\delta(k^+-\frac{x_q'+x_q}{2}P^+)\delta^{(2)}(\bm{k}_\perp-\frac{\bm{k}'_{q\perp}+\bm{k}_{q\perp}}{2}-\frac{x_z'-x_q}{2}\bm{\Delta}_\perp)\\
&(2\pi)^3(x_x'P^+-x_qP^+)\delta(-\bm{k}'_{q\perp}+\bm{k}_{q\perp}+(1-\frac{x_q'+x_q}{2})\bm{\Delta}_\perp)\\
=&\sum_{\lambda,s}\int\frac{d^2\bm{\Delta}_\perp}{(2\pi)^2}\frac{e^{-i\bm{\Delta}_\perp\cdot\bm{b}_\perp}}{16\pi^3}\psi_{\lambda s}^{\uparrow *}(x,\bm{k}_\perp+(1-x)\frac{\bm{\Delta}_\perp}{2})\psi_{\lambda s}^{\uparrow}(x,\bm{k}_\perp-(1-x)\frac{\bm{\Delta}_\perp}{2}).
\end{align*}
}
This result is consistant with the Drell--Yan--West assignment~\cite{Drell:1969km} in the form factor calculations, as demonstrated in the appendix in~\cite{Xiao:2003wf}. For the other twist-two Wigner distributions, similar expressions can be derived but with different combinations of the helicity states. This approach can also be applied to any $N$-particle Fock state.


\end{document}